\documentclass[journal]{IEEEtran}
\IEEEoverridecommandlockouts

\usepackage{graphicx}

\usepackage[font=small]{caption}
\usepackage{comment}
\usepackage[normalem]{ulem}
\usepackage{makecell}
\usepackage{multirow}
\usepackage{caption}
\usepackage{subcaption}
\usepackage{diagbox}
\usepackage{amsmath,amssymb,amsfonts}
\usepackage{amsthm}
\usepackage{soul}
\usepackage{xspace}
\usepackage[most]{tcolorbox}
\usepackage[utf8]{inputenc}
\usepackage[T1]{fontenc}
\usepackage[english]{babel}
\usepackage{balance}

\AtBeginDocument{%
  }

\theoremstyle{plain}
\newtheorem{observation}{Finding}
\newtheorem{theorem}{Theorem}

\usepackage[acronyms,nonumberlist,nopostdot,nomain,nogroupskip,acronymlists={hidden}]{glossaries}
\newglossary[algh]{hidden}{acrh}{acnh}{Hidden Acronyms}
\newacronym{sm}{SM}{Stateful Migration}
\newacronym{sdl}{SDL}{Shared Data Layer}
\newacronym{mose}{MOSE}{Migration Orchestration framework for microServices at the Edge}
\newacronym{pam}{PAM}{Processing-aware Migration}
\newacronym{rbac}{RBAC}{Role-based access control}
\newacronym{ci/cd}{CI/CD}{continuous integration and continuous deployment}
\newacronym{s2i}{S2I}{Source-To-Image}
\newacronym{sa}{SA}{Stand Alone}
\newacronym{sos}{SOS}{Special Order Set}
\newacronym[type=hidden]{open ran}{Open RAN}{Open \gls{ran}}
\newacronym{bts}{BTS}{Base Transceiver Station}
\newacronym{sc-fdma}{SC-FDMA}{Single Carrier-Frequency Division Multiple Access}
\newacronym{papr}{PAPR}{Peak-to-Average Power Ratio}
\newacronym{idrac}{iDRAC}{Integrated Dell Remote Access Controller}
\newacronym{wiot}{WIoT}{Institute for The Wireless Internet of Things}
\newacronym{saas}{SaaS}{Software-as-a-Service}
\newacronym{iaas}{IaaS}{Infrastructure-as-a-Service}
\newacronym{paas}{PaaS}{Platform-as-a-Service}
\newacronym{mrn}{MRN}{Mobile Radio Networks}
\newacronym{ran}{RAN}{Radio Access Network}
\newacronym{o-ran}{O-RAN}{}
\newacronym{cnn}{CNN}{Convolutional Neural Network}
\newacronym{sdn}{SDN}{Software Defined Networking}
\newacronym{nfv}{NFV}{Network Functions Virtualization}
\newacronym{appmgr}{APPMGR}{App Manager}
\newacronym{near-rt}{Near-RT}{Near-real-time}
\newacronym{non-rt}{Non-RT}{Non-real-time}
\newacronym{sla}{SLA}{Service Level Agreement}
\newacronym{3gpp}{3GPP}{3rd Generation Partnership Project}
\newacronym{cu}{CU}{Central Unit}
\newacronym{du}{DU}{Distributed Unit}
\newacronym{2g}{2G}{Second-Generation}
\newacronym{3g}{3G}{Third-Generation}
\newacronym{4g}{4G}{Fourth-generation}
\newacronym{5g}{5G}{Fifth-Generation}
\newacronym{6g}{6G}{Sixth-Generation}
\newacronym{5gc}{5GC}{5G Core}
\newacronym{adc}{ADC}{Analog to Digital Converter}
\newacronym{aerpaw}{AERPAW}{Aerial Experimentation and Research Platform for Advanced Wireless}
\newacronym{ai}{AI}{Artificial Intelligence}
\newacronym{aimd}{AIMD}{Additive Increase Multiplicative Decrease}
\newacronym{am}{AM}{Acknowledged Mode}
\newacronym{amc}{AMC}{Adaptive Modulation and Coding}
\newacronym{ilp}{ILP}{Integer Linear Programming}
\newacronym{milp}{MILP}{Mixed Integer Linear Programming}
\newacronym{amf}{AMF}{Access and Mobility Management Function}
\newacronym{aops}{AOPS}{Adaptive Order Prediction Scheduling}
\newacronym{api}{API}{Application Programming Interface}
\newacronym{apn}{APN}{Access Point Name}
\newacronym{ap}{AP}{Application Protocol}
\newacronym{aqm}{AQM}{Active Queue Management}
\newacronym{asn1}{ASN.1}{Abstract Syntax Notation One}
\newacronym{ausf}{AUSF}{Authentication Server Function}
\newacronym{avc}{AVC}{Advanced Video Coding}
\newacronym{awgn}{AGWN}{Additive White Gaussian Noise}
\newacronym{balia}{BALIA}{Balanced Link Adaptation Algorithm}
\newacronym{bbu}{BBU}{Base Band Unit}
\newacronym{bdp}{BDP}{Bandwidth-Delay Product}
\newacronym{ber}{BER}{Bit Error Rate}
\newacronym{bf}{BF}{Beamforming}
\newacronym{bler}{BLER}{Block Error Rate}
\newacronym{brr}{BRR}{Bayesian Ridge Regressor}
\newacronym{bs}{BS}{Base Station}
\newacronym{bsr}{BSR}{Buffer Status Report}
\newacronym{bss}{BSS}{Business Support System}
\newacronym{ca}{CA}{Carrier Aggregation}
\newacronym{caas}{CaaS}{Connectivity-as-a-Service}
\newacronym{cb}{CB}{Code Block}
\newacronym{cc}{CC}{Congestion Control}
\newacronym{compc}{CC}{Component Carrier}
\newacronym{ccid}{CCID}{Congestion Control ID}
\newacronym{cco}{CC}{Carrier Component}
\newacronym{cdd}{CDD}{Cyclic Delay Diversity}
\newacronym{cdf}{CDF}{Cumulative Distribution Function}
\newacronym{cdn}{CDN}{Content Distribution Network}
\newacronym{cn}{CN}{Core Network}
\newacronym{codel}{CoDel}{Controlled Delay Management}
\newacronym{comac}{COMAC}{Converged Multi-Access and Core}
\newacronym{cord}{CORD}{Central Office Re-architected as a Datacenter}
\newacronym{cornet}{CORNET}{COgnitive Radio NETwork}
\newacronym{cosmos}{COSMOS}{Cloud Enhanced Open Software Defined Mobile Wireless Testbed for City-Scale Deployment}
\newacronym{cots}{COTS}{Commercial Off-the-Shelf}
\newacronym{cp}{CP}{Control Plane}
\newacronym{cpu}{CPU}{Central Processing Unit}
\newacronym{cqi}{CQI}{Channel Quality Information}
\newacronym{cr}{CR}{Cognitive Radio}
\newacronym{cql}{CQL}{Conservative Q-Learning}
\newacronym{cran}{CRAN}{Cloud \gls{ran}}
\newacronym{crs}{CRS}{Cell Reference Signal}
\newacronym{csi}{CSI}{Channel State Information}
\newacronym{csirs}{CSI-RS}{Channel State Information - Reference Signal}
\newacronym{cucp}{CU-CP}{Centralized Unit - Control Plane}
\newacronym{cuup}{CU-UP}{Centralized Unit - User Plane}
\newacronym{d2tcp}{D$^2$TCP}{Deadline-aware Data center TCP}
\newacronym{d2d}{D2D}{Device to Device}
\newacronym{d3}{D$^3$}{Deadline-Driven Delivery} 
\newacronym{dac}{DAC}{Digital to Analog Converter}
\newacronym{dag}{DAG}{Directed Acyclic Graph}
\newacronym{das}{DAS}{Distributed Antenna System}
\newacronym{dash}{DASH}{Dynamic Adaptive Streaming over HTTP}
\newacronym{dc}{DC}{Dual Connectivity}
\newacronym{dccp}{DCCP}{Datagram Congestion Control Protocol}
\newacronym{dce}{DCE}{Direct Code Execution}
\newacronym{dci}{DCI}{Downlink Control Information}
\newacronym{dctcp}{DCTCP}{Data Center TCP}
\newacronym{dl}{DL}{Downlink}
\newacronym{dmr}{DMR}{Deadline Miss Ratio}
\newacronym{dmrs}{DMRS}{DeModulation Reference Signal}
\newacronym{drl}{DRL}{Deep Reinforcement Learning}
\newacronym{drlcc}{DRL-CC}{Deep Reinforcement Learning Congestion Control}
\newacronym{drb}{DRB}{Data Radio Bearer}
\newacronym{drs}{DRS}{Discovery Reference Signal}
\newacronym{dnn}{DNN}{Deep Neural Network}
\newacronym{dqn}{DQN}{Deep Q-Network}
\newacronym{e2e}{E2E}{end-to-end}
\newacronym{e2ap}{E2AP}{E2 Application Protocol}
\newacronym{e2sm}{E2SM}{E2 Service Model}
\newacronym{ecaas}{ECaaS}{Edge-Cloud-as-a-Service}
\newacronym{ecn}{ECN}{Explicit Congestion Notification}
\newacronym{edc}{EDC}{Edge Data Center}
\newacronym{edf}{EDF}{Earliest Deadline First}
\newacronym{embb}{eMBB}{Enhanced Mobile Broadband}
\newacronym{empower}{EMPOWER}{EMpowering transatlantic PlatfOrms for advanced WirEless Research}
\newacronym{enb}{eNB}{evolved Node Base}
\newacronym{endc}{EN-DC}{E-UTRAN-NR Dual Connectivity}
\newacronym{epc}{EPC}{Evolved Packet Core}
\newacronym{eps}{EPS}{Evolved Packet System}
\newacronym{es}{ES}{Edge Server}
\newacronym{etl}{ETL}{Extract, Transform and Load}
\newacronym{etsi}{ETSI}{European Telecommunications Standards Institute}
\newacronym[firstplural=Estimated Times of Arrival (ETAs)]{eta}{ETA}{Estimated Time of Arrival}
\newacronym{eutran}{E-UTRAN}{Evolved Universal Terrestrial Access Network}
\newacronym{faas}{FaaS}{Function-as-a-Service}
\newacronym{fapi}{FAPI}{Functional Application Platform Interface}
\newacronym{fcaps}{FCAPS}{Fault, Configuration, Accounting, Performance and Security}
\newacronym{fdd}{FDD}{Frequency Division Duplexing}
\newacronym{fdm}{FDM}{Frequency Division Multiplexing}
\newacronym{fdma}{FDMA}{Frequency Division Multiple Access}
\newacronym{fed4fire}{FED4FIRE+}{Federation 4 Future Internet Research and Experimentation Plus}
\newacronym{fir}{FIR}{Finite Impulse Response}
\newacronym{fit}{FIT}{Future \acrlong{iot}}
\newacronym{fpga}{FPGA}{Field Programmable Gate Array}
\newacronym{fr1}{FR1}{Frequency Range 1}
\newacronym{fr2}{FR2}{Frequency Range 2}
\newacronym{fs}{FS}{Fast Switching}
\newacronym{fscc}{FSCC}{Flow Sharing Congestion Control}
\newacronym{ftp}{FTP}{File Transfer Protocol}
\newacronym{fw}{FW}{Flow Window}
\newacronym{gbr}{GBR}{Guaranteed Bit Rate}
\newacronym{ge}{GE}{Gaussian Elimination}
\newacronym{gnb}{gNB}{Next Generation Node Base}
\newacronym{gop}{GOP}{Group of Pictures}
\newacronym{gpr}{GPR}{Gaussian Process Regressor}
\newacronym{gpu}{GPU}{Graphics Processing Unit}
\newacronym{gtp}{GTP}{GPRS Tunneling Protocol}
\newacronym{gprs}{GPRS}{General Packet Radio Service}
\newacronym{gtpc}{GTP-C}{GPRS Tunnelling Protocol Control Plane}
\newacronym{gtpu}{GTP-U}{GPRS Tunnelling Protocol User Plane}
\newacronym{gtpv2c}{GTPv2-C}{\gls{gtp} v2 - Control}
\newacronym{gw}{GW}{Gateway}
\newacronym{harq}{HARQ}{Hybrid Automatic Repeat reQuest}
\newacronym{hetnet}{HetNet}{Heterogeneous Network}
\newacronym{hh}{HH}{Hard Handover}
\newacronym{ho}{HO}{Handover}
\newacronym{hol}{HOL}{Head-of-Line}
\newacronym{hsdpa}{HSDPA}{High-Speed Downlink Packet Access}
\newacronym{hsupa}{HSUPA}{High-Speed Uplink Packet Access}
\newacronym{hqf}{HQF}{Highest-quality-first}
\newacronym{hss}{HSS}{Home Subscription Server}
\newacronym{http}{HTTP}{HyperText Transfer Protocol}
\newacronym{ia}{IA}{Initial Access}
\newacronym{iab}{IAB}{Integrated Access and Backhaul}
\newacronym{ic}{IC}{Incident Command}
\newacronym{ietf}{IETF}{Internet Engineering Task Force}
\newacronym{imsi}{IMSI}{International Mobile Subscriber Identity}
\newacronym{imt}{IMT}{International Mobile Telecommunication}
\newacronym{ims}{IMS}{\gls{ip} Multimedia Settings}
\newacronym{iot}{IoT}{Internet of Things}
\newacronym{ip}{IP}{Internet Protocol}
\newacronym{itu}{ITU}{International Telecommunication Union}
\newacronym{kpi}{KPI}{Key Performance Indicator}
\newacronym{kpm}{KPM}{Key Performance Measurement}
\newacronym{ni}{NI}{Network Interfaces}
\newacronym{kvm}{KVM}{Kernel-based Virtual Machine}
\newacronym{los}{LOS}{Line-of-Sight}
\newacronym{ldc}{LDC}{Local Data Center}
\newacronym{lsm}{LSM}{Link-to-System Mapping}
\newacronym{lstm}{LSTM}{Long Short Term Memory}
\newacronym{lte}{LTE}{Long Term Evolution}
\newacronym{lxc}{LXC}{Linux Container}
\newacronym{m2m}{M2M}{Machine to Machine}
\newacronym{mac}{MAC}{Medium Access Control}
\newacronym{manet}{MANET}{Mobile Ad Hoc Network}
\newacronym{mano}{MANO}{Management and Orchestration}
\newacronym{mbr}{MBR}{Maximum Bit Rate}
\newacronym{mc}{MC}{Multi-Connectivity}
\newacronym{mcc}{MCC}{Mobile Cloud Computing}
\newacronym{mchem}{MCHEM}{Massive Channel Emulator}
\newacronym{mcs}{MCS}{Modulation and Coding Scheme}
\newacronym{mdp}{MDP}{Markov Decision Process}
\newacronym{mec}{MEC}{Multi-access Edge Computing}
\newacronym{mec2}{MEC}{Mobile Edge Cloud}
\newacronym{mfc}{MFC}{Mobile Fog Computing}
\newacronym{mgen}{MGEN}{Multi-Generator}
\newacronym{mi}{MI}{Mutual Information}
\newacronym{mib}{MIB}{Master Information Block}
\newacronym{minlp}{MINLP}{Mixed Integer Non-Linear Problem}
\newacronym{miesm}{MIESM}{Mutual Information Based Effective SINR}
\newacronym{mimo}{MIMO}{Multiple Input, Multiple Output}
\newacronym{ml}{ML}{Machine Learning}
\newacronym{mlp}{MLP}{Mixed Linear Programming}
\newacronym{mlr}{MLR}{Maximum-local-rate}
\newacronym[plural=\gls{mme}s,firstplural=Mobility Management Entities (MMEs)]{mme}{MME}{Mobility Management Entity}
\newacronym{mmtc}{mMTC}{Massive Machine-Type Communications}
\newacronym{mmwave}{mmWave}{millimeter wave}
\newacronym{mpdccp}{MP-DCCP}{Multipath Datagram Congestion Control Protocol}
\newacronym{mptcp}{MPTCP}{Multipath TCP}
\newacronym{mr}{MR}{Maximum Rate}
\newacronym{mrdc}{MR-DC}{Multi \gls{rat} \gls{dc}}
\newacronym{mse}{MSE}{Mean Square Error}
\newacronym{mss}{MSS}{Maximum Segment Size}
\newacronym{mt}{MT}{Mobile Termination}
\newacronym{mtd}{MTD}{Machine-Type Device}
\newacronym{mtu}{MTU}{Maximum Transmission Unit}
\newacronym{mumimo}{MU-MIMO}{Multi-user \gls{mimo}}
\newacronym{mvno}{MVNO}{Mobile Virtual Network Operator}
\newacronym{nalu}{NALU}{Network Abstraction Layer Unit}
\newacronym{nas}{NAS}{Network-attached Storage}
\newacronym{ngran}{NG-RAN}{Next Generation - \gls{ran}}
\newacronym{ns3}{ns-3}{Network Simulator 3}
\newacronym{nbiot}{NB-IoT}{Narrow Band IoT}
\newacronym{nf}{NF}{Network Function}
\newacronym{nfvi}{NFVI}{Network Function Virtualization Infrastructure}
\newacronym{nic}{NIC}{Network Interface Card}
\newacronym{nlos}{NLOS}{Non-Line-of-Sight}
\newacronym{now}{NOW}{Non Overlapping Window}
\newacronym{nsm}{NSM}{Network Service Mesh}
\newacronym{nrf}{NRF}{Network Repository Function}
\newacronym{nsa}{NSA}{Non Stand Alone}
\newacronym{nse}{NSE}{Network Slicing Engine}
\newacronym{nssf}{NSSF}{Network Slice Selection Function}
\newacronym{o2i}{O2I}{Outdoor to Indoor}
\newacronym{oai}{OAI}{OpenAirInterface}
\newacronym{oaicn}{OAI-CN}{\gls{oai} \acrlong{cn}}
\newacronym{oairan}{OAI-RAN}{\acrlong{oai} \acrlong{ran}}
\newacronym{oam}{OAM}{Operations, Administration and Maintenance}
\newacronym{ofdm}{OFDM}{Orthogonal Frequency Division Multiplexing}
\newacronym{ofdma}{OFDMA}{Orthogonal Frequency Division Multiple Access}
\newacronym{olia}{OLIA}{Opportunistic Linked Increase Algorithm}
\newacronym{omec}{OMEC}{Open Mobile Evolved Core}
\newacronym{onap}{ONAP}{Open Network Automation Platform}
\newacronym{onf}{ONF}{Open Networking Foundation}
\newacronym{onos}{ONOS}{Open Networking Operating System}
\newacronym{oom}{OOM}{\gls{onap} Operations Manager}
\newacronym{opnfv}{OPNFV}{Open Platform for \gls{nfv}}
\newacronym{orbit}{ORBIT}{Open-Access Research Testbed for Next-Generation Wireless Networks}
\newacronym{os}{OS}{Operating System}
\newacronym{oss}{OSS}{Operations Support System}
\newacronym{pa}{PA}{Position-aware}
\newacronym{pase}{PASE}{Prioritization, Arbitration, and Self-adjusting Endpoints}
\newacronym{pawr}{PAWR}{Platforms for Advanced Wireless Research}
\newacronym{pbch}{PBCH}{Physical Broadcast Channel}
\newacronym{pcell}{PCell}{Primary Cell}
\newacronym{pcef}{PCEF}{Policy and Charging Enforcement Function}
\newacronym{pcfich}{PCFICH}{Physical Control Format Indicator Channel}
\newacronym{pcrf}{PCRF}{Policy and Charging Rules Function}
\newacronym{pdcch}{PDCCH}{Physical Downlink Control Channel}
\newacronym{pdcp}{PDCP}{Packet Data Convergence Protocol}
\newacronym{pdcpc}{PDCP-C}{Packet Data Convergence Protocol - Control Plane}
\newacronym{pdcpu}{PDCP-U}{Packet Data Convergence Protocol - User Plane}
\newacronym{pdsch}{PDSCH}{Physical Downlink Shared Channel}
\newacronym{pdu}{PDU}{Packet Data Unit}
\newacronym{pf}{PF}{Proportional Fair}
\newacronym{pgw}{PGW}{Packet Gateway}
\newacronym{phich}{PHICH}{Physical Hybrid ARQ Indicator Channel}
\newacronym{phy}{PHY}{Physical}
\newacronym{phyu}{PHY-U}{Upper Physical}
\newacronym{phyl}{PHY-L}{Lower Physical}
\newacronym{pmch}{PMCH}{Physical Multicast Channel}
\newacronym{pmi}{PMI}{Precoding Matrix Indicators}
\newacronym{powder}{POWDER}{Platform for Open Wireless Data-driven Experimental Research}
\newacronym{ppo}{PPO}{Proximal Policy Optimization}
\newacronym{ppp}{PPP}{Poisson Point Process}
\newacronym{prach}{PRACH}{Physical Random Access Channel}
\newacronym{prb}{PRB}{Physical Resource Block}
\newacronym{pscell}{PSCell}{Primary cell of the Secondary Node}
\newacronym{psnr}{PSNR}{Peak Signal to Noise Ratio}
\newacronym{pss}{PSS}{Primary Synchronization Signal}
\newacronym{pucch}{PUCCH}{Physical Uplink Control Channel}
\newacronym{pusch}{PUSCH}{Physical Uplink Shared Channel}
\newacronym{qam}{QAM}{Quadrature Amplitude Modulation}
\newacronym{qci}{QCI}{\gls{qos} Class Identifier}
\newacronym{qcqp}{QCQP}{Quadratically Constrained Quadratic Problem}
\newacronym{qoe}{QoE}{Quality of Experience}
\newacronym{qos}{QoS}{Quality of Service}
\newacronym{quic}{QUIC}{Quick UDP Internet Connections}
\newacronym{rach}{RACH}{Random Access Channel}
\newacronym[firstplural=Radio Access Technologies (RATs)]{rat}{RAT}{Radio Access Technology}
\newacronym{rest}{REST}{REpresentational State Transfer}
\newacronym{rcn}{RCN}{Research Coordination Network}
\newacronym{rec}{REC}{Radio Edge Cloud}
\newacronym{red}{RED}{Random Early Detection}
\newacronym{rem}{REM}{Random Ensemble Mixture}
\newacronym{renew}{RENEW}{Reconfigurable Eco-system for Next-generation End-to-end Wireless}
\newacronym{rf}{RF}{Radio Frequency}
\newacronym{rfc}{RFC}{Request for Comments}
\newacronym{rfr}{RFR}{Random Forest Regressor}
\newacronym{ric}{RIC}{RAN Intelligent Controller}
\newacronym{rlc}{RLC}{Radio Link Control}
\newacronym{rl}{RL}{Reinforcement Learning}
\newacronym{rlf}{RLF}{Radio Link Failure}
\newacronym{rlnc}{RLNC}{Random Linear Network Coding}
\newacronym{rmr}{RMR}{RIC Message Routing}
\newacronym{rmse}{RMSE}{Root Mean Squared Error}
\newacronym{rnis}{RNIS}{Radio Network Information Service}
\newacronym{rnib}{RNIB}{Radio Network Information Base}
\newacronym{rr}{RR}{Round Robin}
\newacronym{rrc}{RRC}{Radio Resource Control}
\newacronym{rrm}{RRM}{Radio Resource Management}
\newacronym{rru}{RRU}{Remote Radio Unit}
\newacronym{rs}{RS}{Remote Server}
\newacronym{rsrp}{RSRP}{Reference Signal Received Power}
\newacronym{rsrq}{RSRQ}{Reference Signal Received Quality}
\newacronym{rss}{RSS}{Received Signal Strength}
\newacronym{rssi}{RSSI}{Received Signal Strength Indicator}
\newacronym{rtt}{RTT}{Round Trip Time}
\newacronym{rt}{RT}{Real-time}
\newacronym{ru}{RU}{Radio Unit}
\newacronym{rw}{RW}{Receive Window}
\newacronym{rx}{RX}{Receiver}
\newacronym{s1ap}{S1AP}{S1 Application Protocol}
\newacronym{sack}{SACK}{Selective Acknowledgment}
\newacronym{sap}{SAP}{Service Access Point}
\newacronym{sc2}{SC2}{Spectrum Collaboration Challenge}
\newacronym{scef}{SCEF}{Service Capability Exposure Function}
\newacronym{sch}{SCH}{Secondary Cell Handover}
\newacronym{scoot}{SCOOT}{Split Cycle Offset Optimization Technique}
\newacronym{sctp}{SCTP}{Stream Control Transmission Protocol}
\newacronym{sdap}{SDAP}{Service Data Adaptation Protocol}
\newacronym{sdk}{SDK}{Software Development Kit}
\newacronym{sdm}{SDM}{Space Division Multiplexing}
\newacronym{sdma}{SDMA}{Spatial Division Multiple Access}
\newacronym{sdr}{SDR}{Software-defined Radio}
\newacronym{seba}{SEBA}{SDN-Enabled Broadband Access}
\newacronym{sgsn}{SGSN}{Serving GPRS Support Node}
\newacronym{sgw}{SGW}{Serving Gateway}
\newacronym{si}{SI}{Study Item}
\newacronym{sib}{SIB}{Secondary Information Block}
\newacronym{sinr}{SINR}{Signal to Interference plus Noise Ratio}
\newacronym{sip}{SIP}{Session Initiation Protocol}
\newacronym{siso}{SISO}{Single Input, Single Output}
\newacronym{smf}{SMF}{Session Management Function}
\newacronym{smo}{SMO}{Service Management and Orchestration}
\newacronym{sms}{SMS}{Short Message Service}
\newacronym{smsgmsc}{SMS-GMSC}{\gls{sms}-Gateway}
\newacronym{snr}{SNR}{Signal-to-Noise-Ratio}
\newacronym{son}{SON}{Self-Organizing Network}
\newacronym{sptcp}{SPTCP}{Single Path TCP}
\newacronym{srb}{SRB}{Service Radio Bearer}
\newacronym{srn}{SRN}{Standard Radio Node}
\newacronym{srs}{SRS}{Sounding Reference Signal}
\newacronym{ss}{SS}{Synchronization Signal}
\newacronym{sss}{SSS}{Secondary Synchronization Signal}
\newacronym{st}{ST}{Spanning Tree}
\newacronym{svc}{SVC}{Scalable Video Coding}
\newacronym{tb}{TB}{Transport Block}
\newacronym{tcp}{TCP}{Transmission Control Protocol}
\newacronym{tdd}{TDD}{Time Division Duplexing}
\newacronym{tdm}{TDM}{Time Division Multiplexing}
\newacronym{tdma}{TDMA}{Time Division Multiple Access}
\newacronym{cdma}{CDMA}{Code Division Multiple Access}
\newacronym{tfl}{TfL}{Transport for London}
\newacronym{tfrc}{TFRC}{TCP-Friendly Rate Control}
\newacronym{tft}{TFT}{Traffic Flow Template}
\newacronym{tgen}{TGEN}{Traffic Generator}
\newacronym{tip}{TIP}{Telecom Infra Project}
\newacronym{tm}{TM}{Transparent Mode}
\newacronym{tco}{TCO}{total cost of ownership}
\newacronym{to}{TO}{Telco Operator}
\newacronym{tr}{TR}{Technical Report}
\newacronym{trp}{TRP}{Transmitter Receiver Pair}
\newacronym{ts}{TS}{Traffic Steering}
\newacronym{tti}{TTI}{Transmission Time Interval}
\newacronym{ttt}{TTT}{Time-to-Trigger}
\newacronym{tx}{TX}{Transmitter}
\newacronym{uas}{UAS}{Unmanned Aerial System}
\newacronym{uav}{UAV}{Unmanned Aerial Vehicle}
\newacronym{udm}{UDM}{Unified Data Management}
\newacronym{udp}{UDP}{User Datagram Protocol}
\newacronym{udr}{UDR}{Unified Data Repository}
\newacronym{ue}{UE}{User Equipment}
\newacronym{uhd}{UHD}{\gls{usrp} Hardware Driver}
\newacronym{ul}{UL}{Uplink}
\newacronym{um}{UM}{Unacknowledged Mode}
\newacronym{uml}{UML}{Unified Modeling Language}
\newacronym{upa}{UPA}{Uniform Planar Array}
\newacronym{upf}{UPF}{User Plane Function}
\newacronym{urllc}{URLLC}{Ultra Reliable and Low Latency Communications}
\newacronym{usa}{U.S.}{United States}
\newacronym{usim}{USIM}{Universal Subscriber Identity Module}
\newacronym{usrp}{USRP}{Universal Software Radio Peripheral}
\newacronym{utc}{UTC}{Urban Traffic Control}
\newacronym{vim}{VIM}{Virtualization Infrastructure Manager}
\newacronym{vm}{VM}{Virtual Machine}
\newacronym{vnf}{VNF}{virtual network function}
\newacronym{nr}{NR}{New Radio}
\newacronym{volte}{VoLTE}{Voice over LTE}
\newacronym{vonr}{VoNR}{Voice over NR}
\newacronym{voltha}{VOLTHA}{Virtual OLT HArdware Abstraction}
\newacronym{vr}{VR}{Virtual Reality}
\newacronym{vran}{vRAN}{Virtualized \gls{ran}}
\newacronym{vss}{VSS}{Video Streaming Server}
\newacronym{v2x}{V2X}{Vehicle-to-everything}
\newacronym{wbf}{WBF}{Wired Bias Function}
\newacronym{wf}{WF}{Waterfilling}
\newacronym{wlan}{WLAN}{Wireless Local Area Network}
\newacronym{osm}{OSM}{Open Source \gls{nfv} Management and Orchestration}
\newacronym{pnf}{PNF}{Physical Network Function}
\newacronym{mtc}{MTC}{Machine-type Communications}
\newacronym{osc}{OSC}{O-RAN Software Community}
\newacronym{rc}{RC}{RAN Control}
\newacronym{ar}{AR}{Augmented Reality}
\newacronym{daps}{DAPS}{Dual Active Protocol Stack}
\newacronym{nib}{NIB}{Network Information Base}
\newacronym{isa}{ISA}{Instruction Set Architecture}
\newacronym{abi}{ABI}{Application Binary Interface}
\newacronym{vbs}{vBS}{Virtual Base Station}
\newacronym{up}{UP}{User Plane}
\newacronym{ci}{CI}{Continuous Integration}
\newacronym{cd}{CD}{Continuous Deployment}
\newacronym{nvme}{NVMe}{NVM Express}
\newcommand{\apps}{\mathcal{K}}
\newcommand{\servers}{\mathcal{S}}
\newcommand{\sdl}{\mathrm{SDL}}

\newcommand{\smmr}{\mathrm{SM{-}MR}}
\newcommand{\smmd}{\mathrm{SM{-}MD}}
\glsdisablehyper

\newcommand{\novelty}{CORMO-RAN\xspace}
\newcommand{\problem}{SAL\xspace}

\newcommand{\rev}[1]{{\color{black}#1}}

\usepackage{tikzpagenodes,etoolbox}
\usetikzlibrary{calc}
\usepackage[contents={}]{background}
\AddEverypageHook{%
\ifnumequal{\thepage}{1}{%
 \tikz[remember picture,overlay]{%
     \node[draw,
     minimum width=0.93\textwidth,
     text width=0.92\textwidth,
     font=\footnotesize
     ]
     at ($(current page header area) - (0,5pt)$)
     {%
     This paper has been accepted for publication in {\em IEEE Transactions on Mobile Computing}. This is the authors' accepted version of the manuscript. The final version published by IEEE is “A. Calagna, S. Maxenti, L. Bonati, S. D’Oro, T. Melodia and C. F. Chiasserini, "CORMO-RAN: Energy Efficiency at the Near-RT RIC via Lossless Migration of O-RAN xApps," in {\em IEEE Transactions on Mobile Computing}, doi: 10.1109/TMC.2026.3715058."
     };
 }%
}{}
}

\begin{document}

\title{\rev{\novelty: Energy Efficiency at the Near-RT RIC via Lossless Migration of O-RAN xApps}
\thanks{Antonio Calagna and Carla Fabiana Chiasserini are with the Department of Electronics and Telecommunications, Politecnico di Torino, 10129, Turin, Italy (email: antonio.calagna@polito.it; carla.chiasserini@polito.it). Carla Fabiana Chiasserini is also with CNIT, 43124, Parma, Italy and Chalmers University, SE412-96, Göteborg, Sweden. Stefano Maxenti, Leonardo Bonati, Salvatore D'Oro, and Tommaso Melodia are with the Institute for the Wireless Internet of Things, Northeastern University, Boston, MA 02115, USA (e-mail: maxenti.s@northeastern.edu; l.bonati@northeastern.edu; s.doro@northeastern.edu;  t.melodia@northeastern.edu). This work was partially supported by the U.S. National Telecommunications and Information Administration (NTIA)'s Public Wireless Supply Chain Innovation Fund (PWSCIF) under Award No. 25-60-IF002, by the U.S. National Science Foundation under grant CNS-2112471, by the EC through Grant No.\,101139266 (6G-INTENSE  project), and by the Qatar Research Development and Innovation Council ARG01-0525-230339. The content is solely the responsibility of the authors and does not necessarily represent the official views of Qatar Research Development and Innovation Council.}
}

\author{
\IEEEauthorblockN{Antonio~Calagna,~\IEEEmembership{Member,~IEEE}, Stefano~Maxenti,~\IEEEmembership{Graduate Student Member,~IEEE}, Leonardo~Bonati,~\IEEEmembership{Member,~IEEE}, Salvatore~D'Oro,~\IEEEmembership{Member,~IEEE}, Tommaso~Melodia,~\IEEEmembership{Fellow,~IEEE}, Carla~Fabiana~Chiasserini,~\IEEEmembership{Fellow,~IEEE}}
}

\maketitle

\begin{abstract}
Open \gls{ran} is a key paradigm to attain unprecedented flexibility of the \gls{ran} via disaggregation and \gls{ai}-based applications called xApps.
In dense areas with many active \gls{ran} nodes, compute resources are engineered to support potentially hundreds of xApps monitoring and controlling the \gls{ran} to achieve operator's intents.
However, such resources might become underutilized during low-traffic periods, where most cells are sleeping and, given the reduced \gls{ran} complexity, only a few xApps are needed for its control.
In this paper, we propose \novelty, a data-driven orchestrator that dynamically activates compute nodes based on xApp load to save energy, and performs lossless migration of xApps from nodes to be turned off to active ones while ensuring xApp availability during migration.
\novelty tackles the trade-off among service availability, scalability, and energy consumption while (i)~preserving xApps' internal state to prevent \gls{ran} performance degradation during migration; (ii)~accounting for xApp diversity in state size and timing constraints; and (iii)~implementing several migration strategies and providing guidelines on best strategies to use based on resource availability and requirements.
We prototype \novelty as an rApp, and experimentally evaluate it on an O-RAN private 5G testbed hosted on a Red Hat OpenShift cluster with commercial radio units.
Results demonstrate that \novelty is effective in minimizing energy consumption of the \gls{ric} cluster, yielding up to 64\% energy saving when compared to existing approaches.
\vspace{-5pt}
\end{abstract}


\glsresetall

\section{Introduction}\label{work06:sec:intro}

\begin{figure}[t]
    \centering
    \subfloat[][]{\label{work06:fig:sys_diag}{\includegraphics[width=0.48\columnwidth]{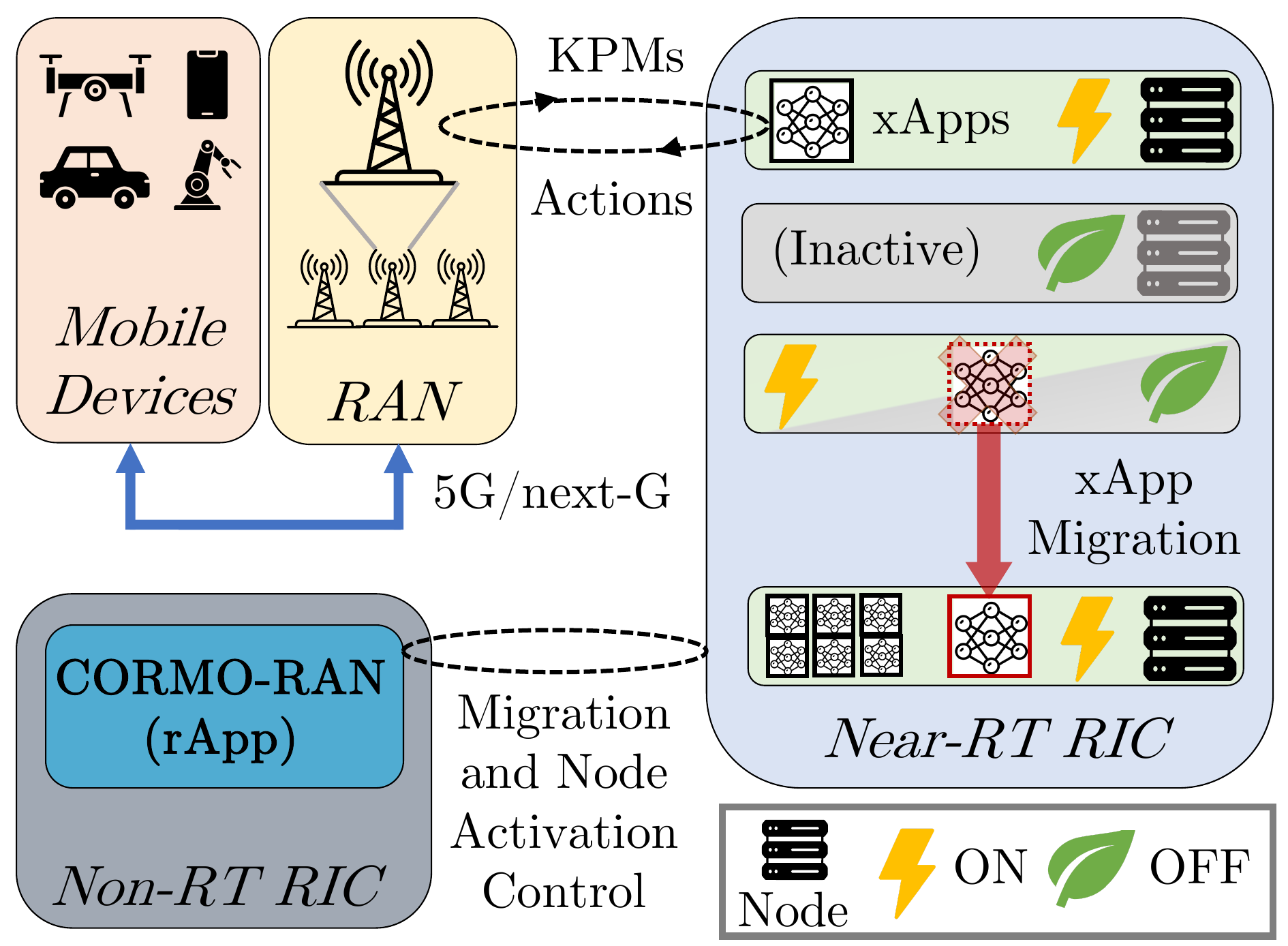}}}
    \hspace{2mm}
    \subfloat[][]{\label{work06:fig:sm_vs_sdl_diagram}{\includegraphics[width=0.48\columnwidth]{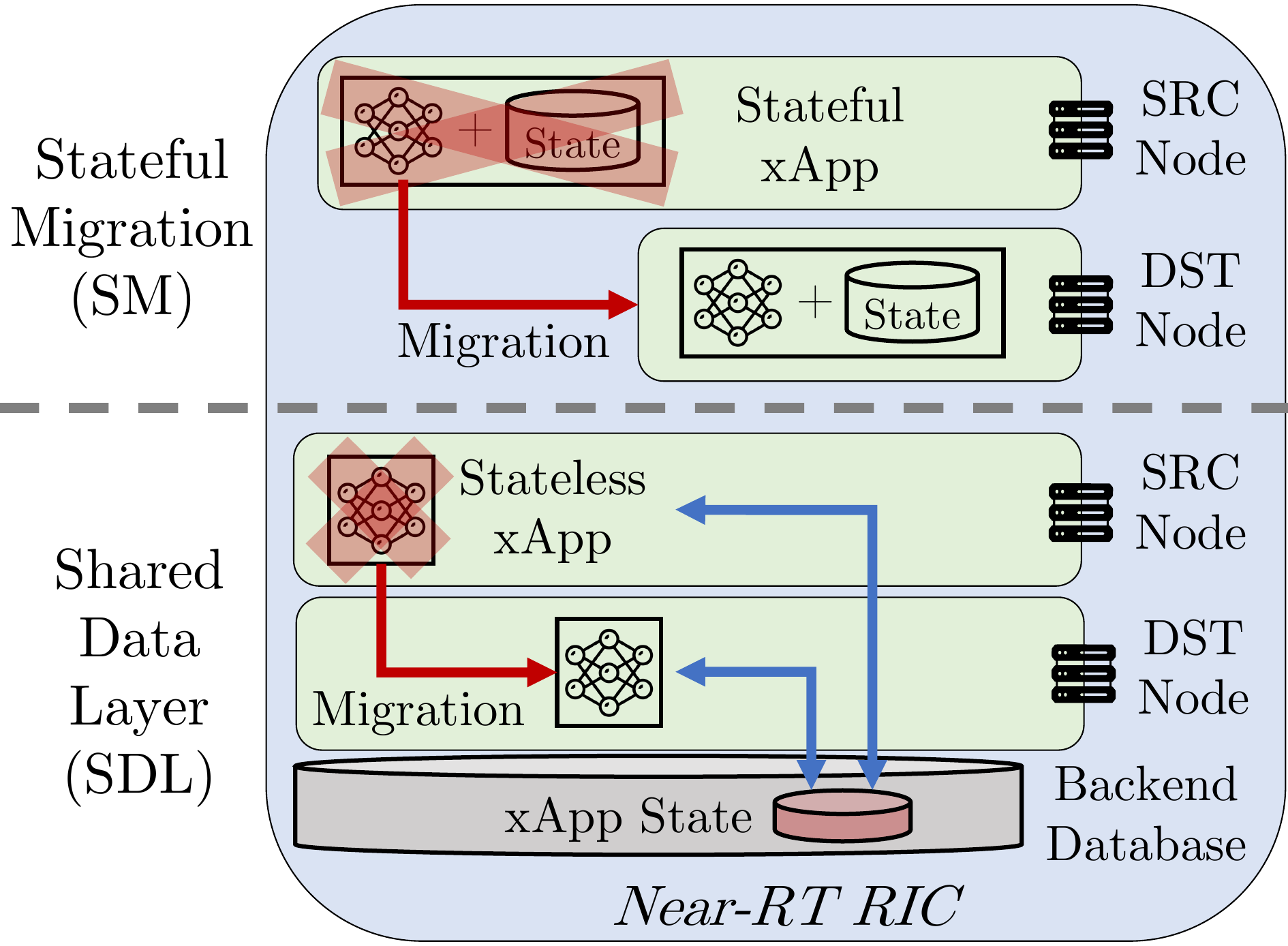}}}
    \caption{(a) Concept representation of CORMO-RAN, and (b) comparison of SM and SDL xApp migration approaches.}
    \label{work06:fig:intro_diagram}
    \vspace{-6mm}
\end{figure}

The Open \gls{ran} paradigm---and its embodiment proposed by the O-RAN ALLIANCE~\cite{oran2018building}---has been heralded as the vehicle to bring unprecedented flexibility to 5G-and-beyond \gls{ran} architectures.
O-RAN promotes enhanced \textit{flexibility} via \gls{ran} disaggregation and virtualization, as well as \textit{adaptability} through data-driven control loops that optimize the \gls{ran} performance~\cite{polese2023understanding}.
Cornerstones of the O-RAN architecture, depicted in Fig.~\ref{work06:fig:sys_diag}, are the \glspl{ric}, which oversee the operations of the \gls{ran} through closed control-loops at different time scales: near-real-time (or near-RT, between $10$\,ms and $1$\,s), and non-real-time (or non-RT, above $1$\,s).
\gls{ran} control is actuated via intelligent applications hosted as microservices on the \glspl{ric}---namely, xApps on the near-RT and rApps on the non-RT \gls{ric}---that leverage \glspl{kpm} coming from the \gls{ran} to perform inference/forecasting or compute policies to optimize network performance.

{\rev{\bf \em Existing Issues.}} xApps enable for the first time self-optimizing and zero-touch cellular networks. However, their contribution to the \gls{ric}'s cluster energy consumption is non-negligible, especially in large deployments with hundreds of xApps~\cite{maxenti2024scaloran}.
Additionally, the number of xApps needed to control the \gls{ran} might vary significantly from peak hours, when traffic demand is high, to nighttime, when most cells might be in energy-saving mode or even turned off. Therefore, even though only a few xApps may be actively controlling RAN elements, many compute nodes would still be active and underutilized, resulting in unnecessary energy consumption.

In this context, microservice migration---i.e., transferring a microservice from a source to a destination node---is a powerful tool to reallocate xApps across different nodes of the same near-RT \gls{ric} cluster to dynamically minimize the number of active nodes and turn off the inactive ones, depending on the network load. 
While \textit{stateless} xApp migration relies on well-established and low-latency techniques that simply deactivate xApps on the source node and recreate them on the destination node, migration of \textit{stateful} xApps is not as trivial.
Indeed, stateful xApps (e.g., used in forecasting, beam tracking and mobility management) need to maintain a history of context-based data to accomplish their tasks.
This data is stored in an internal \textit{state} that must be preserved upon migration to retain control effectiveness and avoid performance degradation.

In this work, we focus on the lossless migration of stateful xApps and consider two approaches (Fig.\ref{work06:fig:sm_vs_sdl_diagram}): \gls{sm}~\cite{calagna2024design}, and the O-RAN \gls{sdl}~\cite{polese2023understanding}.
\gls{sm} migrates xApps together with their state, causing a service disruption referred to as \textit{downtime}, with two variants: \gls{sm}-MR, minimizing resource usage; and \gls{sm}-MD minimizing downtime.
Instead, \gls{sdl} decouples xApps from their state by storing it in a backend database, making xApps virtually stateless from a migration viewpoint. However, this distributed database must guarantee strong consistency, potentially limiting \gls{sdl}'s scalability and feasibility.

\vspace{-2mm}
\begin{figure}[tbh]
    \centering
    \subfloat[][]{\label{work06:fig:sm_vs_sdl_downtime_vs_num_xapps}{\includegraphics[width=0.325\columnwidth]{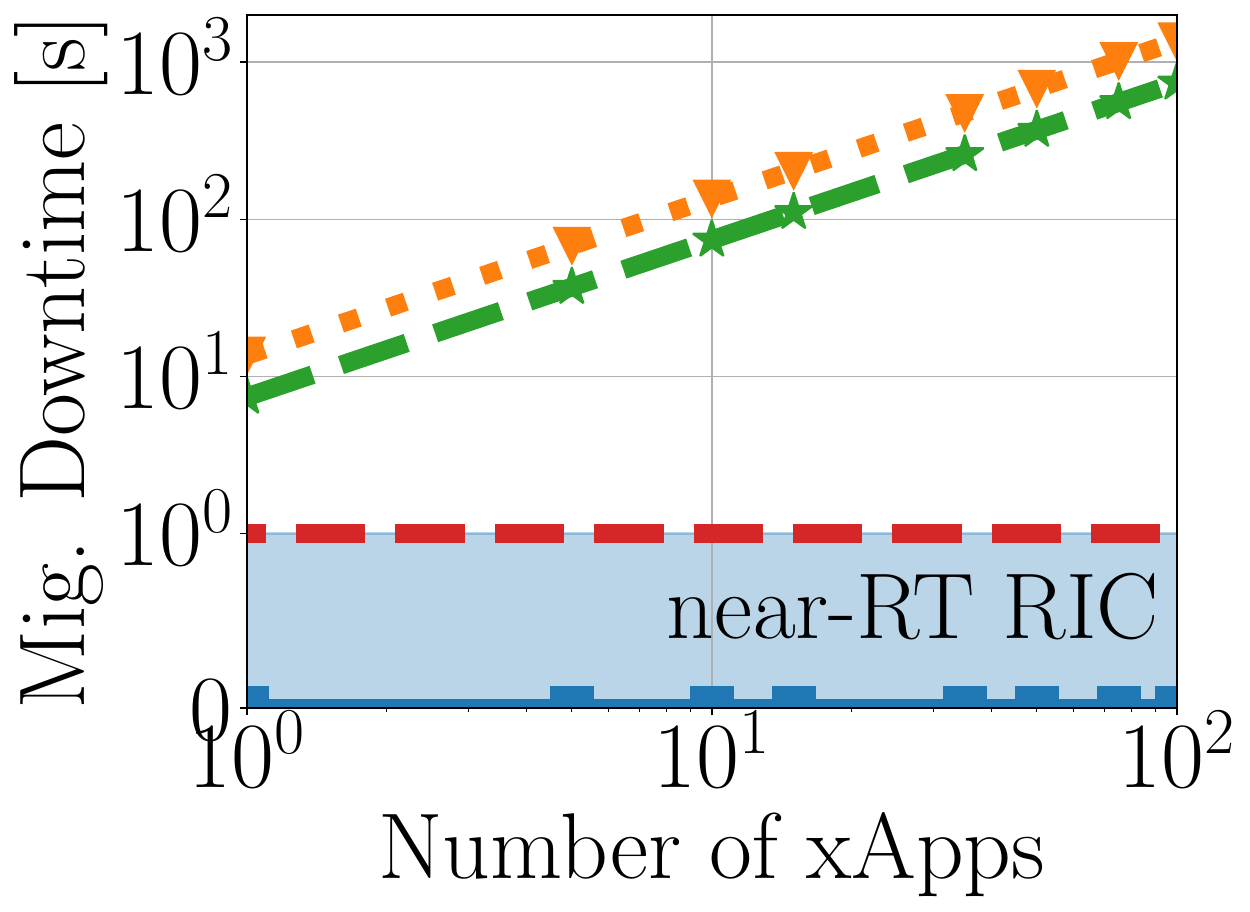}}}
    \subfloat[][]{\label{work06:fig:sm_vs_sdl_svc_dsrp_vs_num_xapps}{\includegraphics[width=0.31\columnwidth]{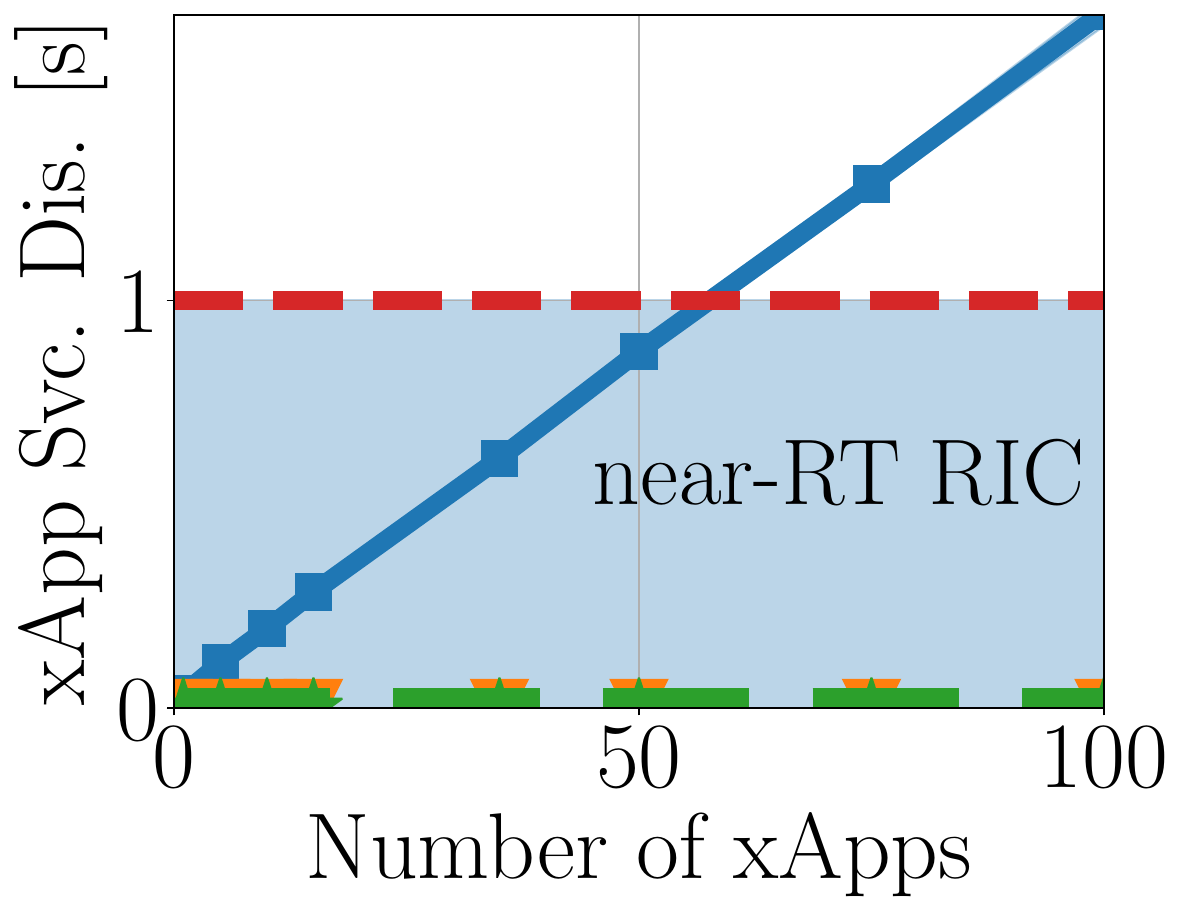}}}
    \subfloat[][]{\label{work06:fig:sm_vs_sdl_energy_vs_num_xapps}{\includegraphics[width=0.32\columnwidth]{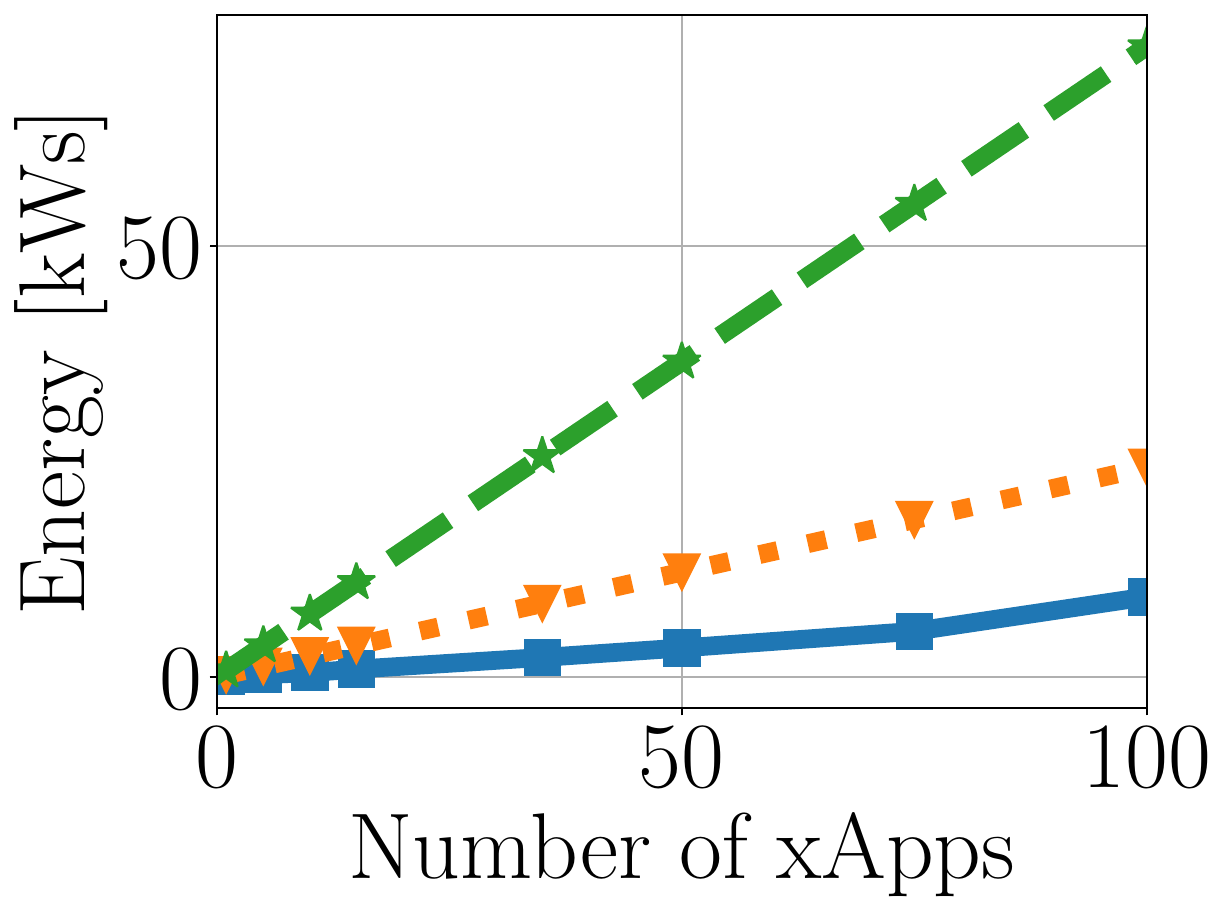}}}
    \newline
    \includegraphics[width=0.7\columnwidth]{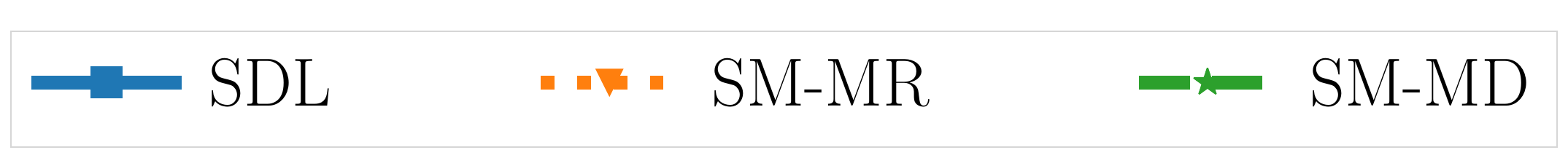}
    \caption{Comparing SDL, SM-MR, and SM-MD: (a) migration downtime, (b) xApp service disruption, and (c) energy usage.}
    \label{work06:fig:sm_vs_sdl_tradeoff}
    \vspace{-2mm}
\end{figure}

To highlight their differences, in Fig.~\ref{work06:fig:sm_vs_sdl_tradeoff} we experimentally compare these strategies for several copies of a same exemplary \gls{drl}-based xApp that receives \glspl{kpm} from the \gls{ran}, stores them as state, and computes a control action. \rev{The red dashed line indicates the near-RT \gls{ric} control loop deadline, which we here set to 1\,s as an example.}
We notice that: (a)~while \gls{sdl} enables zero-downtime migration, \gls{sm} has a downtime that linearly increases with the number of xApps, always violating \rev{any} near-RT \gls{ric} deadline (Fig.~\ref{work06:fig:sm_vs_sdl_downtime_vs_num_xapps}); (b)~contrarily to \gls{sm}, \gls{sdl} has scalability issues, yielding a periodic xApp service disruption that can lead to near-RT \gls{ric} deadline violations (Fig.~\ref{work06:fig:sm_vs_sdl_svc_dsrp_vs_num_xapps}); and (c)~\gls{sdl} yields up to 87\% reduction in energy consumption compared to \gls{sm} (Fig.~\ref{work06:fig:sm_vs_sdl_energy_vs_num_xapps}).
These results show that migrating stateful xApps in O-RAN involves a trade-off among service availability, scalability, and energy consumption.

\rev{{\bf \em Novelty.}
To tackle the above challenges, in this paper we propose \novelty, a data-driven orchestrator that jointly optimizes compute node activation and xApp migration strategies. Differently from existing works, \novelty:
\begin{itemize}
    \item[\emph{(i)}] accounts for stateful xApps whose state must be preserved, and whose control tasks need to be executed within a strict temporal deadline;
    \item[\emph{(ii)}] encompasses both \gls{sm} and \gls{sdl} migration techniques as well as a diverse xApp catalog that captures varying workload use cases;
    \item[\emph{(iii)}] is evaluated on a real near-RT \gls{ric} deployment using a pre-trained, publicly available AI-driven xApp;
    \item[\emph{(iv)}] identifies the feasibility regions of each migration strategy based on traffic load and available resources;
    \item[\emph{(v)}] dynamically computes the optimal allocation of xApps across near-RT \gls{ric} nodes that minimizes the overall system energy consumption.
\end{itemize}
}
Importantly, our work is the first to (i) experimentally characterize performance and trade-offs of state-of-the-art migration techniques in the \gls{o-ran} context; and (ii) to leverage such techniques to minimize the near-RT \gls{ric} energy footprint.
We prototype \novelty as a non-RT \gls{ric} rApp, and leverage it to jointly orchestrate the node activation and migration of xApps on a Red Hat OpenShift cluster. For this, we consider real-world xApps that (i) reflect varying levels of \gls{ran} workload complexity; (ii) are deployed on a cluster of nodes, together with an open-source near-RT \gls{ric}; and (iii) are used to re-configure a private 5G testbed with commercial \glspl{ru} and \glspl{ue}.
Our results demonstrate that \novelty effectively addresses the aforementioned trade-off and enables up to 64\% reduction of the system energy consumption.

\rev{{\bf \em Paper Structure.} The rest of the paper is organized as follows. We first  review relevant related works and emphasize the uniqueness of our study in Sec.~\ref{work06:sec:related_work}. Then, we describe the SM and SDL xApp migration approaches and present the experimental testbed we developed to evaluate them in Sec.~\ref{work06:sec:preliminaries} and Sec.~\ref{work06:sec:testbed}, respectively. We use this testbed to analyze the migration process across diverse classes of xApps in Sec.~\ref{work06:sec:exp_analysis}, and leverage the resulting experimental evidences to formulate our optimization problem and proposed solution, \novelty, in Sec.~\ref{work06:sec:opt_prob}. Finally, we conduct a comprehensive performance evaluation of \novelty in Sec.~\ref{work06:sec:opt_num_eval} and draw our conclusions in Sec.~\ref{work06:sec:conclusions}.
}

\section{Related Work}\label{work06:sec:related_work}
Ongoing efforts and challenges related to sustainable mobile networking are analyzed in~\cite{wang2012survey, masoudi2019green, lopez2022survey, baldini2024toward}. Work in \cite{larsen2023toward} finds the \gls{ran} segment to be the one impacting energy consumption the most (up to 73\%), and \gls{ai} has been identified as a promising solution to minimize such energy consumption~\cite{liang2024energy, larsen2024evolution, kundu2024towards} and improve quality of experience~\cite{ramezanpour2022intelligent, tsampazi2024pandora, dai2025oran}.
For instance, \cite{dryjański2021toward, catalan2024begreen} provide AI-driven solutions to enhance O-RAN energy efficiency through, respectively, traffic steering and cell on/off control. \cite{mungari2025oran} proposes an O-RAN orchestrator that, by semantically sharing xApps across RAN services, aims to maximize the number of the services concurrently deployed while minimizing their overall energy consumption.
Nonetheless, the proliferation of AI-based xApps inevitably contributes to the energy footprint, posing an additional challenge toward network sustainability. 
Indeed, \cite{maxenti2024scaloran} profiles various types of xApps in terms of their energy consumption and demonstrates that scaling up the number of concurrently running xApps leads to a proportional increase in the overall energy usage of the near-RT RIC cluster. Also, it is shown that xApps are a dominant contributor to the overall system energy footprint, thus highlighting the importance of energy-aware orchestration and placement strategies.

As per service migration, \cite{wang2018survey, terneborg2021application} give an overview of current stateful migration techniques along with their \glspl{kpi} and discuss the potential of such techniques in addressing critical mobile scenarios in the general context of edge computing, where service continuity is of utmost importance. \cite{rong2024live, li2024seamless} propose practical solutions for seamless service migration at the network edge, focusing, respectively, on video analytics and real-time rendering applications. To address the lack of a migration model, \cite{calagna2024design} analyzes and captures the practical aspects of stateful migration. \cite{calagna2025mose} introduces MOSE, a novel framework that efficiently implements stateful migration and effectively orchestrates the migration process by fulfilling both network and application KPI targets. Leveraging migration techniques, \cite{wang2019dynamic,mukhopadhyay2022migration,panek2023relocator,afachao2024efficient, adeppady2025efficient} propose solutions to attain an optimal service placement while prioritizing mobile end user requirements.

Regarding xApp state decoupling, although \gls{sdl} is defined as part of the O-RAN specifications~\cite{polese2023understanding}, its implementation details---including the choice of backend database---remain open and are yet to be standardized. In this context, a growing concern regarding the need to rethink traditional database architectures is raised in~\cite{laigner2021distributed,laigner2021data}. Specifically, it is observed that the inherently decentralized data management of microservice architectures poses significant challenges for coordination, as state dependencies and consistency issues are often overlooked, with a non-negligible amount of applications requiring strong consistency guarantees over the shared information they access. \cite{calagna2025enabling} proposes varying architectures and implementations of a holistic data access platform at the edge---sharing the same design principles as SDL---and thoroughly characterizes their performance and trade-offs across a spectrum of scenarios, ranging from loosely controlled loops to latency-critical and compute-demanding use cases.

\rev{
{\bf \em Distinctive Methodology and Contribution.} To the best of our knowledge, our work is the first to experimentally evaluate cutting-edge migration approaches in the O-RAN context to jointly optimize compute node activation and xApp placement while minimizing energy consumption and guaranteeing xApp service availability. Specifically, we: (i) characterize migration techniques to assess their benefits, drawbacks, and performance; (ii) derive a model that captures the fundamental trade-off between downtime, energy consumption, and availability; and (iii) develop algorithms to identify the feasibility regions of each technique and optimize service migration and placement to minimize energy consumption under strict availability constraints. Importantly, although node activation and workload optimization have been investigated in the broader context of edge computing~\cite{gomez2024odesa, avgeris2022enerdge}, our work delivers a fundamentally different perspective that, unlike prior art, captures the unique challenges of the O-RAN context, i.e., by explicitly accounting for its stringent timing requirements and the practical, real-world challenges of xApp migration, e.g., the need to preserve their internal state as a way to guarantee service continuity.
}

\section{Overview of xApp Migration}\label{work06:sec:preliminaries}
This section describes the two main technologies for the lifecycle management of stateful xApps. First, we introduce the concept of stateful migration, along with its \glspl{kpi} (Sec.~\ref{work06:sec:preliminaries:sub:stateful_mig}).\footnote{In this work, \glspl{kpi} refer to migration strategies, while \glspl{kpm} refer to data produced by the \gls{ran} and collected by the near-RT \gls{ric} and xApps.} Then, we provide an overview of the shared data layer approach used in O-RAN~\cite{polese2023understanding} to support data access and sharing among multiple xApps (Sec.~\ref{work06:sec:preliminaries:sub:sdl_mig}).

\begin{figure}[bt]
    \centering
    \includegraphics[width=\columnwidth]{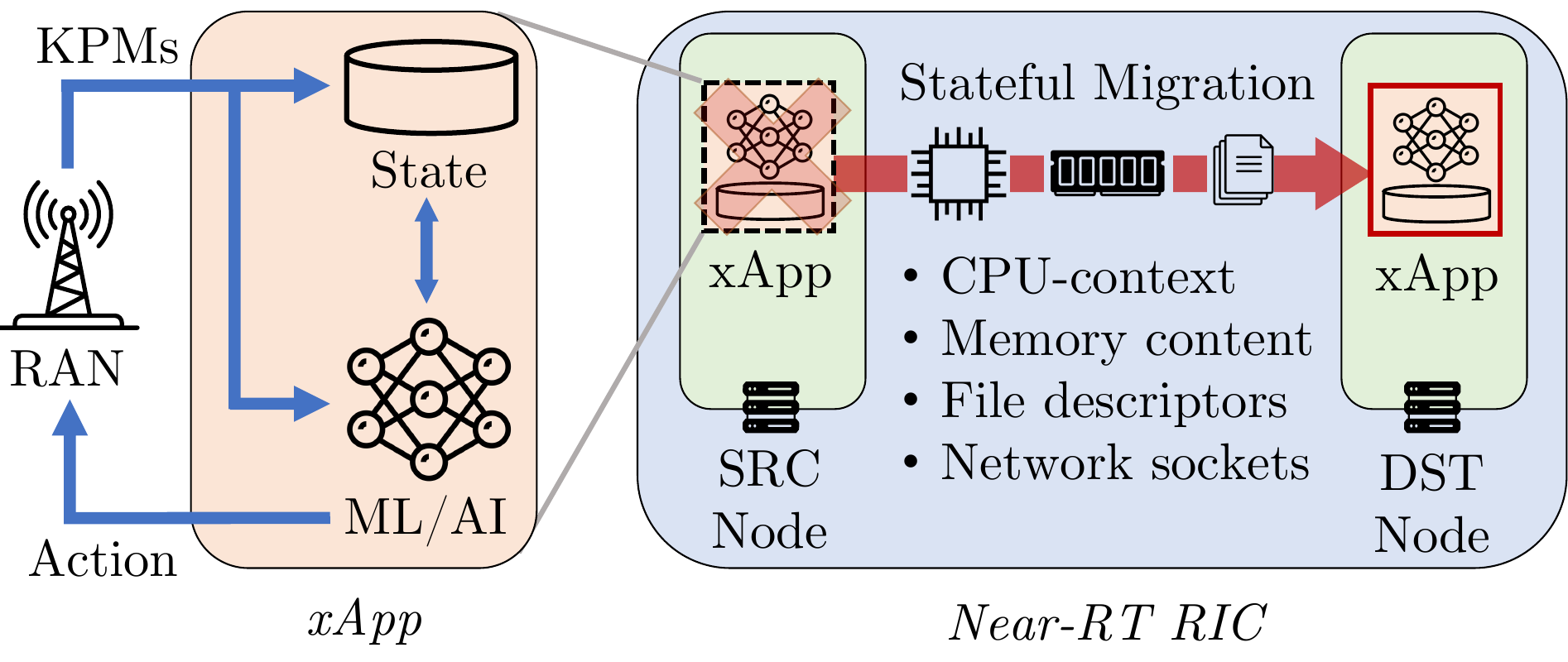}
    \caption{Stateful migration of xApps in O-RAN.}
    \label{work06:fig:sm_diagram}
    \vspace{-3mm}
\end{figure}

\subsection{Stateful Migration (SM)}\label{work06:sec:preliminaries:sub:stateful_mig}
As shown in Fig.~\ref{work06:fig:sm_diagram}, this approach considers the case where the state of the xApps (e.g., context-related metrics) is embedded in the xApp, which runs as a microservice container. 
To preserve the state and ensure service continuity, \gls{sm} relocates the entire container (which includes the state) from the source node to the destination node.
As shown in Fig.~\ref{work06:fig:sm_diagram}, besides the container image, \gls{sm} requires the following pieces of information at the destination node: (i) CPU-context state, e.g., registers, processes tree structure, and namespaces; (ii) memory content, i.e., the pages allocated in the main memory; (iii) network sockets; and (iv) open file descriptors.

\gls{sm} has two variants: SM-MR and SM-MD. The former prioritizes resource minimization during the migration process; the latter focuses on minimizing the migration downtime. \rev{For both variants, we consider to migrate multiple xApps in a sequential way, which reflects the practical limitations of the off-the-shelf, application-agnostic solution we leveraged.} 

SM-MR uses {\em Cold Migration}, consisting of the following steps: (i) collection of the state \rev{checkpoint} at the source node; (ii) transfer of such state from source to destination node; and (iii) restoration of the container state at the destination. To prevent state inconsistency, the container is stopped at the source node throughout these steps, thus causing a service disruption period, i.e., the {\em migration downtime}. 

SM-MD implements the {\em Iterative PreCopy} algorithm and draws on the {\em dirty-page rate} concept, i.e., the number of memory pages per second a container modifies. This strategy consists of: (i) the iterative transfer of dirty pages to the destination node while the container is still running at the source node; and (ii) stopping the container and transferring the remaining dirty pages to the destination node. By minimizing the amount of data to be transferred, SM-MD trades a longer {\em total migration duration} for a shorter downtime.

\begin{figure}[bt!]
    \centering
    \includegraphics[width=\columnwidth]{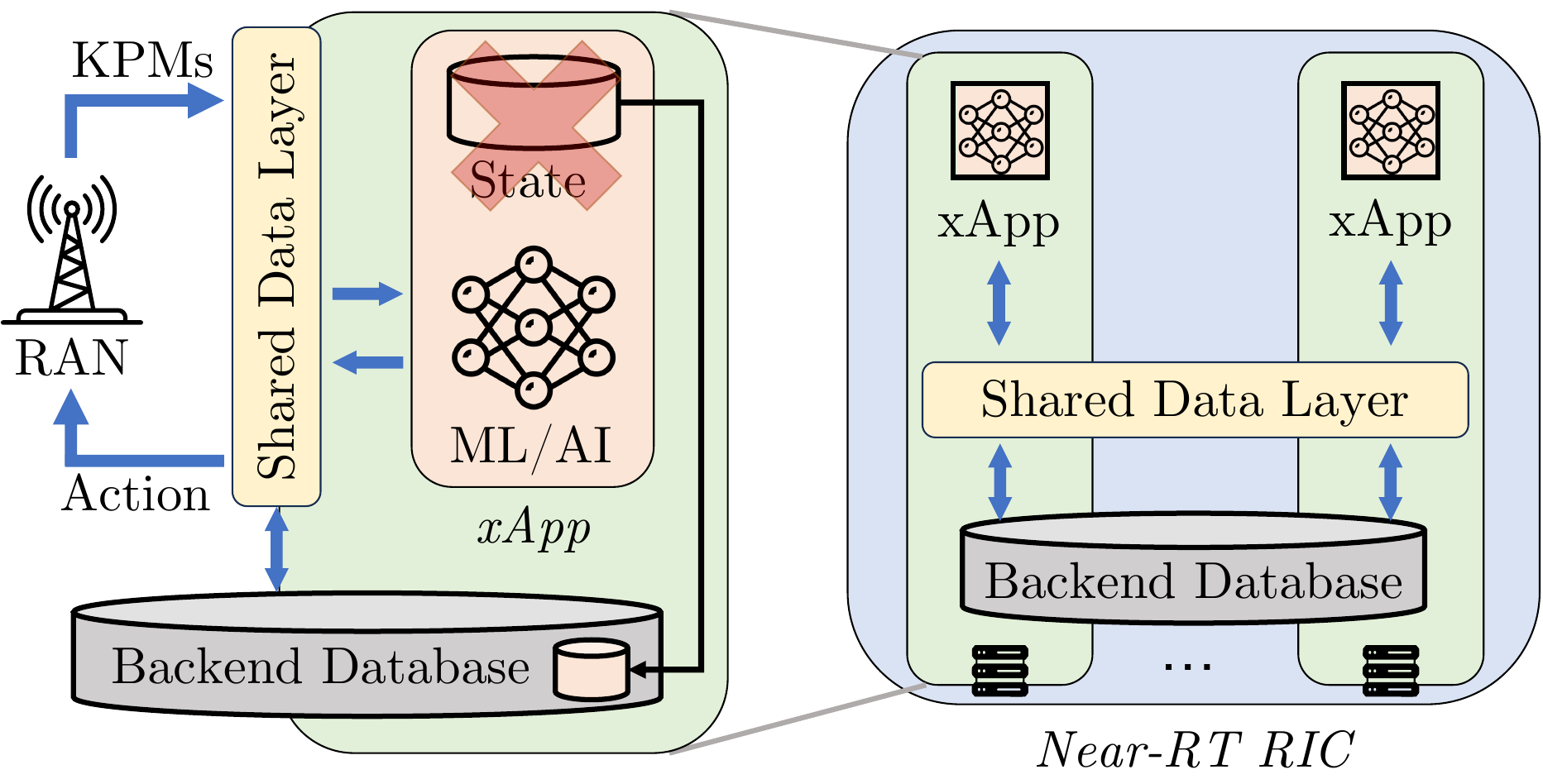}
    \caption{O-RAN shared data layer architecture to decouple xApps from their internal state.}
    \label{work06:fig:sdl_diagram}
    \vspace{-4mm}
\end{figure}

\subsection{Shared Data Layer (SDL)}\label{work06:sec:preliminaries:sub:sdl_mig}
To regulate data production and consumption between xApps, O-RAN introduces a data access platform, called \gls{sdl}, which acts as an abstraction layer between the applications and a backend database where data is stored and shared. As in Fig.~\ref{work06:fig:sdl_diagram}, \gls{sdl} can be used to decouple the xApp from its state, which can be instead stored in the backend database.
From a migration viewpoint, \gls{sdl} effectively transforms stateful xApps into stateless as the state is still present but stored externally in the \gls{sdl}. Therefore, this approach (i) conforms with the requirements of 5G-and-beyond networks~\cite{kulkarni2022cost} and recent microservice-oriented architecture design patterns~\cite{erl2016soa}, both requiring microservices to be stateless to maximize efficiency and scalability; and (ii) enables migration strategies that are zero-downtime by design, i.e., yielding no service disruption to the final users.

Nevertheless, since the state is now outsourced to the backend database, accessing the state incurs in additional delay that might impact xApp performance and timeliness of control policies. Therefore, data access must happen with the lowest possible latency. Also, to avoid the creation of a single point of failure, the backend database must be tolerant to faults and network partitions, e.g., by distributing multiple replicas of its content across the near-RT \gls{ric} nodes. In \novelty we consider a migration strategy where we proactively duplicate the xApp at the destination node and, upon success, we remove it from the source node. Contrarily to \gls{sm}, which needs the xApp instance at the source node to be stopped before resuming execution at the destination host, under \gls{sdl}, the two xApp instances can keep on serving incoming requests from the RAN and updating their internal state while the migration process takes place, yielding no service disruption. Importantly, since the xApp state is shared by the two instances during the migration process, the backend database needs to effectively support concurrent data accesses while guaranteeing strong data consistency to prevent race conditions. 

In summary, \gls{sdl} requires a backend database with: (i) high availability; (ii) high reliability and fault-tolerance; and (iii) strong data consistency. We address such technical challenges by choosing \texttt{etcd}~\cite{etcd} as the near-RT \gls{ric} backend database. Etcd is a distributed reliable key-value store for the most critical data of a distributed system that, by leveraging the Raft~\cite{raft} consensus algorithm, enforces strong data consistency, and tolerance to network partitions and machine failure at the cost of a reduced availability~\cite{cap_theorem}.
We remark that, while other popular state-of-the-art databases such as Redis~\cite{redis} prioritize high availability by favoring eventual consistency guarantees, our work focuses on the more challenging scenario in which strong data consistency must be ensured. This requirement is critical to maintain correctness and coherence of the information shared among multiple xApps, particularly during the coordination of latency-sensitive near-RT RIC control loops. As shown in~\cite{calagna2025enabling}, a comparison between etcd- and Redis-based implementations reveals fundamental trade-offs in terms of scalability, availability, data consistency, and resource usage, with etcd demonstrating superior performance when resilience and strong data consistency are of utmost importance.
\rev{
Compared to other strongly consistent databases such as Zookeeper and Consul, etcd offers a well-established balance of stability, reliability, scalability, and performance---even when operating at multi-gigabyte scales---while avoiding the architectural complexity and latency overhead associated with NewSQL systems~\cite{etcd-comparison, corbett2013spanner, zhou2021foundation}. These characteristics, together with its role as the official Kubernetes' core data store, have made etcd a widely adopted and frequently referenced solution for state-of-the-art distributed systems.
}

To guarantee high reliability, etcd stores data in a multi-version persistent key-value store, preserving the previous version of a key-value pair when its value is updated. As a result, etcd keeps an exact history of its keyspace, which should be periodically compacted to avoid performance degradation and eventual storage space exhaustion. Since compacting old revisions internally fragments etcd by leaving gaps in the backend database, it is also necessary to release this storage space back to the file system through a defragmentation process. Importantly, during defragmentation, the etcd member rebuilds its states and is thus blocked from reading and writing data, yielding service disruption for the xApps. In the following, we refer to the combination of the compaction and defragmentation processes as a {\em maintenance} operation whose periodicity can be controlled to prevent resource exhaustion and etcd performance degradation. Our analysis accounts for the service disruption duration, denoted as {\em defrag downtime}, yielded by each maintenance operation and assesses if, and to what extent, such downtime is compatible with the near-RT \gls{ric} control loop deadlines.

\section{Experimental O-RAN Testbed}\label{work06:sec:testbed}
In this section, we describe the testbed that we developed to evaluate the two approaches above, i.e., \gls{sm} and \gls{sdl}, identify their feasibility region, and determine which approach is best suited to certain operational conditions and compute loads.

{\bf Computing cluster.} We deploy an end-to-end O-RAN system, comprising an open-source near-RT RIC~\cite{near-rt-ric} \rev{from the \gls{osc}} on a Red Hat OpenShift cluster~\cite{bonati2023neutran}. OpenShift~\cite{openshift} is a commercial \rev{container orchestration platform that extends Kubernetes with production-grade security, reliability, resilience, and fault tolerance functionalities, among others.}
\rev{From an infrastructure standpoint, our cluster comprises four Dell R760 compute nodes equipped with 128 Intel Xeon 8462Y+ CPUs, 512\,GB of RAM ($16\times 32$\,GB DDR5 blocks working at 4.8\,GHz frequency) and 960\,GB of storage (NVMe disk operating at a maximum link speed of 16\,gigatransfer/s). Nodes are connected via 10\,Gbps interfaces to a Dell S5248F-ON \gls{sdn} switch that enables fast and reliable communication among them. Besides local storage, cluster nodes are connected to a \gls{nas} that provides persistent storage for the containers and the internal image registry.} 

To gather accurate and comprehensive metrics for our experimental analysis, our testbed integrates Prometheus~\cite{prometheus} and Kepler~\cite{kepler}. Prometheus is a widely adopted Kubernetes monitoring system that facilitates effective cluster-wide metrics aggregation. Kepler, on the other hand, is a renown framework that uses advanced power models to estimate real-time energy consumption at the pod level (i.e., at the Kubernetes fundamental unit). Given the importance of accurately estimating a system carbon footprint~\cite{idle_power_kepler}, Kepler accounts not only for the active computations but also for idle power, i.e., the static node power. As thoroughly discussed in~\cite{amaral2023kepler, centofanti2024impact, akbari2024monitoring}, such idle contribution mainly consists of the power related to hardware components, such as motherboard, fans, network interface cards, and other peripherals, as well as the power consumed by the Kubernetes elements that are necessary for the system to be functional, e.g., the Kubelet and the control plane.

{\bf xApp.} 
To run our experiments, we consider \rev{two publicly available xApps that differ in terms of computational complexity, namely, the DRL-xApp~\cite{xapp-drl-example, bonati2021intelligence} and the KPM-xApp~\cite{xapp-monitor-example}. The former is representative of AI-driven control tasks and implements a pre-trained \gls{drl} agent that, by leveraging \gls{ran} \glspl{kpm}, computes the optimal scheduling policy for the network slices implemented at the base station. The latter is a monitoring xApp that exemplifies less demanding tasks, as it collects and aggregates \glspl{kpm} to derive \gls{ran} performance statistics. Both xApps operate according to an event-driven approach, i.e., executing their logic whenever a \gls{kpm} message of arbitrary size is produced by the \gls{ran}.} To allow for an accurate scalability analysis of the aforementioned migration strategies when hundreds of xApps are deployed on the RIC, we build an E2 agent emulator capable of synthetically generating traffic with varying loads (see Sec.\,\ref{work06:sec:exp_analysis:sub:scenario}). Importantly, the insights and results presented in the following remain valid even when actual RAN nodes are connected to the RIC, and are independent of the specific xApp we use, thus remaining broadly applicable to any kind of \rev{control task}, regardless of its complexity. Since \gls{sm} and \gls{sdl} rely on different xApp architectures, we have modified the above xApps to consider two programmable variants that differ in how the internal state is stored and accessed. \rev{We thus recall that the xApp internal state is defined as the set of all relevant information, e.g., history of context-based data, that the xApp requires to accomplish its tasks and that must be preserved upon migration to ensure control effectiveness and service continuity.} The \gls{sm} variant retains the internal state in a queue of tunable size and stored in the main memory, \rev{along with the other components necessary for xApp execution (e.g., AI model weights)}. Instead, the \gls{sdl} variant leverages etcd APIs to delocalize such state queue onto the database and performs the following steps: (i) interrupt-based watch of the \gls{kpm} key, which is updated every time the \gls{ran} produces a new \gls{kpm} message; (ii) push such message into the state queue; (iii) pop the least recent message from the queue; (iv) produce the control message jointly leveraging the newest message and the queue, and put it on the database so that it can be consumed by the \gls{ran}. \rev{Furthermore, we remark that while SM technology relies on application-agnostic tools and migrates the whole xApp memory content, SDL permits to selectively decouple just the xApp internal state onto the backend database, thus excluding the extra memory content that is independent of the specific xApp-related context.}

{\bf SM.} To implement \gls{sm} we leverage the \gls{mose}~\cite{calagna2025mose}, which consists of two fundamental off-the-shelf tools, namely, CRIU~\cite{criu} and Podman~\cite{podman}. The former is widely considered the key tool to achieve \gls{sm} at a process level, and the latter extends CRIU functionalities to a container level (e.g., xApp containers). Furthermore, \gls{mose} implements the \gls{pam} model~\cite{calagna2024design} to accurately characterize the fundamental migration \glspl{kpi} as a function of the xApp memory usage and dirty-page rate (see Sec.~\ref{work06:sec:preliminaries:sub:stateful_mig}). Leveraging CRIU, Podman, and \gls{pam} model, \gls{mose} configures and orchestrates the migration process to fulfill both the target migration \glspl{kpi} and the vertical's objective, i.e., to  minimize either the migration downtime (\gls{sm}-MD) or the resource consumption in terms of required network bandwidth and CPU usage (\gls{sm}-MR). Also, depending on such objective, we configure the maximum bandwidth used by \gls{mose} as follows: 1\,Gbps (i.e., underutilizing our bandwidth resources) for SM-MR and 5\,Gbps for SM-MD (i.e., saturating our bandwidth resources).

{\bf SDL.} As mentioned in Sec.~\ref{work06:sec:preliminaries:sub:sdl_mig}, to attain migration based on \gls{sdl}, we use etcd to create a distributed backend database. For fault-tolerance purposes, we fix the size of the etcd cluster to three, i.e., the number of the control-plane nodes in our OpenShift cluster. It is worth mentioning that the number of etcd instances is not meant to vary in real-time, as it depends only on the cluster architecture design. To control the etcd maintenance operations, we consider two parameters: (i) the ``snapshot count'', i.e., the number of key-value pairs revisions to retain before compaction; and (ii) the ``maintenance period'', i.e., how often an etcd instance performs compaction and defragmentation. While the latter can be configured in real-time, the snapshot count can be configured only upon etcd cluster bootstrap. Therefore, we set such value to 100 as (i) previous revisions become obsolete, i.e., we only retain the key-value pairs that are needed but we are not interested in their history of changes; and (ii) we observed that this value is the smallest that prevents etcd overload with negligible impact on the overall performance in our testbed.

To conclude, our testbed includes: (i) a real end-to-end \gls{o-ran} system; (ii) a full-fledged computing architecture; (iii) a representative, programmable, and AI-driven xApp; and (iv) a migration framework based on off-the-shelf tools. This setup enables accurate emulation of real-world \gls{o-ran} scenarios and thorough evaluation of migration performance and trade-offs under various strategies and traffic conditions.

\section{Experimental Analysis}\label{work06:sec:exp_analysis}
We use our testbed to experimentally characterize the xApp migration process under \gls{sm} and \gls{sdl}.
We focus on a diverse set of xApp models that capture different use cases (Sec.~\ref{work06:sec:exp_analysis:sub:scenario}). Then, we thoroughly analyze performance and limitations of both approaches, focusing on temporal KPIs (Sec~\ref{work06:sec:exp_analysis:sub:temporal_analysis}) and resource usage (Sec~\ref{work06:sec:exp_analysis:sub:res_analysis}).
All presented results are averaged over 50 repetitions and have a 95\% confidence interval.

\subsection{xApp Reference Scenarios}\label{work06:sec:exp_analysis:sub:scenario}
Although our approach is general, for the sake of illustration, we consider a set $\apps$ of four representative classes of xApp, i.e., $\apps {=} \{A, B, C, D\}$.
Each class represents a realistic \gls{ran} workload scenario and differs in the (i) number of \glspl{kpm} requested from the \gls{ran}, which reflects the size $\omega_{\mathrm{s},k}$ of the E2 RIC Indication (report) messages; (ii) the periodicity $\omega_{\mathrm{p},k}$ of such messages; and (iii) the xApp state size $\rho$.

\begin{table}[t!]
\footnotesize
\caption{Classes of xApp. Each class is also evaluated against different values of state size $\rho {\in} \{1\,\mathrm{MB}, 10\,\mathrm{MB}, 100\,\mathrm{MB}\}$}
\label{work06:tab:xapps_types}
\centering
\begin{tabular}{|c|c|c|c|c|}
\hline
Type/Features & A & B & C & D \\
\hline
Message Size, $\omega_{\mathrm{s},k}$ & 100\,B & 100\,B & 100\,kB & 100\,kB\\
\hline
Message Period, $\omega_{\mathrm{p},k}$ & 1\,s & 100\,ms & 1\,s & 100\,ms\\
\hline
Reference use case & mMTC & IoT & Analytics & UAVs\\
\hline
\end{tabular}
\vspace{-2mm}
\end{table}

As shown in Table~\ref{work06:tab:xapps_types}, each xApp class $k {\in} \apps$ is defined by the 2-tuple  $(\omega_{\mathrm{s},k},\omega_{\mathrm{p},k})$.
Class A addresses scenarios with loose control loops and few \glspl{kpm} (i.e., small message size), typical of control for \gls{mmtc} applications. Class B also features few \glspl{kpm} but with tight control loops, aligning with \gls{iot} telemetry requirements. Class C involves large messages and loose control loops, common in surveillance and analytics applications. Eventually, Class D targets scenarios where control is frequent (e.g., every 100\,ms) and many \glspl{kpm} are processed at the same time (i.e., large message size), which models time-critical applications, e.g., self-driving \glspl{uav}, requiring low latency control.
To consider a wide range of use cases and applications, for each xApp class $k$ we also consider multiple values of state size, i.e., $\rho {\in} \{1\,\mathrm{MB}, 10\,\mathrm{MB}, 100\,\mathrm{MB}\}$.

\subsection{Temporal KPIs Analysis}\label{work06:sec:exp_analysis:sub:temporal_analysis}
{\bf SM.}
We recall that the temporal \glspl{kpi} of the stateful migration process are the migration downtime and the total migration duration, respectively denoted as $T_{\mathrm{D}}^{\mathrm{SM}}$ and $T_{\mathrm{M}}^{\mathrm{SM}}$. These can be characterized via the \gls{pam} model~\cite{calagna2024design} which describes them as functions of the xApp memory usage $M_k$, and the dirty-page rate.
To analyze the dirty-page rate in a way that is independent of the state size, we use its normalized version $r_k$ with respect to the minimum and maximum dirty-page rate values a microservice can achieve. The former is 1\,page/s and the latter is total number of pages allocated in memory per second.
By focusing on the least and most demanding classes, i.e., A and D, we now characterize $M_k$ and $r_k$ and the corresponding values for the \glspl{kpi}.


\begin{figure}[t!]
    \centering
    
    \subfloat[][DRL-xApp]{\label{work06:fig:mem_dpr_vs_state_size_DRL}{\includegraphics[width=0.235\textwidth]{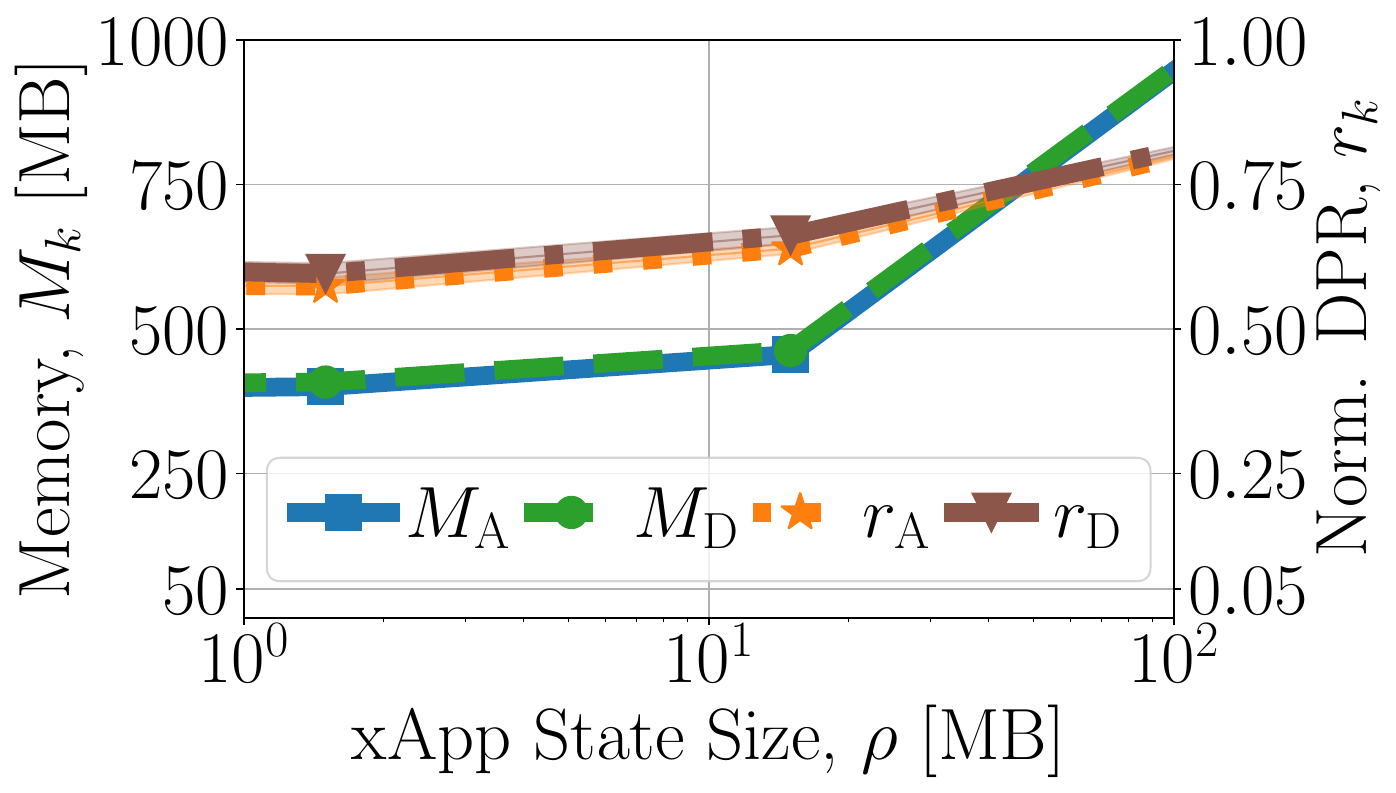}}}
    \hspace{0.5mm}
    \subfloat[][KPM-xApp]{\label{work06:fig:mem_dpr_vs_state_size_KPM}{\includegraphics[width=0.235\textwidth]{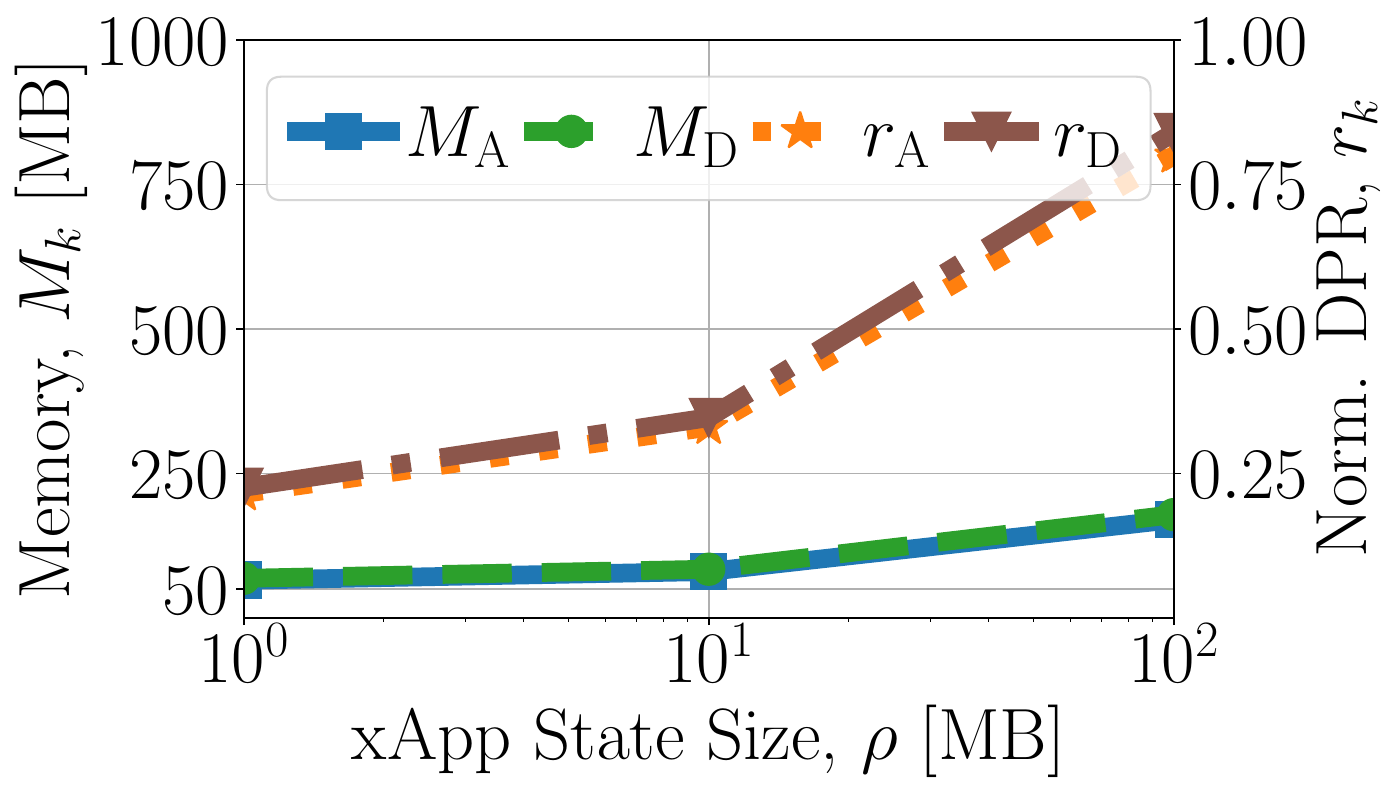}}}

    \caption{\rev{xApp memory usage $M_k$ and normalized dirty-page rate $r_k$ across DRL- and KPM- xApps and varying classes $k$.}}
    \label{work06:fig:mem_dpr_vs_state_size}
    \vspace{-4mm}
\end{figure}

\begin{figure}[t!]
    \centering
    {\includegraphics[width=1\columnwidth]{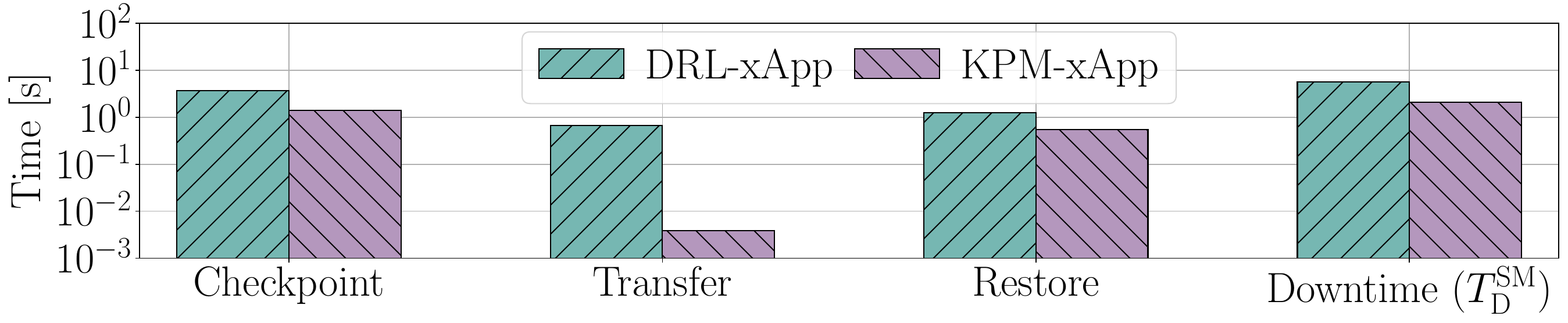}}
    \caption{\rev{Migration downtime components across DRL- and KPM- xApps under the SM-MD strategy and for $\rho{=}1$\,MB.}}
    \label{work06:fig:sm_downtime_breakdown}
    \vspace{-2mm}
\end{figure}

Fig.~\ref{work06:fig:mem_dpr_vs_state_size} shows $M_k$ and $r_k$ for varying state size $\rho$ and xApp classes. We notice that the values of $M_k$ of the DRL-xApp are significantly higher than $\rho$ and they are independent of $k$, as $M_k$ is affected by neither the message size nor the message frequency. Also, $r_k$ takes large values, which indicates that most of the xApp memory content changes every second and these variations are independent of the xApp class. This behavior is because (i) the xApps's \gls{ai} model consumes more memory than that used to store the \glspl{kpm} received over E2; and (ii) the execution of \gls{ai} models requires frequent allocation/release of memory pages to handle tensors~\cite{gao2020estimating}. Despite these results have been obtained by using the DRL-based xApp architecture from \cite{bonati2021intelligence}, they can be extended to general AI-based xApps, whose models and workload characteristics may vary, but still be dominant in terms of memory consumption. \rev{Looking at the results for the KPM-xApp (Fig.~\ref{work06:fig:mem_dpr_vs_state_size_KPM}), we observe that the values of $M_k$ are considerably lower than those of the DRL-xApp. Also, $r_k$ is now strongly correlated with $\rho$, thus ranging from small to large values. This behavior is due to the absence of the AI model, which makes the xApp state the dominant contribution to the overall memory usage. Nevertheless, despite such difference in behavior, both $M_k$ and $r_k$ remain independent of the traffic class $k$ as in the DRL-xApp case, yielding that frequency and size of the KPM messages have no impact on the way in which dynamic memory is managed by the operating system.

\begin{observation}[Relevant xApp features]\label{work06:obs:xapp_memory_dpr}
    \rev{Regardless of the xApp nature, its memory usage $M_k$ and dirty-page rate $r_k$ depend primarily on the state size $\rho$ and not the traffic class $k$.}
\end{observation}
}

\begin{figure}[t!]
    \centering
    \subfloat[][]{\label{work06:fig:sm_down_vs_num_xapps_MR}{\includegraphics[width=0.24\textwidth]{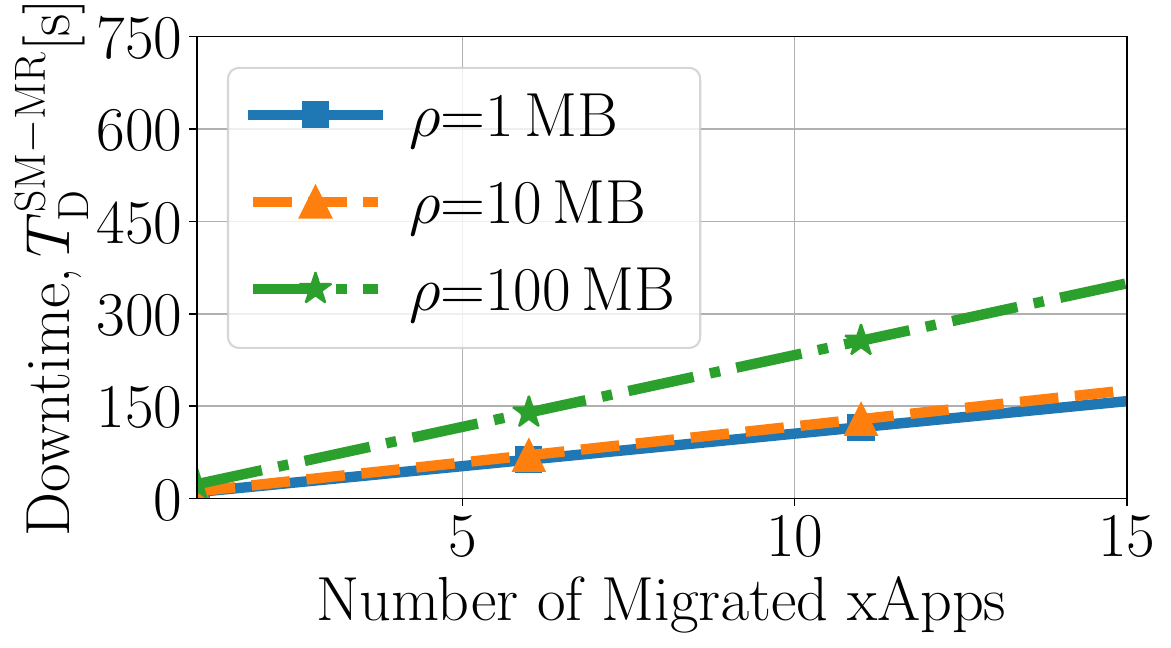}}}
    \subfloat[][]{\label{work06:fig:sm_down_vs_num_xapps_MD}{\includegraphics[width=0.24\textwidth]{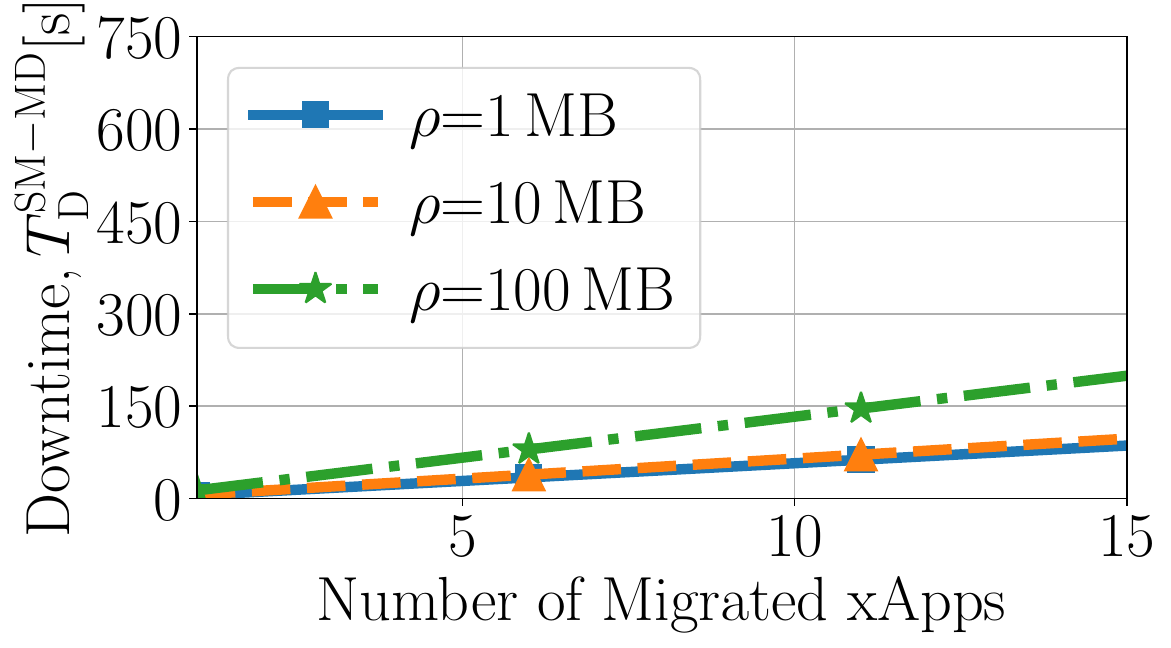}}}
    \\
    \subfloat[][]{\label{work06:fig:sm_mig_vs_num_xapps_MR}{\includegraphics[width=0.24\textwidth]{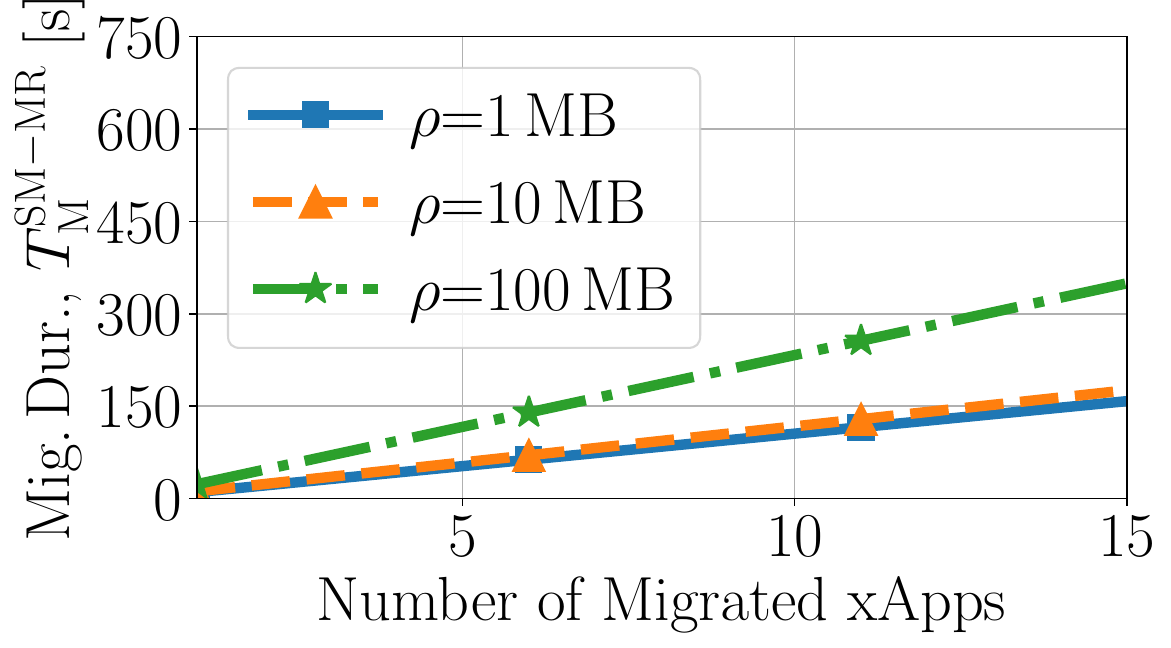}}}
    \subfloat[][]{\label{work06:fig:sm_mig_vs_num_xapps_MD}{\includegraphics[width=0.24\textwidth]{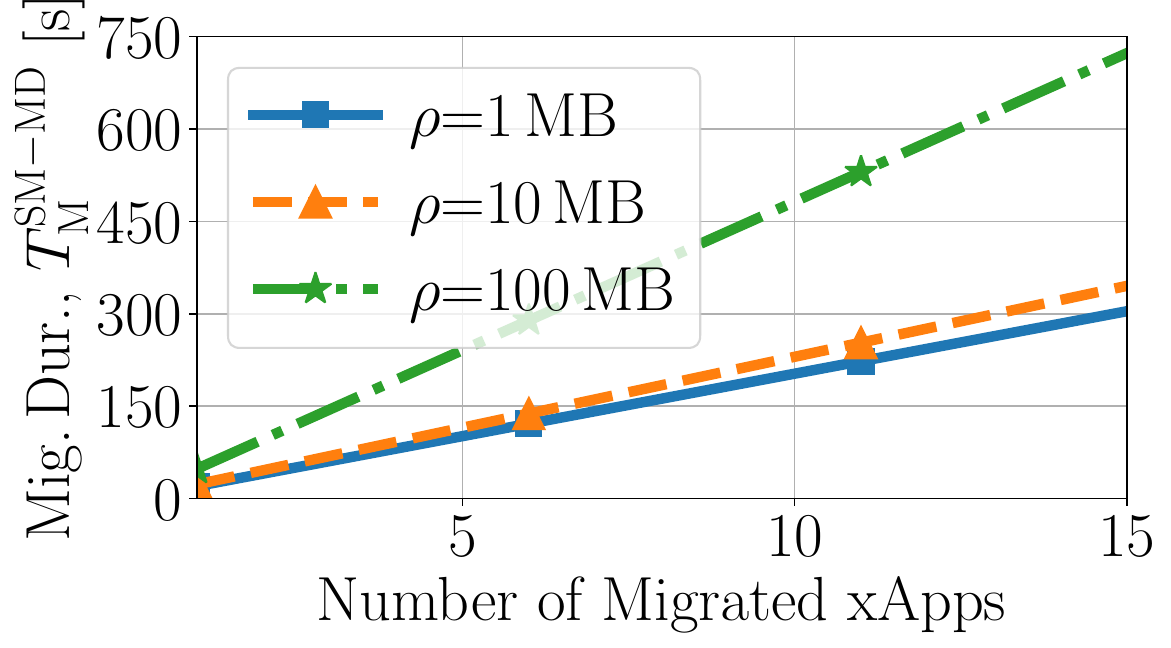}}}
    \caption{\rev{Stateful migration KPIs for varying xApp state size $\rho$.}}
    \label{work06:fig:sm_kpis}
    \vspace{-4mm}
\end{figure}

\rev{
Before evaluating the migration KPIs under varying load conditions, we first analyze the migration downtime into its fundamental components and assess whether SM is compatible with near-RT RIC control loop deadlines. As discussed in Sec.~\ref{work06:sec:preliminaries:sub:stateful_mig}, we recall that such downtime primarily consists of three main stages, namely, checkpoint, transfer, and restore of the xApp state. We characterize each stage separately by migrating a single DRL-/KPM-xApp under $\rho{=}1$\,MB and SM-MD, i.e., the strategy that minimizes downtime. Fig.~\ref{work06:fig:sm_downtime_breakdown} shows that migrating a KPM-xApp yields an approximately $3\times$ lower value of $T_{\mathrm{D}}^{\mathrm{SM}}$ compared to the DRL-xApp, with highest gap on the network transfer contribution. This is due to the fact that, as shown in Fig.~\ref{work06:fig:mem_dpr_vs_state_size}, the KPM-xApp features much lower values of $M_k$ and $r_k$, yielding lower processing complexity and a lighter data transfer from source to destination node. However, consistently with the evidences in~\cite{calagna2024design}, even when migrating a lightweight KPM-xApp under SM-MD, the latency overhead introduced by checkpoint and restore operations dominates the overall migration downtime, which is in the order of 2\,s and thus incompatible with any near-RT RIC control loop deadline. Despite this, \gls{sm} remains a fundamental technique to consider, particularly in scenarios where xApp state decoupling via \gls{sdl} is impractical or infeasible, as discussed in the following. Since \gls{sm} is a one-time event, it can be flexibly scheduled at those times where load is low and/or temporary service disruption has minimal impact on network performance and is tolerable. 
To account for this and for potential future developments that reduce the processing complexity, we keep our model of the migration KPIs general and allow tunable values for the maximum acceptable migration downtime. 

\begin{observation}[Stateful migration feasibility]\label{work06:obs:sm_feasibility}
    \rev{Regardless of the nature and features of the xApp, current technical limitations of the available migration tools render the SM strategy  incompatible with any near-RT RIC control loop deadline.}
\end{observation}

From now on, we focus our evaluation on the DRL-xApp, which realistically represents most demanding and AI-driven use cases. All considerations are in fact independent of the nature of the xApp and thus generalize to the case of simpler xApps.
}
Fig.~\ref{work06:fig:sm_kpis} depicts the \rev{cumulative} $T_{\mathrm{D}}^{\mathrm{SM}}$ and $T_{\mathrm{M}}^{\mathrm{SM}}$ as functions of the number of xApps being sequentially migrated and the value of xApp state size $\rho$, respectively. Results demonstrate that SM-MR yields $T_{\mathrm{D}}^{\mathrm{SM}}{=}T_{\mathrm{M}}^{\mathrm{SM}}$ while SM-MD achieves a lower $T_{\mathrm{D}}^{\mathrm{SM}}$ at the cost of a higher value of $T_{\mathrm{M}}^{\mathrm{SM}}$. Also, it can be observed that (i) both \glspl{kpi} depend on $\rho$; (ii) regardless of the \gls{sm} strategy, the dependency of the \glspl{kpi} on the number of xApps can be described by a linear function. 

\rev{
\begin{observation}[Stateful migration KPIs]\label{work06:obs:sm_kpis}
    Although the migration downtime and the migration duration depend on the stateful migration strategy and value of state size, both linearly increase with the number of migrated xApps.
\end{observation}
}

{\bf SDL.}
We now investigate the performance and resource usage of the xApp migration process with SDL. Specifically, we (i) assess the impact of \gls{sdl} on the migration \glspl{kpi} and the xApp resource usage; (ii) characterize the service disruption due to etcd maintenance; and (iii) analyze the etcd resource usage in terms of power consumption and CPU, memory, and disk usage for varying system configurations. We recall that the \gls{sdl} strategy decouples the stateful component of each xApp from the xApp itself as the state is stored in the backend database. Therefore, under \gls{sdl}, stateful xApps are treated as stateless from the migration viewpoint, enabling a zero-downtime migration process (see Sec.~\ref{work06:sec:preliminaries:sub:sdl_mig}).

\begin{figure}[t!]
    \centering
    \subfloat[][]{\label{work06:fig:sdl_mig_dur_vs_num_xapps}{\includegraphics[width=0.48\columnwidth]{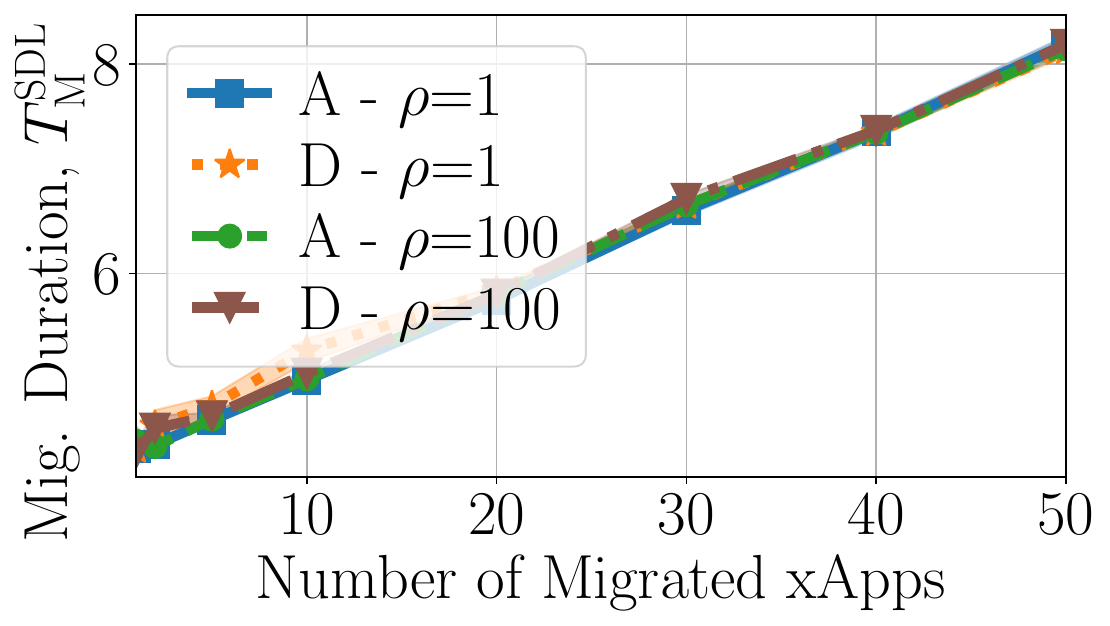}}}
    \hspace{1mm}
    \subfloat[][]{\label{work06:fig:put_vs_num_xapps}{\includegraphics[width=0.49\columnwidth]{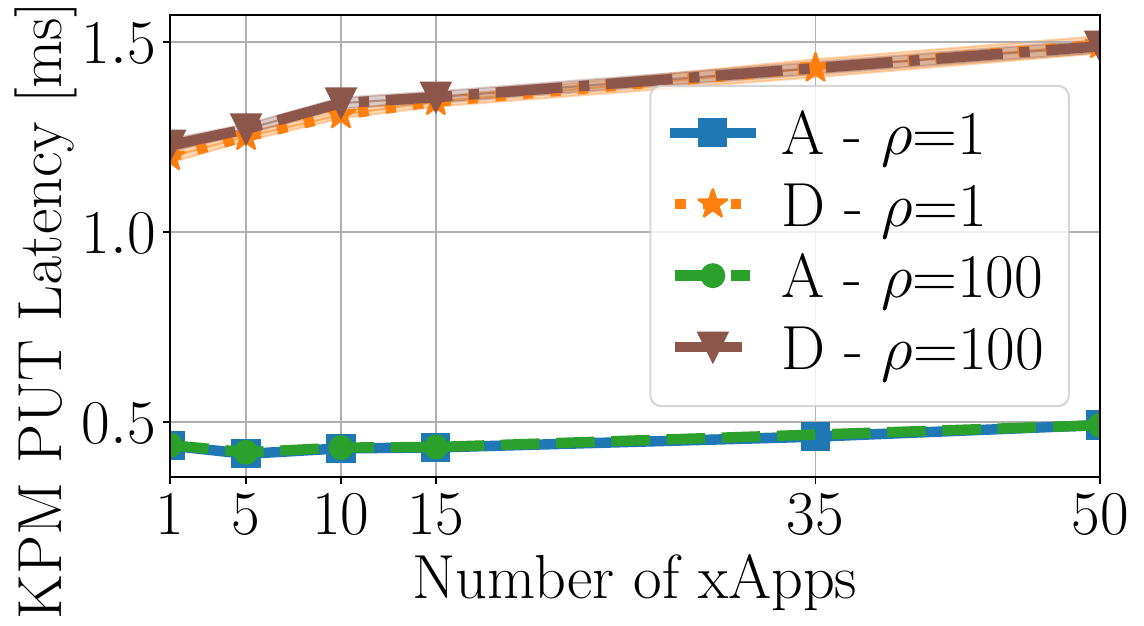}}}
    \\
    \subfloat[][]{\label{work06:fig:cdf_vs_latency}{\includegraphics[width=\columnwidth]{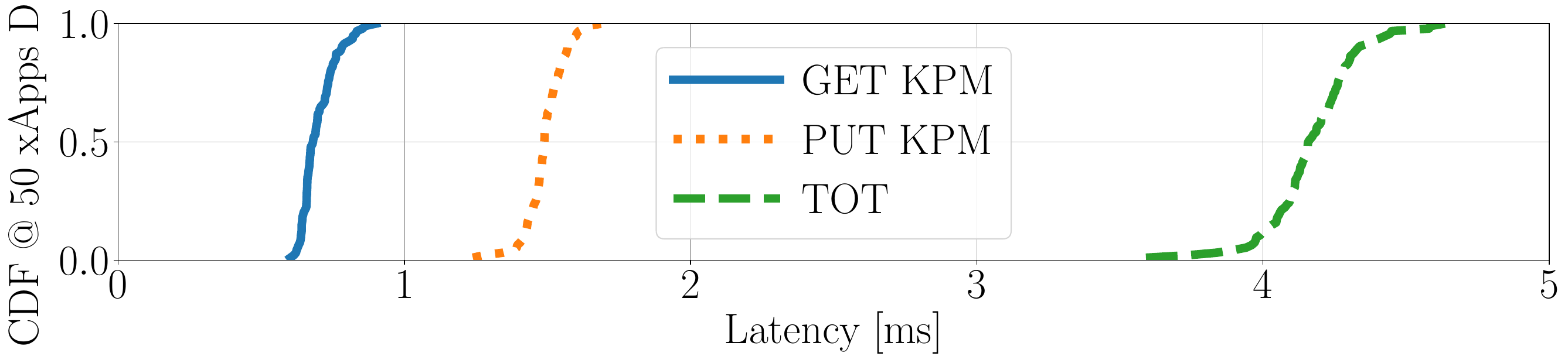}}}
    \caption{\rev{Stateless migration performance analysis: (a) migration duration, (b) average etcd KPM PUT latency, (c) etcd latency CDFs.}}
    \label{work06:fig:stateless_migration}
    \vspace{-4mm}
\end{figure}


Fig.~\ref{work06:fig:sdl_mig_dur_vs_num_xapps} shows the migration duration $T_{\mathrm{M}}^{\mathrm{SDL}}$ as a function of the number of migrated xApps. Results highlight that $T_{\mathrm{M}}^{\mathrm{SDL}}$ is independent of both $k$ and $\rho$. Indeed, xApps are virtually stateless under \gls{sdl} and $T_{\mathrm{M}}^{\mathrm{SDL}}$ corresponds to the time needed to instantiate new xApps, which mostly depends on the amount of memory to be allocated, that is now independent of the xApp class and state size. For the same reason, $T_{\mathrm{M}}^{\mathrm{SDL}}$ is up to two orders of magnitude lower than $T_{\mathrm{M}}^{\mathrm{SM}}$ (Fig.~\ref{work06:fig:sm_kpis}).

\begin{observation}[SDL migration KPIs]\label{work06:obs:sdl_kpis}
    Under \gls{sdl}, xApps are virtually stateless migration-wise. The migration duration (i) is independent of both the xApp class and the state size; and (ii) grows with the number of xApps being migrated linearly.
\end{observation}

\begin{figure*}[t!]
    \centering
    \subfloat[][$\rho=1\,\mathrm{MB}$, $\nu=1\,\mathrm{s}$]{\label{work06:fig:defrag_s1_m1_vs_traffic}{\includegraphics[width=0.25\textwidth]{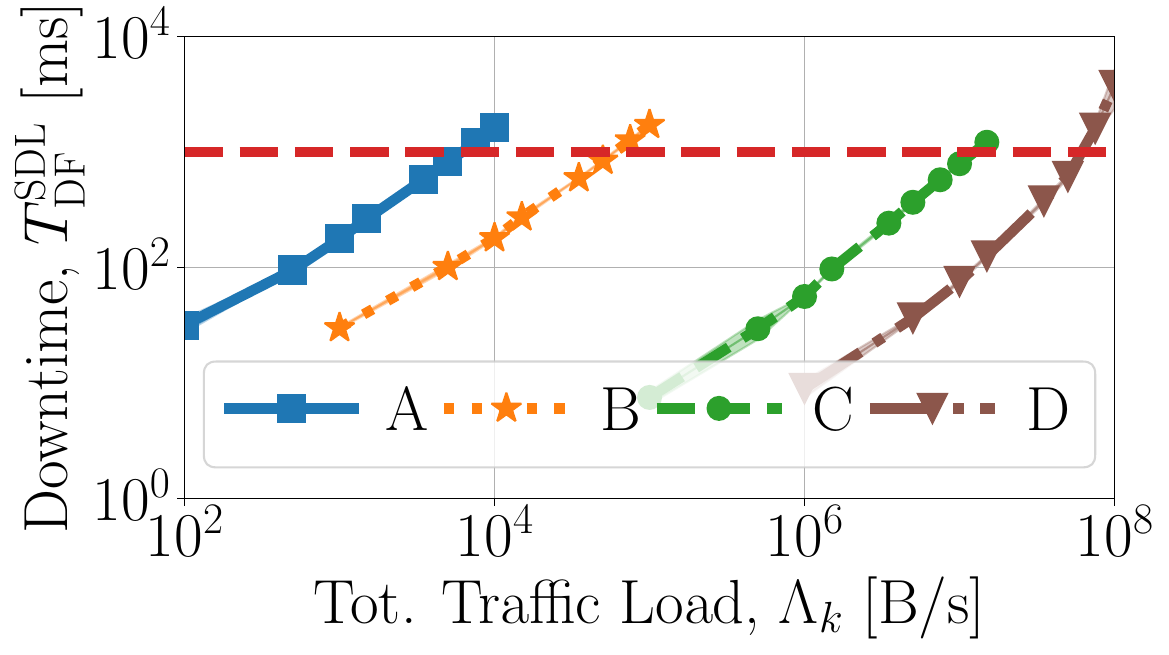}}}
    \subfloat[][$\rho=10\,\mathrm{MB}$, $\nu=1\,\mathrm{s}$]{\label{work06:fig:defrag_s10_m1_vs_traffic}{\includegraphics[width=0.25\textwidth]{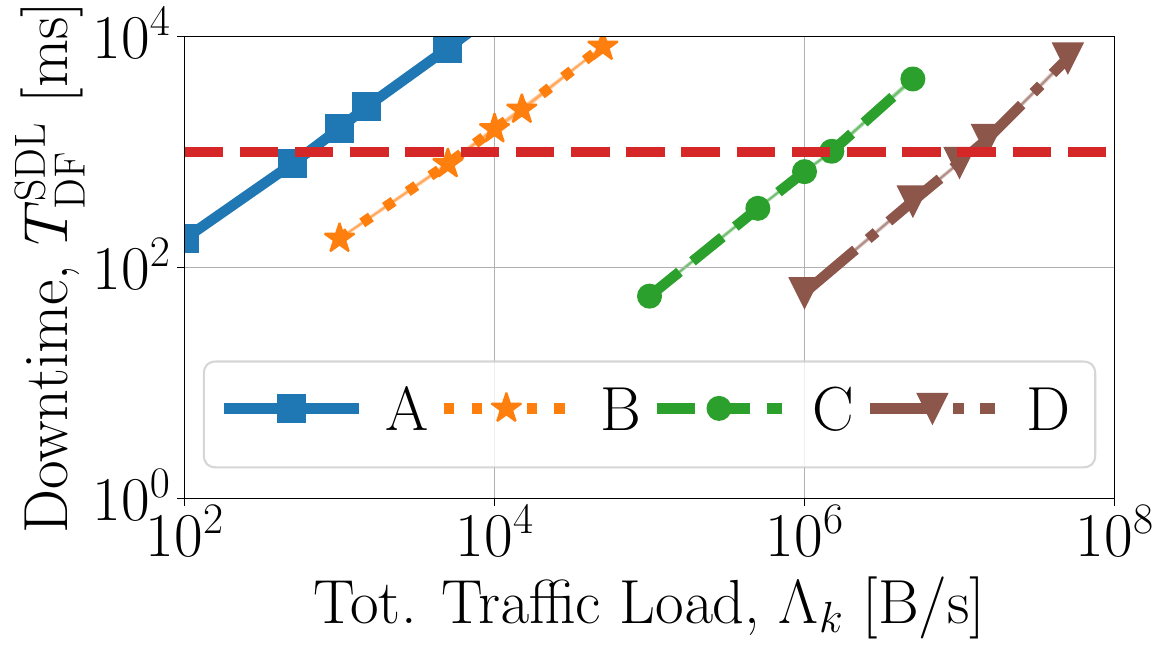}}}
    \subfloat[][$\rho=1\,\mathrm{MB}$, $\nu=120\,\mathrm{s}$]{\label{work06:fig:defrag_s1_m120_vs_traffic}{\includegraphics[width=0.25\textwidth]{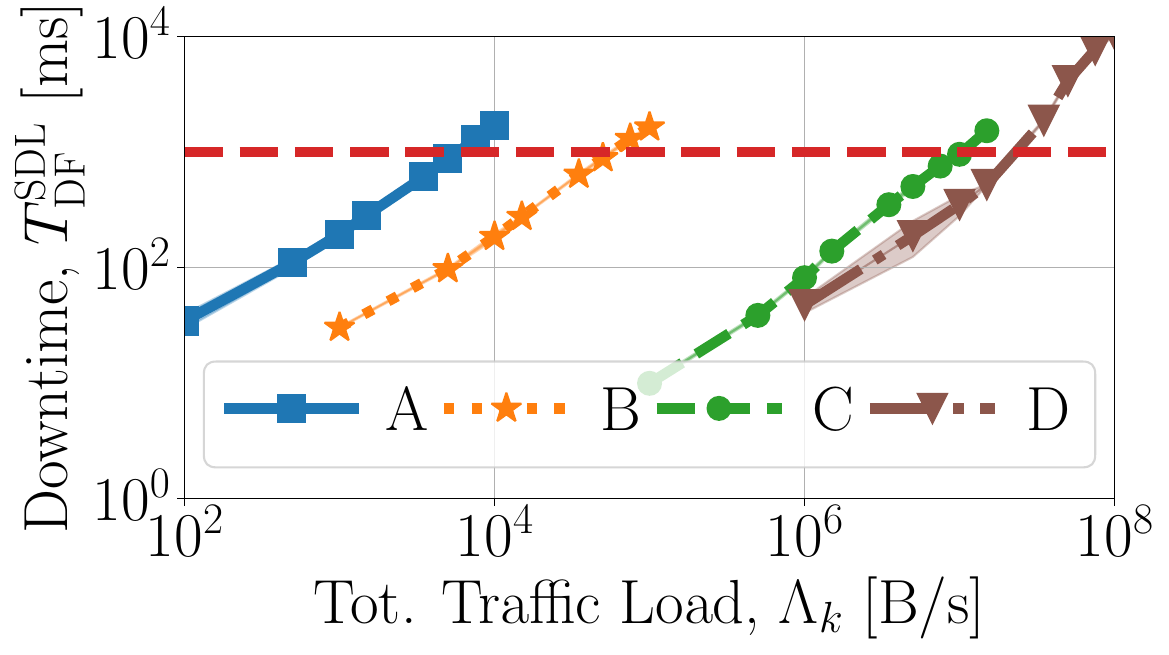}}}
    \subfloat[][$\rho=10\,\mathrm{MB}$, $\nu=120\,\mathrm{s}$]{\label{work06:fig:defrag_s10_m120_vs_traffic}{\includegraphics[width=0.25\textwidth]{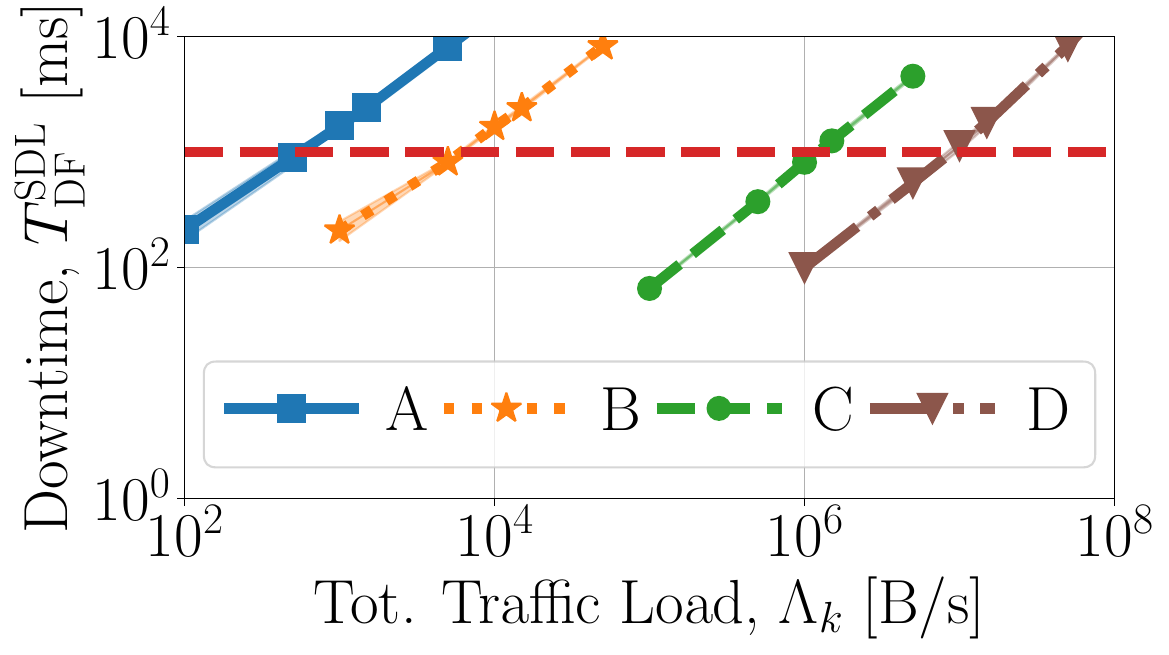}}}
    \caption{Etcd defrag downtime for varying classes of xApp, values of xApp state size $\rho$ and etcd maintenance period $\nu$.}
    \label{work06:fig:defrag_downtime_vs_traffic}
    \vspace{-3mm}
\end{figure*}

As discussed in Sec.~\ref{work06:sec:preliminaries:sub:sdl_mig}, etcd is a reliable and robust backend database solution to allow \gls{sdl} in effectively decoupling the xApp from its internal state. Now, we also demonstrate experimentally that etcd meets the strict timing requirements of the near-RT \gls{ric}. \rev{To this end, we characterize the latency overhead that an xApp experiences when reading and writing key-value pairs from/to etcd as part of its control loop. First, we measure the average latency of a write operation (commonly referred to as {\em PUT} latency) for a single KPM key-value pair. Then, we delve into the entire xApp control loop, from KPM generation to control message production, and provide a breakdown of the total latency overhead.}

Fig.~\ref{work06:fig:put_vs_num_xapps} reports the \rev{KPM} PUT latency as a function of the number of xApps for different xApp classes and state size $\rho$ values. Results show that PUT latency is (i) increasing with the number of xApps due the larger number of requests to access the database; (ii) independent of the xApp state size that is stored on etcd; and (iii) dependent on the xApp class. We recall that class A is characterized by small and infrequent messages, while class D puts a higher pressure on etcd by generating large and frequent messages. Importantly, the PUT latency never exceeds two milliseconds even in extreme scenarios with many xApps of class D.

\begin{table}[t]
\footnotesize
\caption{\rev{Breakdown of the etcd communication latency overhead within an xApp control loop under high traffic load, i.e., 50 xApps of type D}}
\label{tab:sdl_breakdown_analysis}
\renewcommand{\arraystretch}{1.1}
\centering
\begin{tabular}{|
    p{3.3cm}|
    >{\centering\arraybackslash}p{1.7cm}|
    >{\centering\arraybackslash}p{1cm}|
    >{\centering\arraybackslash}p{1cm}|}
\hline
\bf Steps / Duration [ms]
    & \bf 95\% C.I.
    & \bf p95
    & \bf p99
    \\ \hline

\bf 1) Watch (GET) KPM
    & $0.699 \pm 0.012$
    & $0.824$
    & $0.865$
    \\ 

\bf 2--3) Push/Pop (PUT) KPM
    & $1.491 \pm 0.014$
    & $1.600$
    & $1.640$
    \\

\bf 4) Push (PUT) Ctrl Msg
    & $0.492 \pm 0.005$
    & $0.527$
    & $0.539$
    \\ \hline

\bf Total
    & $4.137 \pm 0.033$
    & $4.396$
    & $4.551$
    \\ \hline
\end{tabular}
\vspace{-4mm}
\end{table}

\rev{To assess etcd feasibility under any near-RT \gls{ric} control loop deadline ranging from 10\,ms to 1\,s, we now analyze the latency of the entire xApp control loop. As described in Sec.~\ref{work06:sec:testbed}, for each KPM message prodcued by the RAN, the xApp performs the following steps: (1) read the KPM message (GET); (2-3) update the state queue (KPM PUT of size $\omega_{\mathrm{s},k}$); (4) produce and push the control message (CTRL PUT of size fixed to 100\,B). We focus on a high traffic load scenario featuring 50 concurrently running xApps of class D. Fig.~\ref{work06:fig:cdf_vs_latency} reports the latency CDFs for the GET KPM and PUT KPM operations as well as the cumulative control loop latency overhead. Similarly, Table~\ref{tab:sdl_breakdown_analysis} analyzes the latency of each step, including the 95\,\% confidence interval and 95th- and 99th-percentile values. Results demonstrate that (i) the KPM PUT operations are slower than KPM GET due to the strong consistency guarantees, (ii) the KPM PUT operations are also slower than CTRL PUT ones because of their larger payloads, and (iii) the latency distributions---both for individual steps and for the full control loop---exhibit tight tails. These findings reveal than even in such high demand scenario, the total latency overhead never exceeds five milliseconds, which is compatible even with the tightest near-RT RIC control loop deadline of 10\,ms.
}

\begin{observation}[Etcd feasibility]\label{work06:obs:sdl_feasibility}
    Regardless of the xApp class and its state size, the communication latency introduced by etcd is negligible with respect to the near-RT control loop deadlines, making etcd a suitable solution for \gls{sdl}'s backend database.
\end{observation}

As discussed in Sec.~\ref{work06:sec:preliminaries:sub:sdl_mig}, etcd needs periodic maintenance operations, i.e., compaction and defragmentation of stale key-value pairs. Let $\nu$ be the maintenance period. The defragmentation of an etcd instance makes that instance unavailable every $\nu$ seconds. Therefore, to assess the impact of etcd maintenance on performance and resource usage, we now consider $\nu$ as a parameter for our analysis.

We start by investigating the defrag downtime $T_{\mathrm{DF}}^{\mathrm{SDL}}$ as a function of the total traffic load $\Lambda_k$ directed towards the etcd database. \rev{We define $\Lambda_k {=} N_k {\cdot} \omega_{\mathrm{s}, k}/ \omega_{\mathrm{p}, k}$, where $N_k$ is the number of concurrently active xApps of class $k$. We found $\Lambda_k$ to be the best auxiliary metric to compactly capture---and jointly encompass---both the number of xApps and the class-specific xApp features that influence the etcd workload, thus providing the clearest visualization of our results. Since each xApp class $k$ yields a fixed configuration of $\omega_{\mathrm{s},k}$ and $\omega_{\mathrm{p},k}$, analyzing $T_{\mathrm{DF}}^{\mathrm{SDL}}$ as a function of $\Lambda_k$ is equivalent to observing its relation with respect to the total number of xApps $N_k$.}

Fig.~\ref{work06:fig:defrag_downtime_vs_traffic} reports $T_{\mathrm{DF}}^{\mathrm{SDL}}$ as a function of $\Lambda_k$ for varying classes of xApp, state size $\rho$, and maintenance period $\nu$. The red dashed line in all plots underlines \rev{the exemplary} 1\,s control loop deadline. Some relevant findings on $T_{\mathrm{DF}}^{\mathrm{SDL}}$ can be highlighted: (i) regardless of the xApp class, its trend with respect to $\Lambda_k$ can be well approximated by a linear relation; (ii) it is strongly influenced by $\rho$, denoting a positive correlation; (iii) the dependency on $\nu$ is negligible, with the only exception of xApps class D, for which $T_{\mathrm{DF}}^{\mathrm{SDL}}$ increases with $\nu$ due to the significant load etcd is subject to; and (iv) $T_{\mathrm{DF}}^{\mathrm{SDL}}$ may be incompatible with the \rev{arbitrary near-RT \gls{ric} control loop deadline}, which hints at scalability issues for \gls{sdl}.

\begin{observation}[Defrag downtime]\label{work06:obs:sdl_scalability}
    For all classes of xApp, the defrag downtime increases linearly with the total \rev{number of xApps}. It substantially increases with the xApp state size while the dependency on the maintenance period is not as strong. Also, given the near-RT \gls{ric} threshold on such downtime, scalability limits of the \gls{sdl} approach emerge. 
\end{observation}

\subsection{Resource Usage Analysis}\label{work06:sec:exp_analysis:sub:res_analysis}

\begin{figure}[bt!]
    \centering
    \subfloat[][xApp non-SDL]{\label{work06:fig:tcp_cpu_power_vs_state_size}{\includegraphics[width=0.24\textwidth]{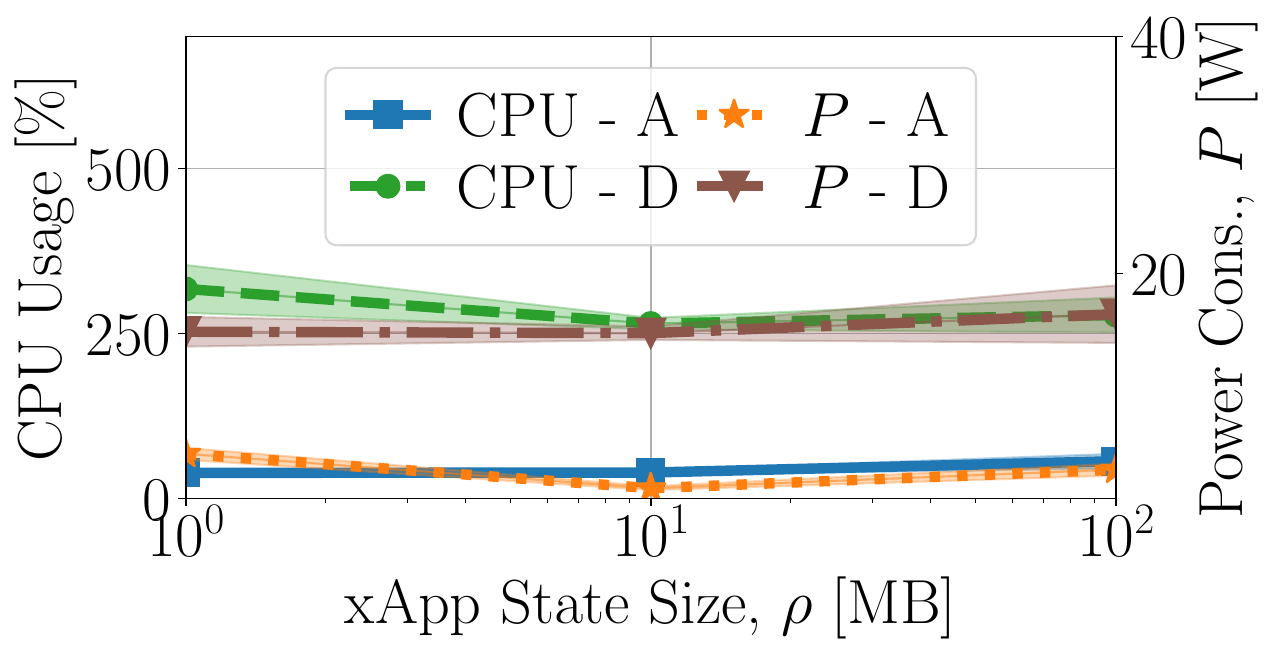}}}
    \hspace{0.2mm}
    \subfloat[][xApp SDL]{\label{work06:fig:etcd_cpu_power_vs_state_size}{\includegraphics[width=0.24\textwidth]{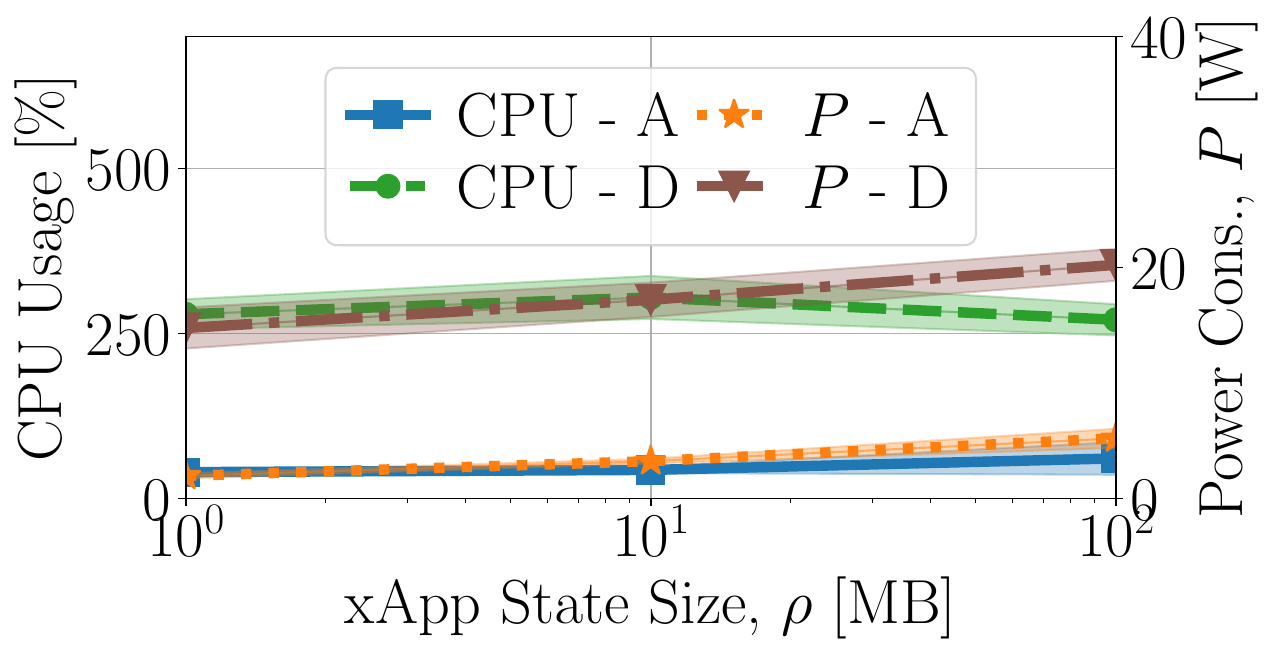}}}
    \caption{xApp resource usage, for both (a) non-SDL, and (b) SDL options and varying classes of xApp.}
    \label{work06:fig:xapp_sm_vs_sdl}
    \vspace{-3mm}
\end{figure}

{\bf xApp.}
We now analyze \gls{sdl}'s impact on xApp resource utilization in terms of CPU and power consumption.
Fig.~\ref{work06:fig:tcp_cpu_power_vs_state_size} and~\ref{work06:fig:etcd_cpu_power_vs_state_size} compare the case where the xApp allocates its state in memory (xApp non-\gls{sdl}), with that where the xApp uses \gls{sdl} to put its state on the backend database (xApp \gls{sdl}). Both figures report CPU and power consumption as functions of the xApp state size $\rho$ for varying xApp classes. We notice that the values of CPU and power consumption in both cases are comparable, suggesting that the way the xApp retains its state has no significant impact on its resource consumption. Results also underline that CPU and power consumption are independent of the xApp state size but strongly depend on the xApp class. In fact, the AI algorithm of an xApp of class D produces an inference on the input metrics every 100\,ms, yielding a higher resource consumption than an xApp of class A, which, instead, does that every 1\,s. 

\begin{observation}[xApp resource consumption]\label{work06:obs:xapp_res_usage}
    Regardless of the xApp state size, the use of \gls{sdl} has negligible impact on the xApp resource consumption. Furthermore, the resource consumption strongly depends on the xApp class and the frequency with which its AI algorithm is executed.
\end{observation}

\begin{figure}[ht!]
    \centering
    {\includegraphics[width=1\columnwidth]{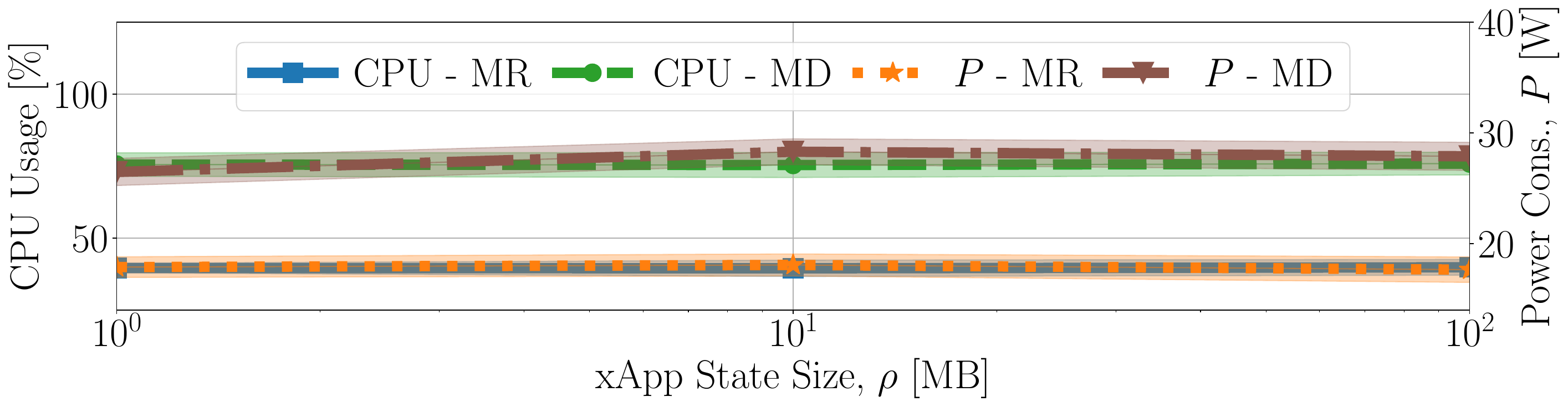}}
    \caption{Resource usage for both SM-MR and SM-MD.}
    \label{work06:fig:mose_cpu_power_vs_state_size}
    \vspace{-2mm}
\end{figure}

{\bf SM.}
Fig.~\ref{work06:fig:mose_cpu_power_vs_state_size} shows \rev{the instantaneous} CPU usage and power consumption as functions of the xApp state size for both SM-MR and SM-MD. As discussed in Sec.~\ref{work06:sec:preliminaries:sub:stateful_mig}, the additional complexity introduced by SM-MD to attain a lower downtime with respect to SM-MR yields higher CPU and power consumption. Remarkably, regardless of the \gls{sm} strategy, \rev{the instantaneous} CPU usage and power consumption remain constant as the state size grows. \rev{Indeed, the dependence on state size emerges in the SM KPIs (Finding~\ref{work06:obs:sm_kpis}), yielding that, as the state size increases, this same instantaneous resource usage must be sustained for a longer duration.}

\begin{observation}[Resource usage]\label{work06:obs:sm_res_usage}
    Stateful migration \rev{instantaneous} CPU and power usage are independent of the xApp state size and they are functions of the selected SM strategy.
\end{observation}

\begin{figure*}[t!]
    \centering
    \subfloat[][$\rho=1\,\mathrm{MB}$, $\nu=1\,\mathrm{s}$]{\label{work06:fig:cpu_vs_load_rho1_nu1}{\includegraphics[width=0.25\textwidth]{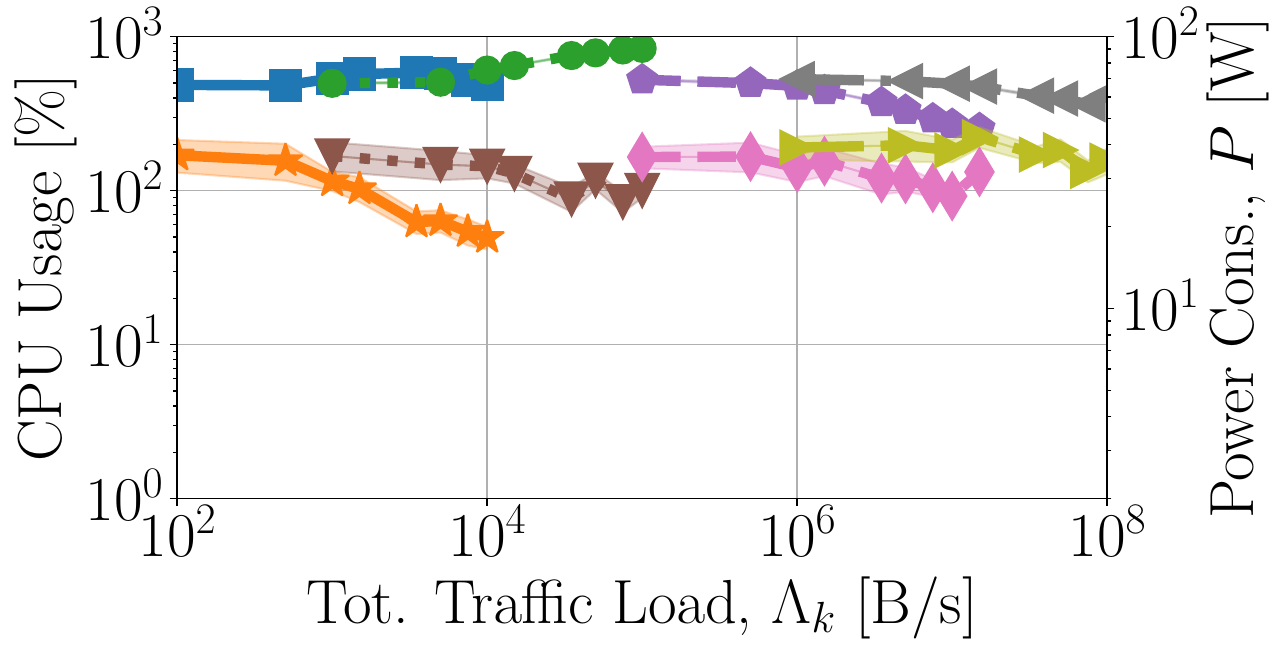}}}
    \subfloat[][$\rho=10\,\mathrm{MB}$, $\nu=1\,\mathrm{s}$]{\label{work06:fig:cpu_vs_load_rho10_nu1}{\includegraphics[width=0.25\textwidth]{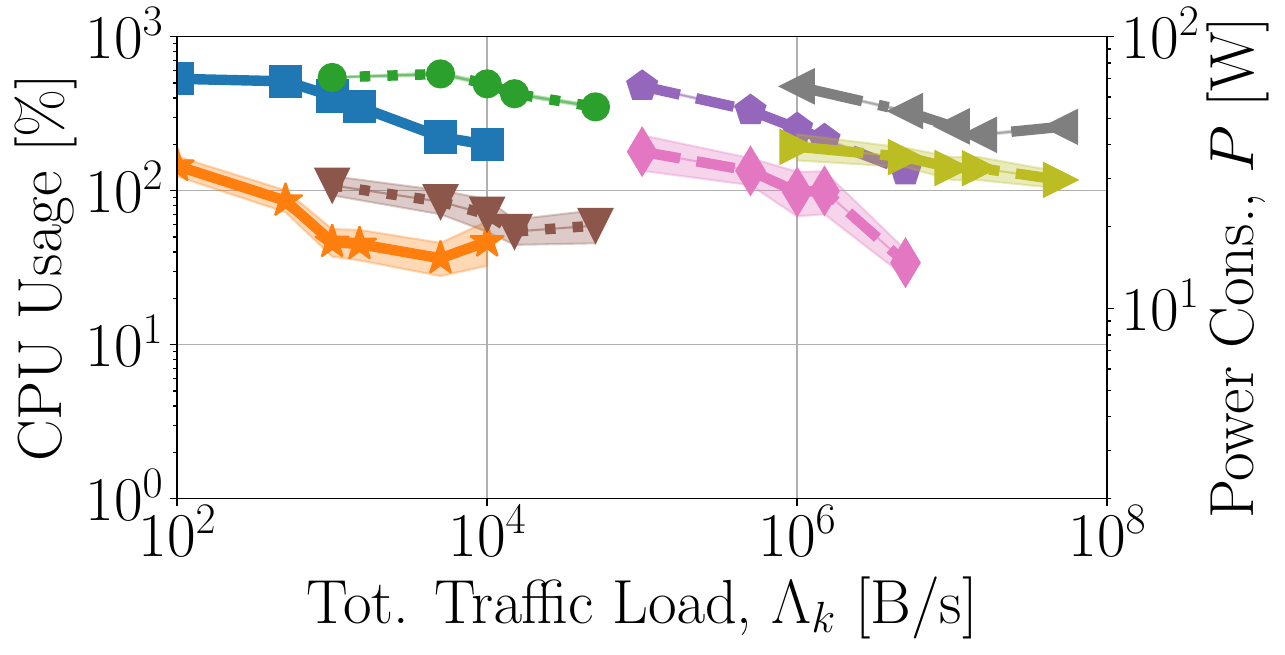}}}
    \subfloat[][$\rho=1\,\mathrm{MB}$, $\nu=120\,\mathrm{s}$]{\label{work06:fig:cpu_vs_load_rho1_nu120}{\includegraphics[width=0.25\textwidth]{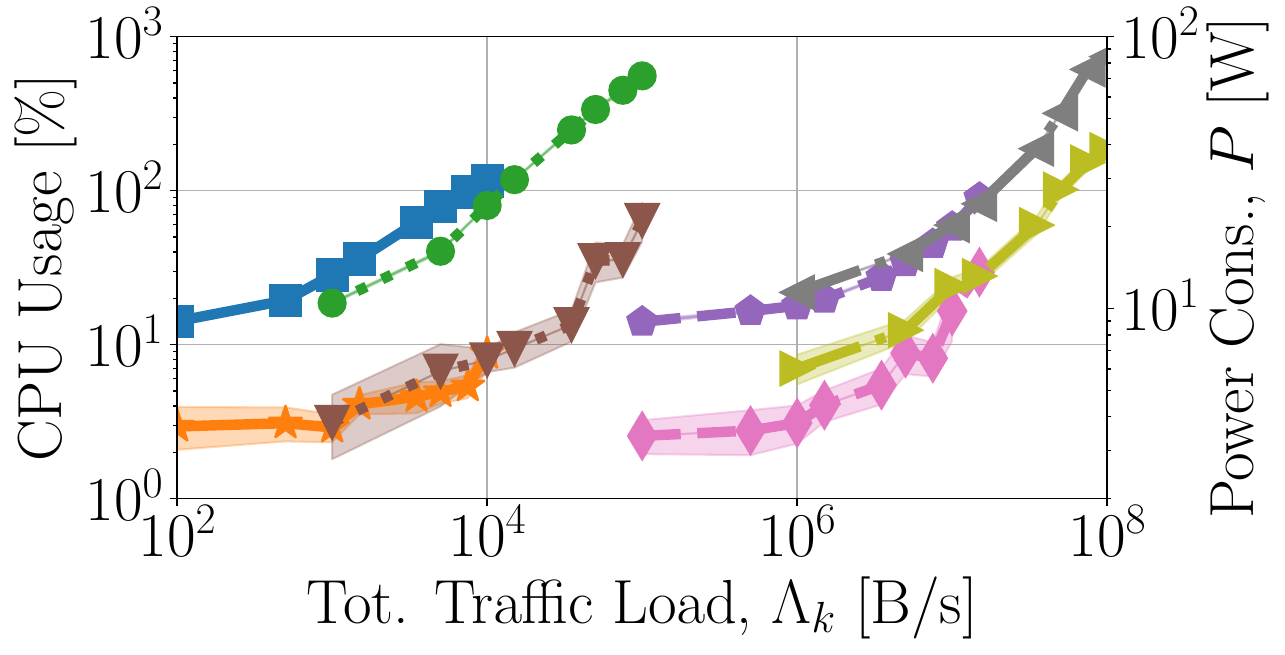}}}
    \subfloat[][$\rho=10\,\mathrm{MB}$, $\nu=120\,\mathrm{s}$]{\label{work06:fig:cpu_vs_load_rho10_nu120}{\includegraphics[width=0.25\textwidth]{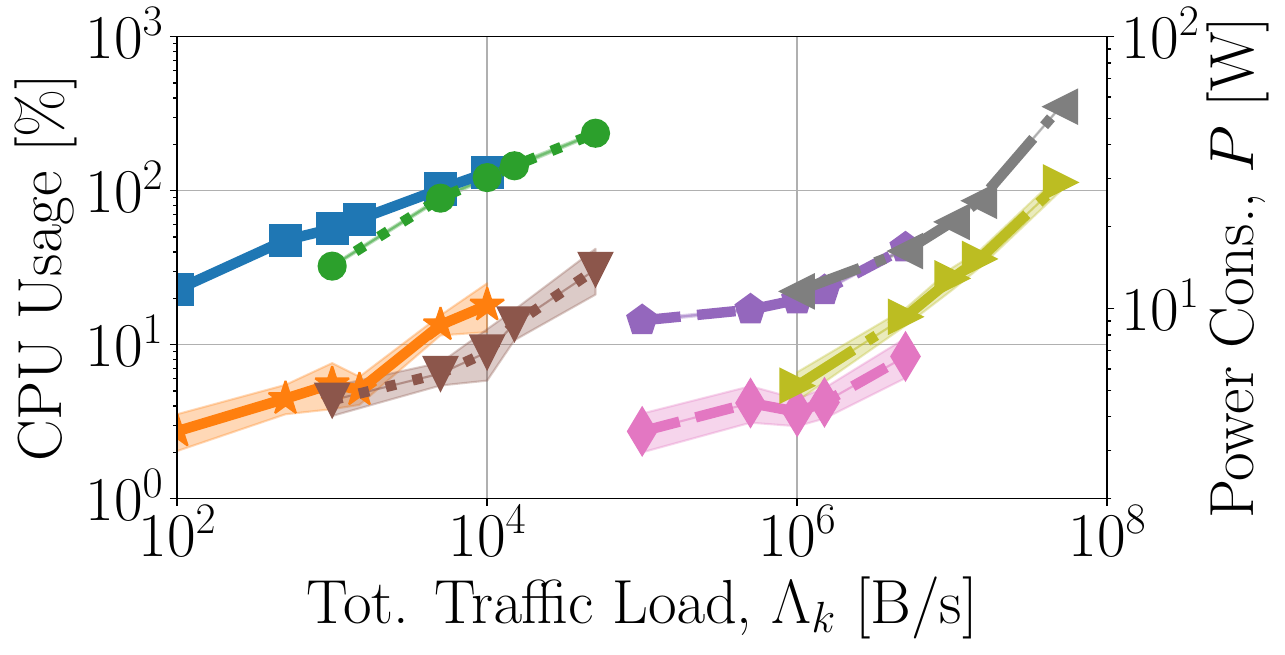}}}
    \\
    \vspace{1ex} \includegraphics[width=0.8\textwidth]{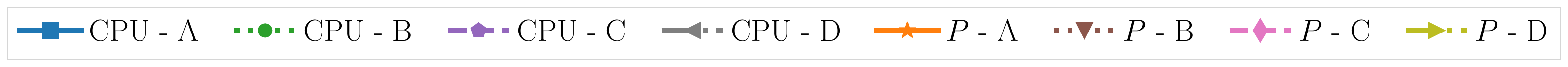}
    \caption{\rev{Etcd CPU and power consumption for varying xApp classes, values of xApp state size $\rho$, and maintenance period $\nu$.}}
    \label{work06:fig:cpu_vs_load_vs_rho_vs_nu}
    \vspace{-2mm}
\end{figure*}
\begin{figure*}[t!]
    \centering
    \subfloat[][$\rho=1\,\mathrm{MB}$, $\nu=1\,\mathrm{s}$]{\label{work06:fig:mem_vs_load_rho1_nu1}{\includegraphics[width=0.25\textwidth]{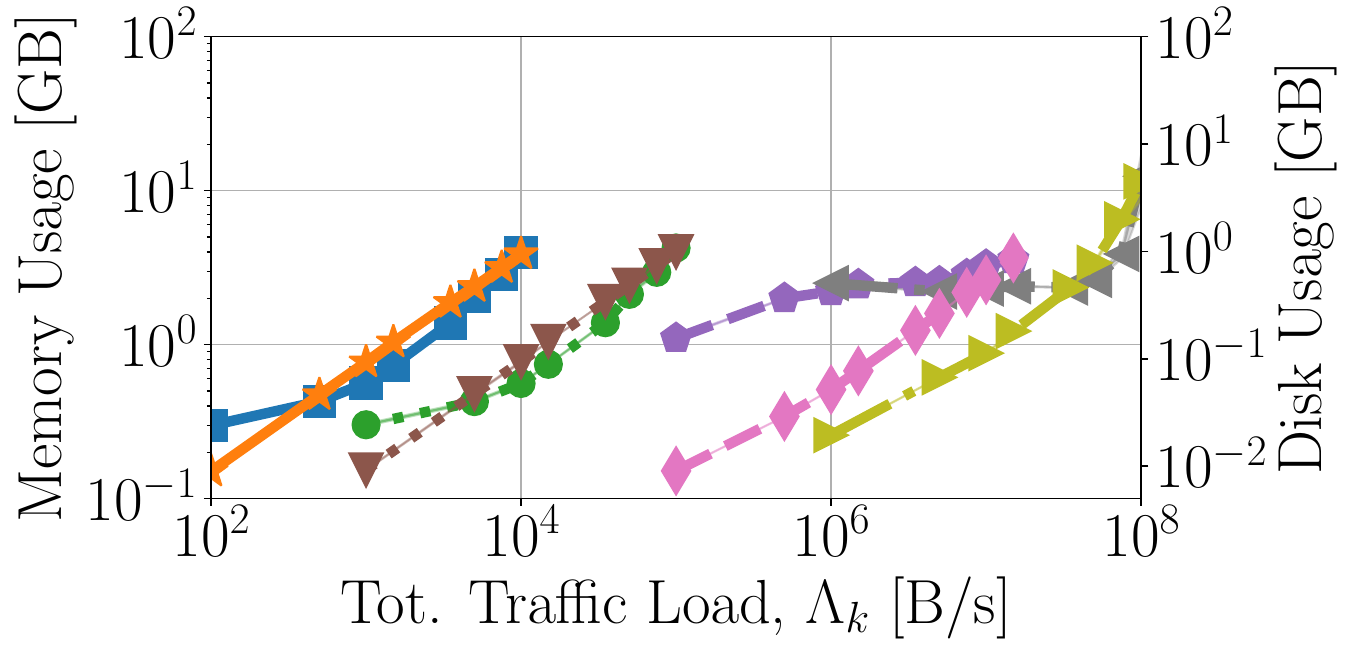}}}
    \subfloat[][$\rho=10\,\mathrm{MB}$, $\nu=1\,\mathrm{s}$]{\label{work06:fig:mem_vs_load_rho10_nu1}{\includegraphics[width=0.25\textwidth]{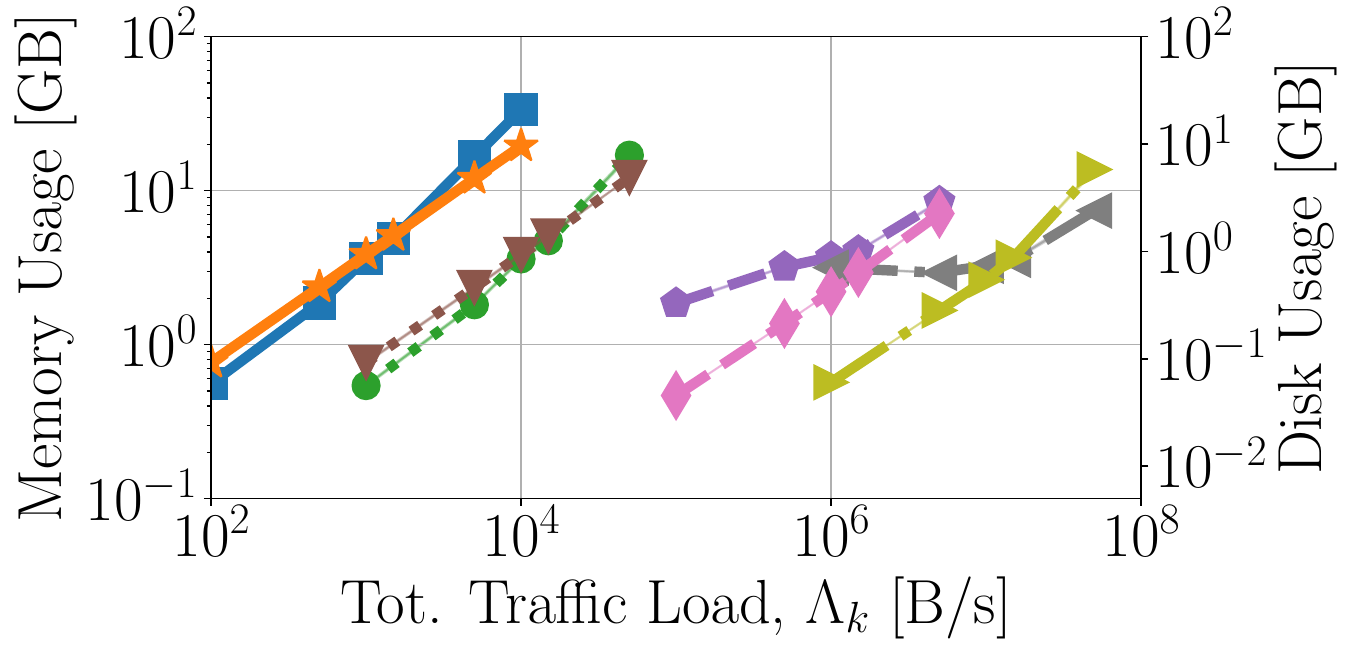}}}
    \subfloat[][$\rho=1\,\mathrm{MB}$, $\nu=120\,\mathrm{s}$]{\label{work06:fig:mem_vs_load_rho1_nu120}{\includegraphics[width=0.25\textwidth]{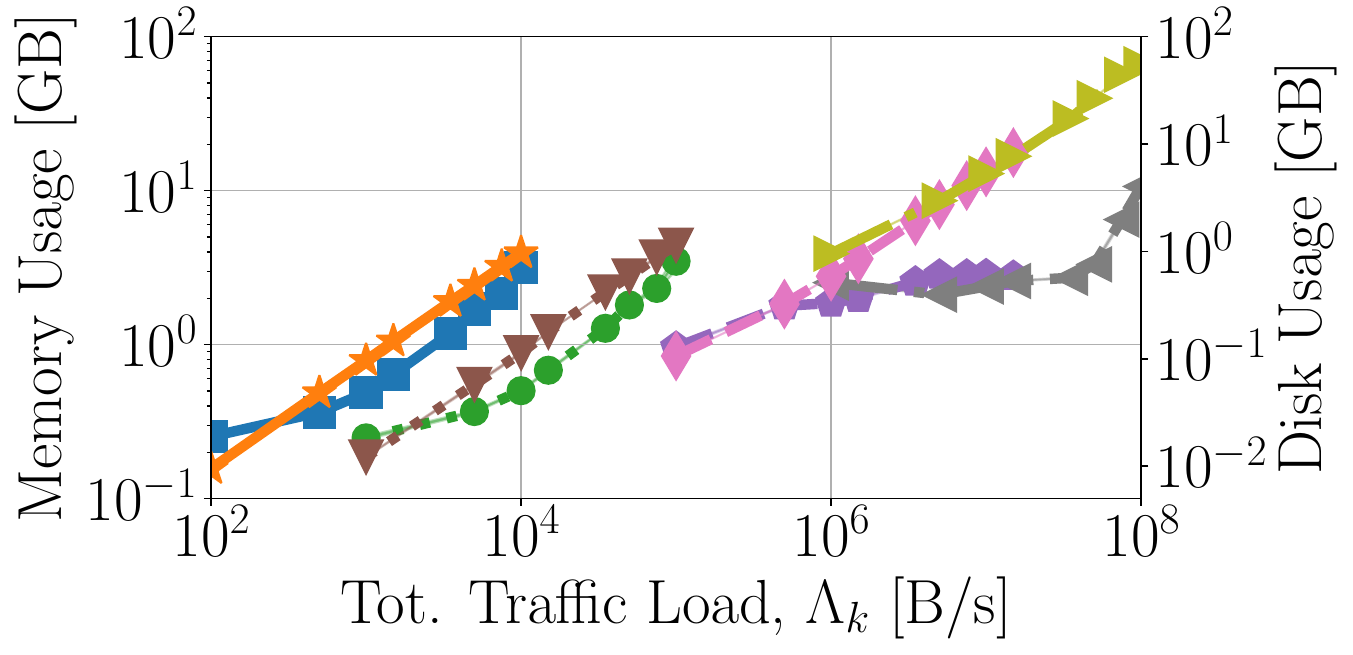}}}
    \subfloat[][$\rho=10\,\mathrm{MB}$, $\nu=120\,\mathrm{s}$]{\label{work06:fig:mem_vs_load_rho10_nu120}{\includegraphics[width=0.25\textwidth]{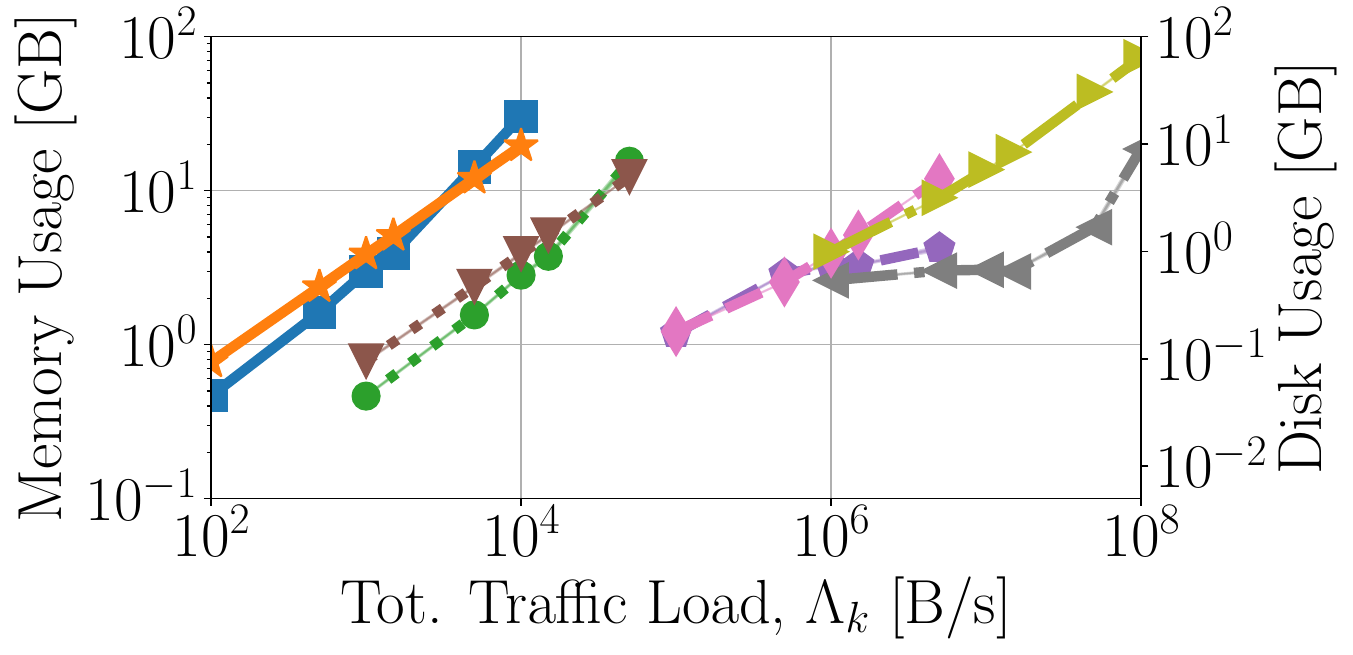}}}
    \\
    \vspace{1ex} \includegraphics[width=0.8\textwidth]{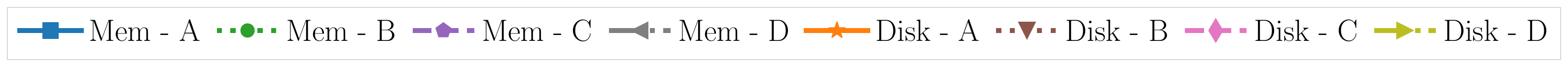}
    \caption{\rev{Etcd memory and disk usage for varying xApp classes, values of xApp state size $\rho$, and maintenance period $\nu$.}}
    \label{work06:fig:mem_vs_load_vs_rho_vs_nu}
    \vspace{-3mm}
\end{figure*}

{\bf SDL.}
Finally, we examine the impact of the maintenance period $\nu$ on etcd's resource consumption. Fig.~\ref{work06:fig:cpu_vs_load_vs_rho_vs_nu} depicts CPU usage and power consumption of etcd as a function of the total traffic load $\Lambda_k$ for different xApp classes, values of state size $\rho$, and $\nu$. As expected, lower values of $\nu$ imply more frequent etcd maintenance operations, yielding an increase on CPU usage and power consumption that is up to two orders of magnitude for low values of $\Lambda_k$ (e.g., comparing Fig.~\ref{work06:fig:cpu_vs_load_rho1_nu1} and Fig.~\ref{work06:fig:cpu_vs_load_rho1_nu120}). On the contrary, no significant impact on resource consumption is observed when $\rho$ increases, as the amount of state size being retained does not affect CPU or power consumption. Moreover, when $\nu{=}1$\,s, both CPU usage and power consumption exhibit a slightly decreasing trend with respect to $\Lambda_k$. This is because etcd saturates due to: (i) the frequent maintenance operations that make etcd instances unavailable; and (ii) the increasingly high number of key-value pairs being stored/accessed by the xApps. On the other hand, when $\nu{=}120$\,s (i.e., when etcd is not saturating), CPU usage and power consumption grow with $\Lambda_k$. Remarkably, regardless of the values of $\nu$ and $\rho$, the dependency of CPU usage and power consumption upon $\Lambda_k$ can be well approximated by a linear relation. 

\begin{observation}[Etcd CPU and power usage]\label{work06:obs:sdl_res_usage_cpu}
    Etcd \rev{instantaneous} CPU and power consumption substantially decreases with the maintenance period but is practically independent of the state size. For all xApp classes, both CPU and power consumption exhibit a linear relationship with \rev{the total number of xApps}.
\end{observation}

Similarly, Fig.~\ref{work06:fig:mem_vs_load_vs_rho_vs_nu} depicts the etcd memory and disk usage versus the total traffic load $\Lambda_k$ and for varying xApp classes, values of state size $\rho$, and maintenance period $\nu$. First, we notice that the impact of both $\rho$ and $\nu$ on the results is not negligible and depends on the type of xApp. Indeed, despite increasing values of $\rho$ and $\nu$ yield a general increase on memory and disk usage, two exceptions can be observed: (i) when xApps of class D are considered, the value of $\rho$ has negligible impact on the memory and disk usage; and (ii) when the xApps are of class A and they feature a state size $\rho{=}1$\,MB, varying the value of $\nu$ makes no significant difference in memory and disk usage. Secondly, focusing on the configurations that do not violate the near-RT \gls{ric} deadline (see Fig.~\ref{work06:fig:defrag_downtime_vs_traffic}), results show that the dependency of both memory and disk usage on the total traffic load can be well approximated by a linear relation regardless of $\rho$ and $\nu$. 

\begin{observation}[Etcd memory and disk usage]\label{work06:obs:sdl_res_usage_mem}
    Etcd memory and disk utilization depend on the xApp classes, their state size, and the value of the maintenance period. In general, both memory and disk usage exhibit a linear relation with respect to the total traffic load, i.e., the number of xApps.
\end{observation}

\section{Problem Formulation}\label{work06:sec:opt_prob}

Our findings show that achieving lossless migration of stateful xApps is non-trivial due to a variety of trade-offs involving resource utilization, scalability, and service availability. Cloud-native technologies allow to dynamically activate compute nodes but do not consider the strict requirements of O-RAN systems described above.
For this reason, we propose \novelty, an energy-aware framework that jointly optimizes compute nodes activation and lossless xApp migration while guaranteeing uninterrupted xApp control.
To integrate \novelty within the \gls{o-ran} architecture, we prototyped \novelty as an rApp running on the non-RT \gls{ric}, which is a component of the \gls{smo} framework and it is in charge of handling all orchestration, management and automation procedures to monitor and control \gls{ran} components.

\subsection{System Model} \label{work06:sec:sys_model}
We consider a compute cluster of nodes, each consisting of a server.
Let $\servers$ be the set of servers. The cluster hosts the near-RT RIC along with a total number $N_k$ of xApps for each class $k$. Consistently with our testbed (see Sec.~\ref{work06:sec:testbed}), we consider resource-constrained and identical servers with respect to CPU, memory and disk availability. However, we remark that the notation can be easily extended to heterogeneous deployments, making \novelty independent of the specific cluster architecture and resource capabilities. For each server $s{\in}\servers$, we define a binary indicator $\alpha_s$ that identify servers that can be turned off to save energy (i.e., $\alpha_s{=}1$), and those that must be on always (i.e., $\alpha_s{=}0$) such as master servers, or servers hosting the near-RT \gls{ric} and other fundamental services. We introduce a binary variable $\mu_s$ to identify which server is active (i.e., $\mu_s{=}1$) or turned off (i.e., $\mu_s{=}0$). We let $\boldsymbol{\mu}{=}(\mu_s)_{s{\in}\servers}$ denote the server activation policy, and let $\mu_s{=}0$ only if $\alpha_s{=}1$.

We consider a timeslot-based optimization problem where the joint server activation and xApp migration problem is solved periodically at discrete time intervals of $\Delta T$ hours.
\rev{Let the superscript $0$ denote the system state at the beginning of the current timeslot.}
For each xApp class $k{\in}\apps$, $n^0_{k,s}{\in}\mathbb{N}^+_0$ is a non-negative integer parameter to indicate how many xApps of class $k$ are running on server $s$ at the beginning of the timeslot. We also consider $\boldsymbol{\mu}^0_s{=}(\mu_s^0)_{s\in\servers}$ where $\mu_s^0{\in}\{0,1\}$ indicates whether server $s$ is active ($\mu_s^0{=}1$) at the beginning of the timeslot, or not.

For a given cluster status (e.g., defined by the number $n^0_{k,s}$ of xApps already deployed on each server $s$ and its activation status $\mu_s^0$), the goal of CORMO-RAN is to determine both the server activation policy $\boldsymbol{\mu}$ and the xApp migration policy $\mathbf{x}$. The latter is defined as $\mathbf{x}{=}(x_{k,s,s'})_{k\in\apps, (s,s')\in\servers^2}$ where $x_{k,s,s'}{\in}\mathbb{N}^+_0$ is used to indicate how many xApps of class $k$ are being reallocated from $s$ to $s'$. If $s{\neq} s'$, $x_{k,s,s'}$ represents the number of xApps that are being migrated; if $s{=}s'$, $x_{k,s,s}$ represents the number of xApps that remain on $s$.

In addition to migration, we consider both deployment of new xApps as well as undeployment.
Let $n_k^\mathrm{-}$ and $n_k^\mathrm{+}$ be the number of xApps of class $k$ to be undeployed and deployed, respectively. Without loss of generality, we introduce a virtual server $\tilde{s}{\in}\servers$ hosting all xApps to be deployed. Thus, we set $n^0_{k,\tilde{s}}{=}n_k^\mathrm{+}$ for all $k{\in}\apps$. Also, $\tilde{s}$ has infinite computational resources and zero energy consumption, as this server does not contribute to any utility or cost, but it is only used to simplify the notation while retaining generality.
Since xApps to be undeployed become irrelevant to \gls{ran} operations, at the beginning of each slot we remove a total of $n^-_{k}$ from all servers in $\servers{\setminus}\{\tilde{s}\}$.
In this way, $\sum_{s'\in\servers\setminus\{s\}} x_{k,s,s'}$ represents the total number of xApps of class $k$ to be migrated from $s$. 

\begin{table*}[t]
\footnotesize
\caption{Experimental parameter settings for SDL, under $\rho{=}1$, $\nu{=}1$ (upper), and xApp resource consumption (lower)}
\label{work06:tab:param_sdl_xapp}
\renewcommand{\arraystretch}{1.1}
\addtolength{\tabcolsep}{-0.2em}
\centering
\begin{tabular}{r|ccccccccc|}
\cline{2-10} 
\multicolumn{1}{l|}{} & \multicolumn{1}{c|}{$\delta^{\sdl}_{E,k}$\,[W]} & \multicolumn{1}{c|}{$\delta^{\sdl}_{{\mathrm{CPU}},k}$} & \multicolumn{1}{c|}{$\delta^{\sdl}_{{\mathrm{MEM}},k}$\,[GB]} & \multicolumn{1}{c|}{$\delta^{\sdl}_{{\mathrm{DISK}},k}$\,[GB]} & \multicolumn{1}{c|}{$b^{\sdl}_{E,k}$\,[W]} & \multicolumn{1}{c|}{$b^{\sdl}_{{\mathrm{CPU}},k}$} & \multicolumn{1}{c|}{$b^{\sdl}_{{\mathrm{MEM}},k}$\,[GB]} & \multicolumn{1}{c|}{$b^{\sdl}_{{\mathrm{DISK}},k}$\,[GB]} & \multicolumn{1}{c|}{$\sigma_{x}$\,[ms]}\\ \hline

\multicolumn{1}{|r|}{A} & \multicolumn{1}{c|}{-0.18} & \multicolumn{1}{c|}{-0.00} & \multicolumn{1}{c|}{0.04} & \multicolumn{1}{c|}{0.01} & \multicolumn{1}{c|}{32.35} & \multicolumn{1}{c|}{5.32} & \multicolumn{1}{c|}{0.20} & \multicolumn{1}{c|}{0.00} & \multicolumn{1}{c|}{16.62} \\ \hline

\multicolumn{1}{|r|}{B} & \multicolumn{1}{c|}{-0.09} & \multicolumn{1}{c|}{0.03} & \multicolumn{1}{c|}{0.04} & \multicolumn{1}{c|}{0.01} & \multicolumn{1}{c|}{33.60} & \multicolumn{1}{c|}{5.57} & \multicolumn{1}{c|}{0.17} & \multicolumn{1}{c|}{0.00} & \multicolumn{1}{c|}{17.07} \\ \hline

\multicolumn{1}{|r|}{C} & \multicolumn{1}{c|}{-0.10} & \multicolumn{1}{c|}{-0.03} & \multicolumn{1}{c|}{0.02} & \multicolumn{1}{c|}{0.00} & \multicolumn{1}{c|}{35.48} & \multicolumn{1}{c|}{4.97} & \multicolumn{1}{c|}{1.82} & \multicolumn{1}{c|}{0.00} & \multicolumn{1}{c|}{7.71} \\ \hline

\multicolumn{1}{|r|}{D} & \multicolumn{1}{c|}{-0.06} & \multicolumn{1}{c|}{-0.01} & \multicolumn{1}{c|}{0.08} & \multicolumn{1}{c|}{0.03} & \multicolumn{1}{c|}{40.20} & \multicolumn{1}{c|}{5.00} & \multicolumn{1}{c|}{1.04} & \multicolumn{1}{c|}{0.00} & \multicolumn{1}{c|}{11.62} \\ \hline
\end{tabular}
\\[1em]
\begin{tabular}{r|ccc|}
\cline{2-4}
\multicolumn{1}{l|}{} & \multicolumn{1}{c|}{$p_{E,k}$\,[W]} & \multicolumn{1}{c|}{$p_{{\mathrm{CPU}},k}$} & \multicolumn{1}{c|}{$p_{{\mathrm{MEM}},k}$\,[GB]} \\ \hline

\multicolumn{1}{|r|}{A} & \multicolumn{1}{c|}{3.43} & \multicolumn{1}{c|}{0.47} & \multicolumn{1}{c|}{0.52} \\ \hline

\multicolumn{1}{|r|}{B} & \multicolumn{1}{c|}{16.48} & \multicolumn{1}{c|}{2.86} & \multicolumn{1}{c|}{0.52} \\ \hline

\multicolumn{1}{|r|}{C} & \multicolumn{1}{c|}{3.43} & \multicolumn{1}{c|}{0.47} & \multicolumn{1}{c|}{0.52} \\ \hline

\multicolumn{1}{|r|}{D} & \multicolumn{1}{c|}{16.48} & \multicolumn{1}{c|}{2.86} & \multicolumn{1}{c|}{0.52} \\ \hline
\end{tabular}
\end{table*}

\begin{table*}[htb]
\footnotesize
\caption{Experimental parameter settings for idle near-RT RIC consumption and SM resource usage $\forall k, \rho, \nu$}
\label{work06:tab:param_idle_sm_res}
\renewcommand{\arraystretch}{1.1}
\centering
\begin{tabular}{|cccccccccc|}
\cline{2-5}
\hline
\multicolumn{1}{|c|}{$\delta_M^{\mathrm{SDL}}$\,[s]} & \multicolumn{1}{|c|}{$b_M^{\mathrm{SDL}}$\,[s]} & \multicolumn{1}{|c|}{$b_{\mathrm{CPU}}^{\mathrm{SM-MR}}$\,[s]} & \multicolumn{1}{|c|}{$b_{\mathrm{CPU}}^{\mathrm{SM-MD}}$\,[s]} & \multicolumn{1}{|c|}{$b_E^{\mathrm{SM-MR}}$\,[W]} & \multicolumn{1}{|c|}{$b_E^{\mathrm{SM-MD}}$\,[W]} & \multicolumn{1}{|c|}{$q_{E_s}$\,[W]} & \multicolumn{1}{c|}{$q_{{\mathrm{CPU}}_s}$} & \multicolumn{1}{|c|}{$q_{{\mathrm{MEM}}_s}$\,[GB]} & \multicolumn{1}{c|}{$q_{{\mathrm{DISK}}_s}$\,[GB]}\\ \hline
\multicolumn{1}{|c|}{0.08} & \multicolumn{1}{|c|}{4.27} & \multicolumn{1}{|c|}{0.40} & \multicolumn{1}{|c|}{0.76} & \multicolumn{1}{|c|}{17.87} & \multicolumn{1}{|c|}{27.56} & \multicolumn{1}{|c|}{120} & \multicolumn{1}{|c|}{0.1} & \multicolumn{1}{|c|}{5.7} & \multicolumn{1}{|c|}{3.2} \\ \hline
\end{tabular}
\vspace{-3mm}
\end{table*}

\begin{table}[ht]
\footnotesize
\caption{Experimental parameter settings for SM KPIs $\forall k, \nu$}
\label{work06:tab:param_sm_kpis}
\renewcommand{\arraystretch}{1.1}
\centering
\begin{tabular}{r|cccc|}
\cline{2-4}
\multicolumn{1}{l|}{} & \multicolumn{1}{c|}{$\delta_D^{\mathrm{SM-MR}}$\,[s]} & \multicolumn{1}{c|}{$\delta_D^{\mathrm{SM-MD}}$\,[s]} & \multicolumn{1}{c|}{$\delta_M^{\mathrm{SM-MD}}$\,[s]} \\ \hline
\multicolumn{1}{|l|}{$\rho=1\,\mathrm{MB}$} & \multicolumn{1}{c|}{10.55} & \multicolumn{1}{c|}{5.74} & \multicolumn{1}{c|}{20.28} \\ \hline
\multicolumn{1}{|l|}{$\rho=10\,\mathrm{MB}$} & \multicolumn{1}{c|}{11.73} & \multicolumn{1}{c|}{6.49} & \multicolumn{1}{c|}{23.02} \\ \hline
\multicolumn{1}{|l|}{$\rho=100\,\mathrm{MB}$} & \multicolumn{1}{c|}{23.3} & \multicolumn{1}{c|}{13.3} & \multicolumn{1}{c|}{48.2} \\ \hline
\end{tabular}
\vspace{-3mm}
\end{table}

\textbf{Temporal \glspl{kpi}.}~Finding~\ref{work06:obs:xapp_memory_dpr} suggests that memory usage and dirty-page rate are dominated by \gls{ai} execution and depend on the xApp state size $\rho$. \rev{Finding~\ref{work06:obs:sm_feasibility} indicates that due to technical limitations, the SM strategy is incompatible with the near-RT RIC control loop deadline but still worth to consider.} Since Findings~\ref{work06:obs:sm_kpis} and~\ref{work06:obs:sdl_kpis} suggest a linear relationship, the migration downtime and the total migration duration are:
\vspace{-2mm}
\begin{tcolorbox}
[colback=red!0!white,colframe=blue!35!black,title={From Experimental Findings~\ref{work06:obs:xapp_memory_dpr}, \ref{work06:obs:sm_feasibility}, \ref{work06:obs:sm_kpis}, \ref{work06:obs:sdl_kpis}},toprule=0mm,titlerule=0mm]
\vspace{-4mm}
\begin{align}
    T_{\mathrm{D}_{k,s}}^{\tau} &= \delta_{\mathrm{D}}^{\tau} \cdot \sum_{s'\in\servers\setminus\{s\}} x_{k,s,s'} + b_{\mathrm{D}}^{\tau}
    \label{work06:eq:downtime_sm}
    \\
    T_{\mathrm{M}_{k,s}}^{\tau} &= \delta_{\mathrm{M}}^{\tau} \cdot \sum_{s'\in\servers\setminus\{s\}} x_{k,s,s'} + b_{\mathrm{M}}^{\tau},
    \label{work06:eq:mig_dur_sm}
\end{align}
\vspace{-4mm}
\end{tcolorbox}

\noindent
where $\tau{\in}\{\sdl,\smmr,\smmd\}$ and $\delta^\tau_{\mathrm{D}}$, $\delta^\tau_{\mathrm{M}}$ are the slopes of the linear approximation we have experimentally measured from Fig.~\ref{work06:fig:sm_kpis} for SM, and Fig.~\ref{work06:fig:sdl_mig_dur_vs_num_xapps} for SDL, while $b_{\mathrm{D}}^{\tau}$ and $b_{\mathrm{M}}^{\tau}$ are the intercept for the two \glspl{kpi}. The values of all parameters are summarized in Tables~\ref{work06:tab:param_sdl_xapp}, \ref{work06:tab:param_idle_sm_res} and~\ref{work06:tab:param_sm_kpis}. It is worth mentioning that Finding~\ref{work06:obs:sdl_kpis} provides experimental evidence that xApps behave as stateless under \gls{sdl}, which results in zero-downtime migration, i.e., $T_{\mathrm{D}_{k,s}}^{\sdl}{=}0 \,\, \forall k,s$. Moreover, Fig.~\ref{work06:fig:sm_kpis} shows that $b_{\mathrm{M}}^{\smmd}{=}b_{\mathrm{M}}^{\smmr}{=}0$ and $\delta^{\smmr}_{\mathrm{D}} {=} \delta^{\smmr}_{\mathrm{M}}$.
\rev{Further, although our model is derived under sequential migrations of multiple xApps, the linear formulation would remain valid in the case of parallel strategies---albeit with slopes and intercepts acquiring different physical interpretations. This makes our model (i) adaptive to future SM developments that support parallel migrations, (ii) compatible with any configuration of the Kubernetes scheduler to handle batches of parallel deployments, and (iii) broadly applicable, as capturing the worst-case sequential behavior ensures that any faster parallel approach naturally falls within the bounds demonstrated in our following evaluation.}

Note that the time necessary to instantiate new xApps does not depend on the specific migration strategy as the state is always empty upon instantiation. Therefore, the time to instantiate new xApps can be computed by using Fig.~\ref{work06:fig:sdl_mig_dur_vs_num_xapps} (i.e., which corresponds to the time needed to migrate a virtually stateless xApp in \gls{sdl}) and is defined as:
\begin{tcolorbox}
[colback=red!0!white,colframe=blue!35!black,title={From Experimental Finding~\ref{work06:obs:sdl_kpis}},toprule=0mm,titlerule=0mm,bottom=0mm]
\vspace{-4mm}
    \begin{align}
        \tilde{T}_{k,s} &= \delta_{\mathrm{M}}^{\sdl} \cdot x_{k,\tilde{s},s} + b_{\mathrm{M}}^{\sdl}.
        \label{work06:eq:new_xapps}
    \end{align}
\vspace{-4mm}
\end{tcolorbox}

\textbf{\gls{sdl} feasibility.}~As we pointed out in Finding~\ref{work06:obs:sdl_feasibility} and~\ref{work06:obs:sdl_scalability}, etcd is indeed a valid solution for the \gls{sdl} backend database but it is subject to scalability limits as the defrag downtime may exceed the near-RT \gls{ric} \rev{control loop deadline}.
To capture this aspect, we model the defrag downtime as a linear function of the total number $N_k$ of xApps, with $\sigma_k$ being the slope we experimentally measure from Fig.~\ref{work06:fig:defrag_downtime_vs_traffic}, i.e.,
\vspace{-2mm}
\begin{tcolorbox}
[colback=red!0!white,colframe=blue!35!black,title={From Experimental Findings~\ref{work06:obs:sdl_feasibility}, \ref{work06:obs:sdl_scalability}},toprule=0mm,titlerule=0mm]
\vspace{-2mm}
    \begin{equation}
    T_{\mathrm{DF}}^{\sdl} = \sum_{k\in\apps} \sigma_{k} N_k \, .
    \label{work06:eq:defrag_downtime}
    \end{equation}
\vspace{-4mm}
\end{tcolorbox}

Moreover, we denote $T_{\mathrm{active}}$ as the time an xApp is active within the maintenance period $\nu$. For etcd to be a feasible lossless xApp migration strategy in O-RAN, it must always avoid permanent service disruption, i.e., $    T_{\mathrm{active}}{=}\nu {-}T_{\mathrm{DF}}^{\sdl}{>}0$.

\textbf{Resource Consumption.}~To model the resource consumption associated to a server $s$ we consider three contributions: (i) the idle consumption; (ii) the load-based resource consumption, which scales linearly with the number of xApps hosted by $s$~\cite{fan2007power}; and (iii) the resource consumption required to execute the specific migration strategy. 

The general resource consumption model for any server $s{\in}\servers{\setminus}\{\tilde{s}\}$ with respect to migration strategy $\tau$ is:
\vspace{-2mm}
\begin{tcolorbox}
[colback=red!0!white,colframe=blue!35!black,title={From Experimental Finding~\ref{work06:obs:xapp_res_usage}},toprule=0mm,titlerule=0mm]
\vspace{-4mm}
    \begin{align}
        R_{\chi_s}^{\tau} &= \mu_s q_{\chi_s} + \sum_{k\in\apps} \sum_{s'\in\servers} p_{\chi_k}  x_{k,s,s'} + \tilde{R}_{\chi_s}^{\tau} \label{eq:resources_total}
    \end{align}
\vspace{-4mm}
\end{tcolorbox}

\noindent
where $\chi{\in}\{\mathrm{CPU},\mathrm{MEM},\mathrm{DISK}\}$ is the type of resource, used to indicate CPU, memory and disk resources, respectively.

The first term in~\eqref{eq:resources_total} represents the idle consumption $q_{\chi_s}$ when the server is active (i.e., $\mu_s{=}1$). The second term considers the load-based consumption observed in Finding~\ref{work06:obs:xapp_res_usage}, where
$p_{\chi_k}$ is the slope of the linear approximation evaluated experimentally. 
Disk resources leveraged by our xApps (Sec.~\ref{work06:sec:testbed}) are negligible, yielding $p_{{\mathrm{DISK}},k}{=}0$. The other values for $q_{\chi_s}$ and $p_{\chi_k}$ are summarized in Tables~\ref{work06:tab:param_idle_sm_res} and \ref{work06:tab:param_sdl_xapp}. The third element captures the intrinsic resource consumption of both \gls{sm} and \gls{sdl} on each server $s$ defined as:
\vspace{-2mm}
\begin{tcolorbox}
[colback=red!0!white,colframe=blue!35!black,title={From Experimental Findings~\ref{work06:obs:sm_res_usage}, \ref{work06:obs:sdl_res_usage_cpu}, \ref{work06:obs:sdl_res_usage_mem}},toprule=0mm,titlerule=0mm]
\vspace{-4mm}
    \begin{align}
        \tilde{R}_{\chi_s}^{\sdl} &= \frac{1}{|\servers|} \cdot \sum_{k\in\apps} \left(\delta^{\sdl}_{\chi_k} N_k + b^{\sdl}_{\chi_k}\right) \label{eq:tilde_r_sdl} \\
        \tilde{R}_{\chi_s}^{\smmr} &= b^{\smmr}_{\chi} \label{eq:tilde_r_smmr}\\ 
        \tilde{R}_{\chi_s}^{\smmd} &= b^{\smmd}_{\chi} \label{eq:tilde_r_smmd}
    \end{align}
\vspace{-6mm}
\end{tcolorbox}

Accordingly, the \gls{sdl} resource consumption, modeled in \eqref{eq:tilde_r_sdl}, is equally distributed across all servers and linearly depend on the total number $N_k$ of xApps, with slope $\delta^{\sdl}_{\chi_k}$ and intercept $b^{\sdl}_{\chi_k}$. 
Also, the CPU consumption for \gls{sm} is practically constant regardless of the value of state size, and only depends on the specific \gls{sm} strategy being employed. Moreover, the consumption of memory and disk resources are negligible, i.e.,  $b^{\tau}_{\mathrm{MEM}}{=}b^{\tau}_{\mathrm{DISK}}{=}0$ for $\tau{\in}\{\smmr,\smmd\}$.

\textbf{Energy Consumption.}~To evaluate energy consumption of each migration strategy, we need to consider the energy consumed by resource utilization due to xApp execution, as well as the energy caused by the migration process itself. 
From Finding~\ref{work06:obs:sm_res_usage}, the energy consumption caused by \gls{sm} is:
\begin{tcolorbox}
[colback=red!0!white,colframe=blue!35!black,title={From Experimental Finding~\ref{work06:obs:sm_res_usage}},toprule=0mm,titlerule=0mm]
\vspace{-4mm}
    \begin{align}
        E_s^{\tau} &= b^{\tau}_E \sum_{k\in\apps} T^{\tau}_{\mathrm{M}_{k,s}}\,, \hspace{0.2cm} \tau{\in}\{\smmr,\smmd\} \label{work06:eq:energy_sm}
    \end{align}
\vspace{-4mm}
\end{tcolorbox}
\noindent
where, $T^{\tau}_{\mathrm{M}_{k,s}}$ is defined in~\eqref{work06:eq:mig_dur_sm}, and $b^{\tau}_E$ represents the measured constant \rev{power} consumption as reported in Tables~\ref{work06:tab:param_sdl_xapp} and~\ref{work06:tab:param_idle_sm_res}.

With respect to \gls{sdl}, Findings~\ref{work06:obs:sdl_res_usage_cpu} and~\ref{work06:obs:sdl_res_usage_mem} show that the energy associated to \gls{sdl} linearly depends on the total number $N_k$ of xApps. Similarly to \eqref{eq:tilde_r_sdl}, this energy cost is distributed across the $|\servers|$ servers, and the \gls{sdl} energy cost per server $s$ is:
\begin{tcolorbox}
[colback=red!0!white,colframe=blue!35!black,title={From Experimental Findings~\ref{work06:obs:sdl_res_usage_cpu}, \ref{work06:obs:sdl_res_usage_mem}},toprule=0mm,titlerule=0mm]
\vspace{-4mm}
    \begin{align}
        E^{\sdl}_s &= \frac{\Delta T}{|\servers|} \cdot \sum_{k\in\apps} \left(\delta^{\sdl}_{\mathrm{E}_k} N_k + b^{\sdl}_{\mathrm{E}_k}\right)
        \label{work06:eq:energy_sdl}
    \end{align}
\vspace{-4mm}
\end{tcolorbox}
\noindent
where $\delta^{\sdl}_{\chi_k}$ and $b^{\sdl}_{\chi_k}$ are reported in Table~\ref{work06:tab:param_sdl_xapp}.

In~\eqref{work06:eq:energy_sdl}, the cost to maintain \gls{sdl} is continuous over the entire optimization interval $\Delta T$ as the states of the xApps need to be continuously updated in the backend database. This substantially differs from \gls{sm} where the cost of maintaining the state is incurred only for the duration of the migration process. However, we also notice that the migration process prevents servers from being turned off before the migrated xApps are activated on the destination server, yielding an extra active time that is in the order of a few seconds for \gls{sdl}, but reaches several hundreds of seconds for \gls{sm}.
Hence, the total energy consumption of the system is
\begin{multline}
    E_{s} = E^\tau_{s} + \!\! \sum_{k\in\apps} \left(T^\tau_{\mathrm{M}_{k,s}}+\tilde{T}_{k,s}\right) \cdot \left(q_{\mathrm{E}_s} + \sum_{k\in\apps} p_{\mathrm{E}_k} n^0_{k,s} \right) +\\
    + \! \left[\Delta T - \!\! \sum_{k\in\apps} \! \left(T^\tau_{\mathrm{M}_{k,s}} \!\! +\tilde{T}_{k,s}\right)\right] \! \cdot \! \left(\mu_s q_{\mathrm{E}_s} \!\! + \!\!\sum_{k\in\apps} \sum_{s'\in\servers} p_{\mathrm{E}_k} x_{k,s',s}\right) \, ,
    \label{work06:eq:energy_server}
\end{multline}
\noindent
where $E^\tau_{s}$ is defined in \eqref{work06:eq:energy_sm} or  \eqref{work06:eq:energy_sdl} based on the migration strategy $\tau{\in}\{\sdl,\smmr,\smmd\}$ being selected. The second term in \eqref{work06:eq:energy_server} accounts for the energy consumed during the migration process, and the third term accounts for the energy consumed by the server to execute the xApps it hosts. 

\subsection{Formulating the Problem} \label{work06:sec:problem_form}
We can now formulate the joint Server Activation and Lossless stateful xApp migration (\problem) problem:
\begin{tcolorbox}
[colback=red!0!white,colframe=red!35!black,titlerule=0mm,bottom=0mm]
\vspace{-4mm}
\begin{align}
    \min_{\mathbf{x}, \boldsymbol{\mu}} \hspace{0.2cm} & \sum_{s\in\servers} E_s \label{work06:eq:problem} \tag{\problem} \\
    \mathrm{s.t.:}
    &  \sum_{s'\in\servers} x_{k,s,s'} = n^0_{k,s} \hspace{0.2cm} \forall (k,s)\in\apps\times\servers \label{work06:eq:con2} \\
    &  \sum_{k\in\apps} \sum_{s\in\servers} x_{k,s,\tilde{s}} = 0 \hspace{0.2cm}  \label{work06:eq:con3} \\
    &  \sum_{s\in\servers\setminus\{\tilde{s}\}} x_{k,\tilde{s},s} = n^0_{a,\tilde{s}} \hspace{0.2cm} \forall k\in\apps \label{work06:eq:con4} \\
    &  \sum_{k\in\apps} \sum_{s'\in\servers\setminus\{s\}} x_{k,s,s'} \leq M \mu^0_s \hspace{0.2cm} \forall s\in\servers \label{work06:eq:con5} \\
    & \sum_{k\in\apps} \sum_{s'\in\servers} x_{k,s',s} \leq M \mu_s \hspace{0.2cm} \forall s\in\servers \label{work06:eq:con6} \\
    & \mu_s \leq \sum_{k\in\apps} \sum_{s'\in\servers} x_{k,s',s}  \hspace{0.2cm} \forall s\in\servers \label{work06:eq:con6_bis} \\
    & R_{\chi_s} \leq R_{\chi_s}^{\mathrm{MAX}} \mu_s \hspace{0.2cm} \forall s\in\servers  \label{work06:eq:con7} \\
    & \mu_s \geq 1 - \alpha_s \hspace{0.2cm} \forall s\in\servers \label{work06:eq:con9} \\
    & \sum_{k\in\apps} T^{\tau}_{D_{k,s}} \leq T^{\mathrm{max}}_{D_s} \hspace{0.2cm} \forall s\in\servers \label{work06:eq:con8} \\
    & T_{\mathrm{DF}}^{\sdl} {<} T^{\mathrm{max}}_{\mathrm{DF}} \hspace{0.2cm} \forall s\in\servers \label{work06:eq:con10} \\
    & T_{\mathrm{active}} > 0 \hspace{0.2cm} \forall s\in\servers \label{work06:eq:con11}
\end{align}
\vspace{-4mm}
\end{tcolorbox}

\noindent
where $E_s(\cdot)$ is defined in~\eqref{work06:eq:energy_server}, $\chi{\in}\{\mathrm{CPU}, \mathrm{MEM}, \mathrm{DISK}\}$, and $\tau{\in}\{\sdl, \smmr, \smmd\}$.
Constraint~\eqref{work06:eq:con2} ensures that we migrate only active xApps, and that we allocate all required xApps (those in the virtual server and those already deployed). Constraints~\eqref{work06:eq:con3} and~\eqref{work06:eq:con4} ensure that no xApps remain on the virtual server.
Constraint~\eqref{work06:eq:con5} imposes that xApps are instantiated on active servers only. Constraints~\eqref{work06:eq:con6} and~\eqref{work06:eq:con6_bis} ensure that we migrate xApps only from active servers and we shut down inactive servers, where $M$ is any large number such that $M{>}\sum_{k\in\apps}\sum_{s\in\servers}\sum_{s'\in\servers} x_{k,s,s'}$. Constraint~\eqref{work06:eq:con7} enforces resource constraints on each server.
Constraint~\eqref{work06:eq:con9} makes sure that we shut down only servers that can be turned off (i.e., with $\alpha_s{=}1$). Constraint~\eqref{work06:eq:con8} imposes that the downtime due to xApps being migrated to $s$ for any migration strategy $\tau$ is below a tolerable threshold $T^{\mathrm{max}}_{D_k}$. Finally, Constraints~\eqref{work06:eq:con10} and~\eqref{work06:eq:con11} enforce \gls{sdl} feasibility \rev{with respect to an arbitrary near-RT RIC control loop deadline $T^{\mathrm{max}}_{\mathrm{DF}}$, chosen, for instance, to ensure service continuity of the xApp with the tightest timing constraints.}

\begin{theorem}\label{th:np_hard}
Problem \eqref{work06:eq:problem} is NP-hard.
\end{theorem}
\noindent
\begin{proof}
The~\eqref{work06:eq:problem} problem is a mixed integer quadratic programming (MIQP) problem as it involves both binary ($\boldsymbol{\mu}$) and integer ($\mathbf{x}$) variables. 
It is well-known that the general decision version of MIQPs is NP-complete~\cite{pia2017mixed}. Being Problem \eqref{work06:eq:problem} a MIQP, we can build a polynomial-time reduction to the general formulation of MIQP in \cite{pia2017mixed}, which proves that Problem \eqref{work06:eq:problem} is NP-hard by reduction.
\end{proof}

\subsection{Solving the \problem Problem}\label{work06:sec:opt_solution}
Although \problem problem is NP-hard, it can be solved optimally via branch-and-bound (B\&B) where the original problem is transformed into its linear-programming relaxation and is iteratively solved by exploring the branches and assessing the integrality (and binary) constraints of variables. This process can also be made more efficient using cutting planes that exclude inefficient branches.
It has been shown \cite{pia2023approximation} that polynomial-time $\epsilon$-approximation algorithms for MIQP exist. How to build such polynomial approximation for the \problem problem is out of the scope of this paper, but, as shown in Sec.\,\ref{work06:sec:opt_num_eval}, the \problem problem can still be optimally solved within 1 second even in the case of 100 xApps to be migrated.

\section{\novelty Evaluation}\label{work06:sec:opt_num_eval}
To evaluate \novelty and compute an optimal solution to the \problem Problem, we use MATLAB and Gurobi on a server with Intel Xeon E5-2680 with 28 cores and 16\,GB of RAM.

We consider a cluster of four nodes, hosting the near-RT RIC components as well as a varying number of xApps, and, for each value, we consider 75\% of them to be of class $k$ \rev{(which corresponds to the dominant class) and the remaining 25\% to be evenly distributed among the other classes.} To be consistent with our testbed in Sec.~\ref{work06:sec:testbed}, we set $R^{\mathrm{max}}_{\mathrm{CPU}_s}{=}128$ (virtual) CPU cores, $R^{\mathrm{max}}_{\mathrm{MEM}_s}{=}125\,\mathrm{GB}$, and $R^{\mathrm{max}}_{\mathrm{DISK}_s}{=}250\,\mathrm{GB}$ and consider realistic values for the temporal parameters: $\Delta T{=}1\,\mathrm{h}$, i.e., running \novelty optimization cycles on an hourly basis, $T^{\mathrm{max}}_{D_k}{=}300\,\mathrm{s}$, i.e., the arbitrary maximum stateful migration downtime that can be tolerated, and $T^{\mathrm{max}}_{\mathrm{DF}}{=}1\,\mathrm{s}$, i.e., the near-RT deadline that must not be exceeded while performing periodic \gls{sdl} maintenance. \rev{It is worth mentioning that the choice of $\Delta T$ depends on how rapidly traffic fluctuates across the cells controlled by the near-RT \gls{ric}. While we consider \novelty optimization cycles to run on an hourly basis and as an rApp within the non-RT \gls{ric}, the ultimate decision of when and whether to trigger \novelty is driven by traffic dynamics and is thus left to the network operator. Indeed, our formulation is compatible not only with periodic executions but also with event-based triggering, e.g., upon detecting workload increases or any other relevant condition identified through continuous monitoring operations.}

\begin{figure}[hbt!]
    \centering
    \subfloat[][]{\label{work06:fig:opt_mipgap}{\includegraphics[width=0.48\columnwidth]{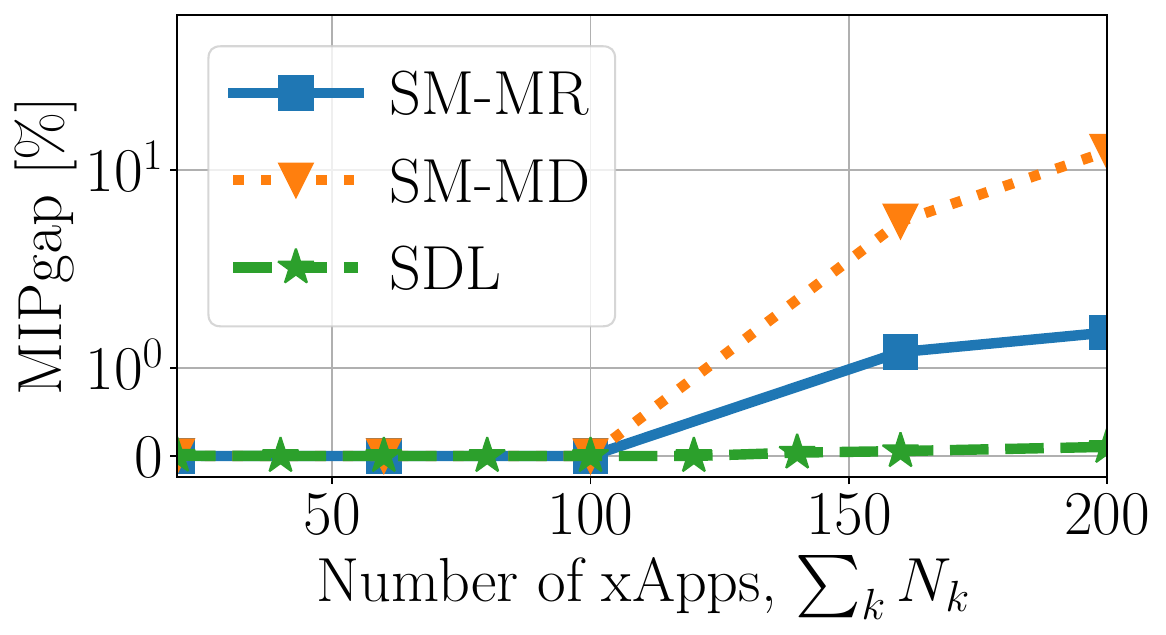}}}
    \hspace{1mm}
    \subfloat[][]{\label{work06:fig:opt_runtime}{\includegraphics[width=0.48\columnwidth]{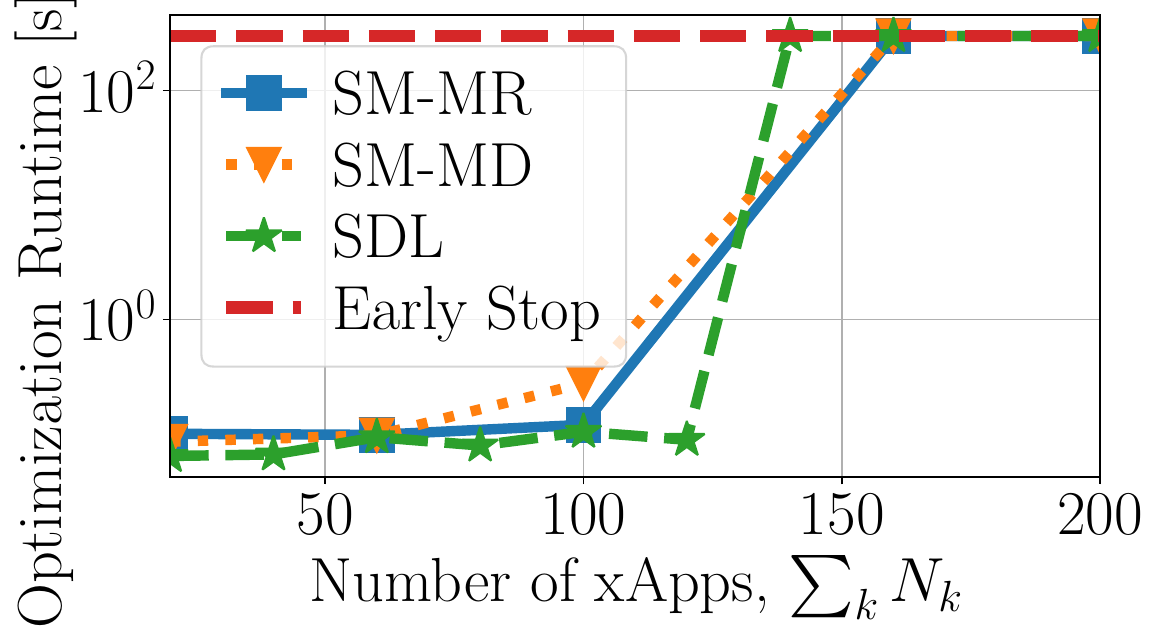}}}
    \caption{\novelty performance: (a) MIPgap and (b) runtime for $k{=}\mathrm{A}$, $\rho{=}1\,\mathrm{MB}$, and $\nu{=}1\,\mathrm{s}$.}
    \label{work06:fig:opt_performance}
    \vspace{-2mm}
\end{figure}

Fig.~\ref{work06:fig:opt_performance} shows the optimization performance for varying number of xApps and for the following exemplary configuration: dominant xApp class $k{=}\mathrm{A}$, state size $\rho{=}1\,\mathrm{MB}$, and maintenance period $\nu{=}1\,\mathrm{s}$.
Results demonstrate that up to about 120 xApps \problem can be solved optimally and within 1 second, regardless of the migration strategy being used. As the complexity of the scenario increases, i.e., the number of xApps grows above 120, the optimization runtime reaches the early stop deadline, i.e., 300\,s, but still yielding a reasonably small MIPgap (up to 10\% in the case of SM-MD and 200 xApps). We thus conclude that, despite being NP-hard, \problem can be solved optimally without algorithmic approximations.

\begin{figure*}[t!]
    \centering
    \subfloat[][$\mathrm{A},\rho{=}1,\nu{=}1$]{\label{work06:fig:energy_kA_rho1_nu1}{\includegraphics[width=0.23\textwidth]{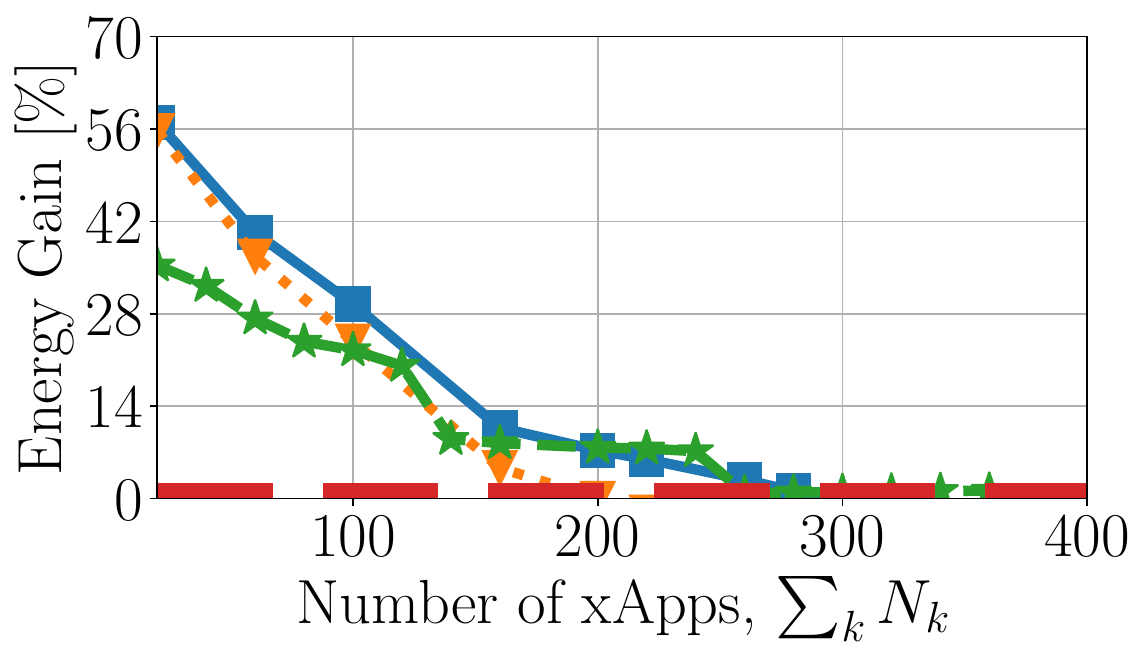}}}
    \hspace{2mm}
    \subfloat[][$\mathrm{A},\rho{=}1,\nu{=}1$]{\label{work06:fig:ratio_kA_rho1_nu1}{\includegraphics[width=0.23\textwidth]{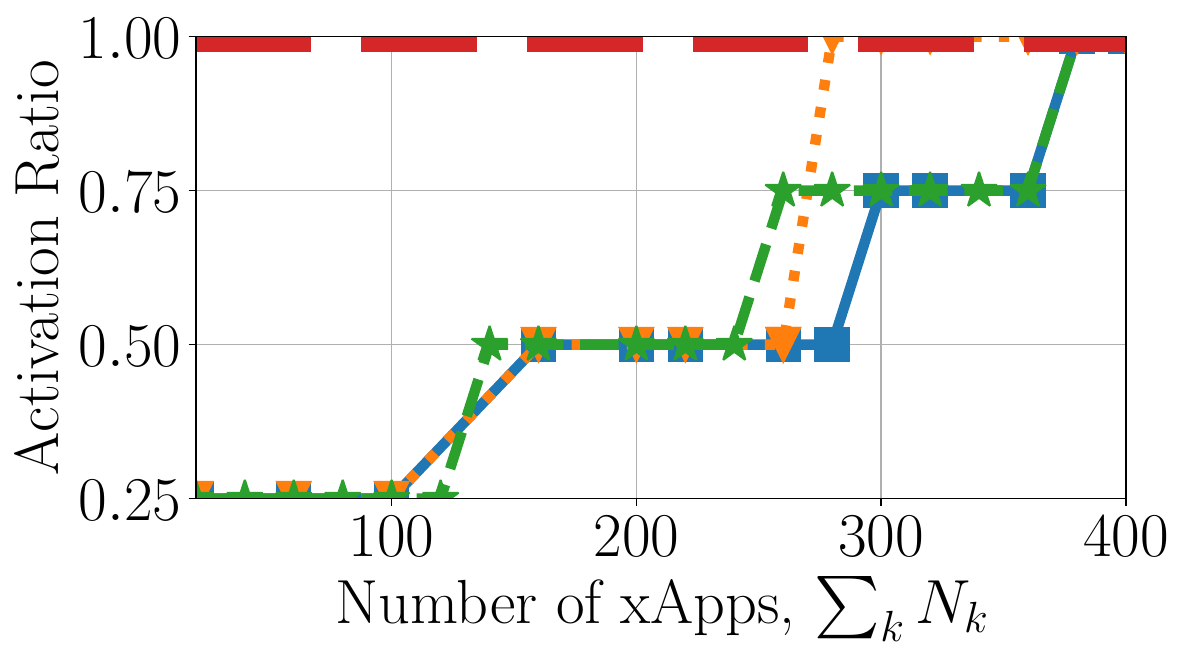}}}
    \hspace{2mm}
    \subfloat[][$\mathrm{A},\rho{=}1,\nu{=}120$]{\label{work06:fig:energy_kA_rho1_nu120}{\includegraphics[width=0.23\textwidth]{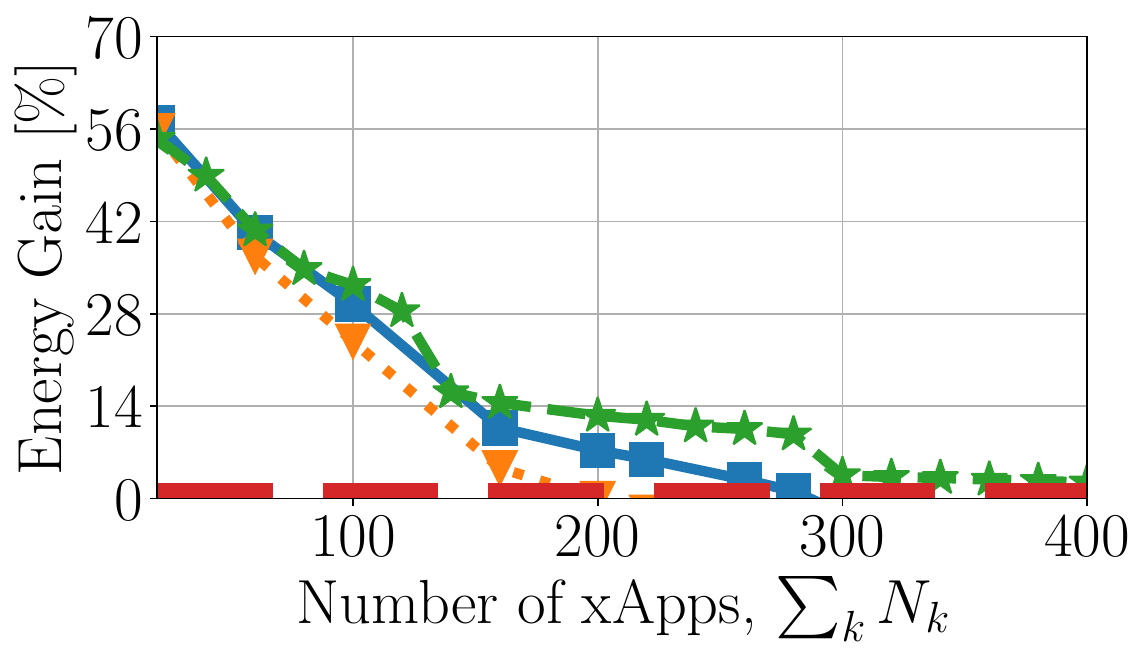}}}
    \hspace{2mm}
    \subfloat[][$\mathrm{A},\rho{=}1,\nu{=}120$]{\label{work06:fig:ratio_kA_rho1_nu120}{\includegraphics[width=0.23\textwidth]{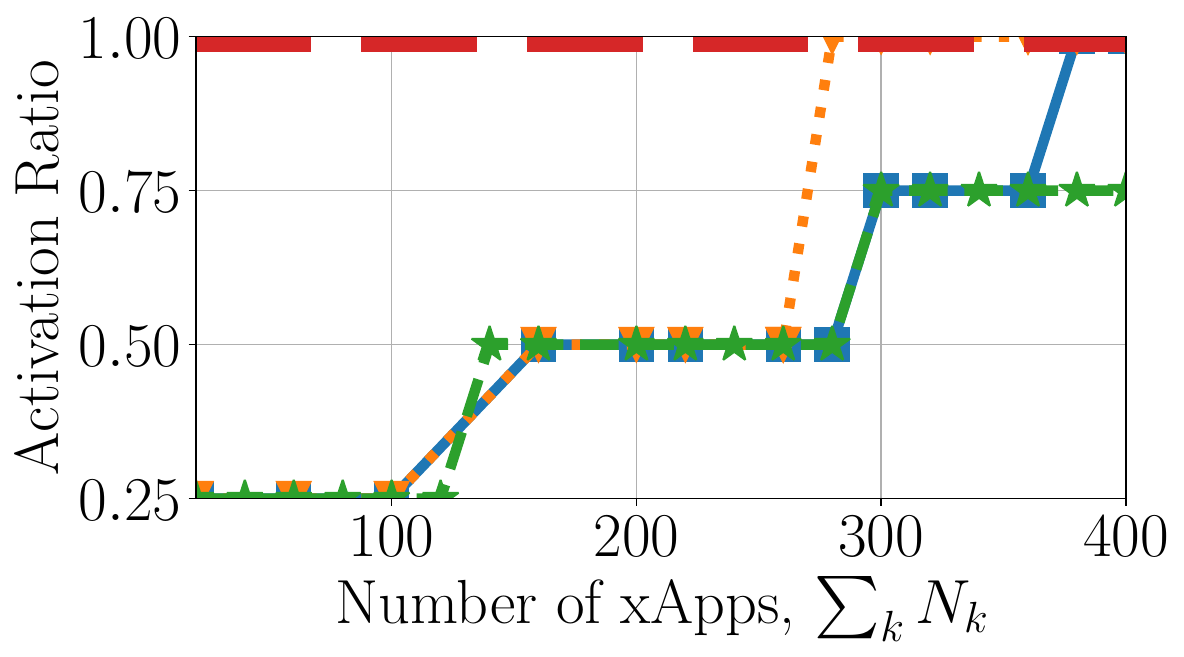}}}
    \\
    \subfloat[][$\mathrm{A},\rho{=}100,\nu{=}1$]{\label{work06:fig:energy_kA_rho100_nu1}{\includegraphics[width=0.23\textwidth]{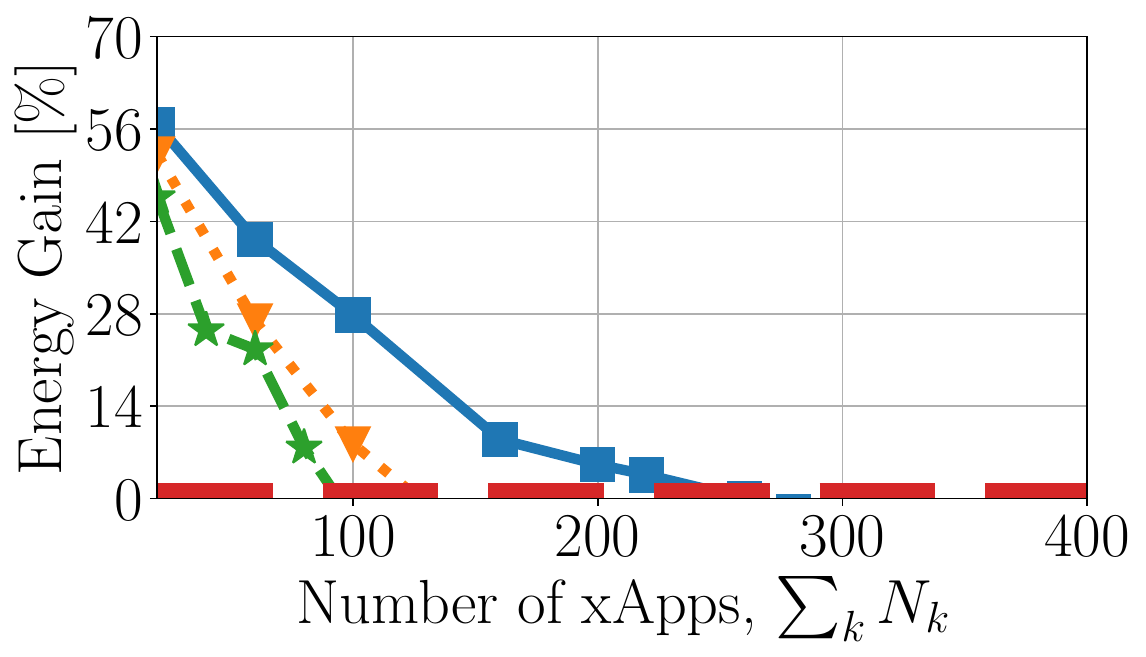}}}
    \hspace{2mm}
    \subfloat[][$\mathrm{A},\rho{=}100,\nu{=}1$]{\label{work06:fig:ratio_kA_rho100_nu1}{\includegraphics[width=0.23\textwidth]{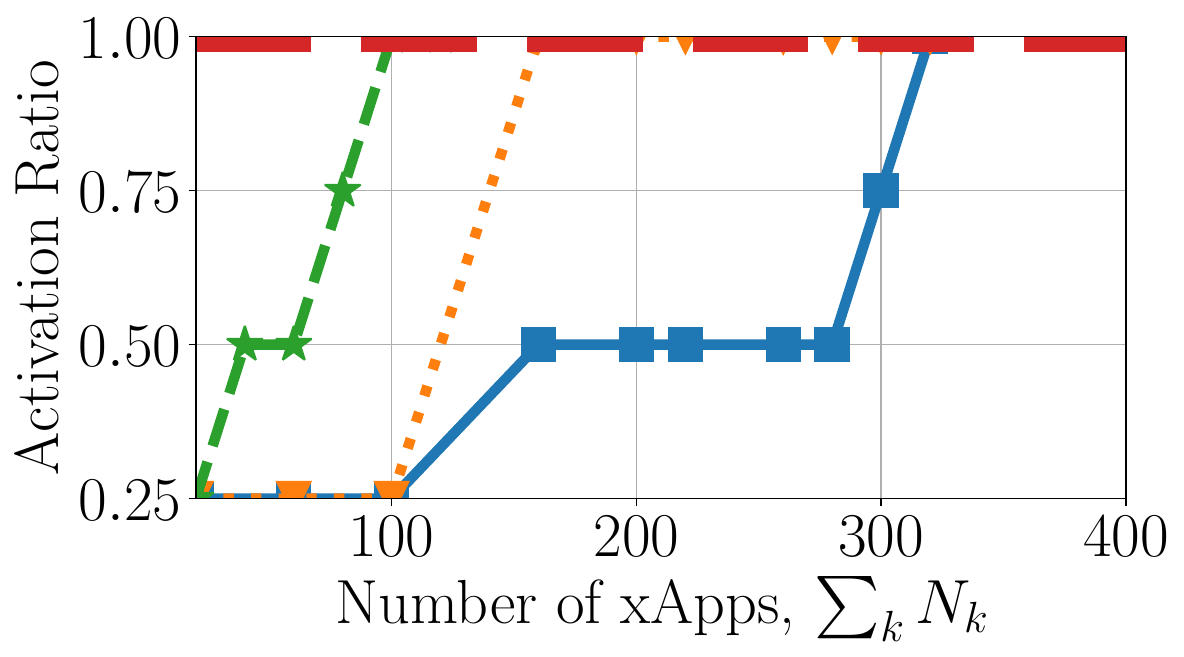}}}
    \hspace{2mm}
    \subfloat[][$\mathrm{B},\rho{=}1,\nu{=}1$]{\label{work06:fig:energy_kB_rho1_nu1}{\includegraphics[width=0.23\textwidth]{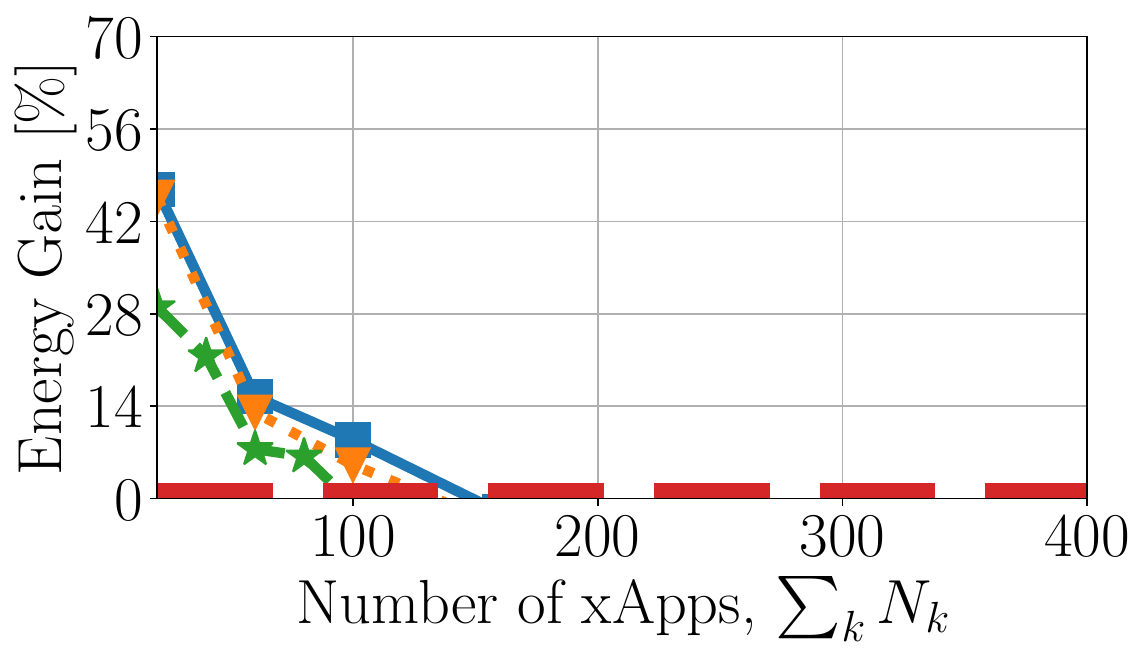}}}
    \hspace{2mm}
    \subfloat[][$\mathrm{B},\rho{=}1,\nu{=}1$]{\label{work06:fig:ratio_kB_rho1_nu1}{\includegraphics[width=0.23\textwidth]{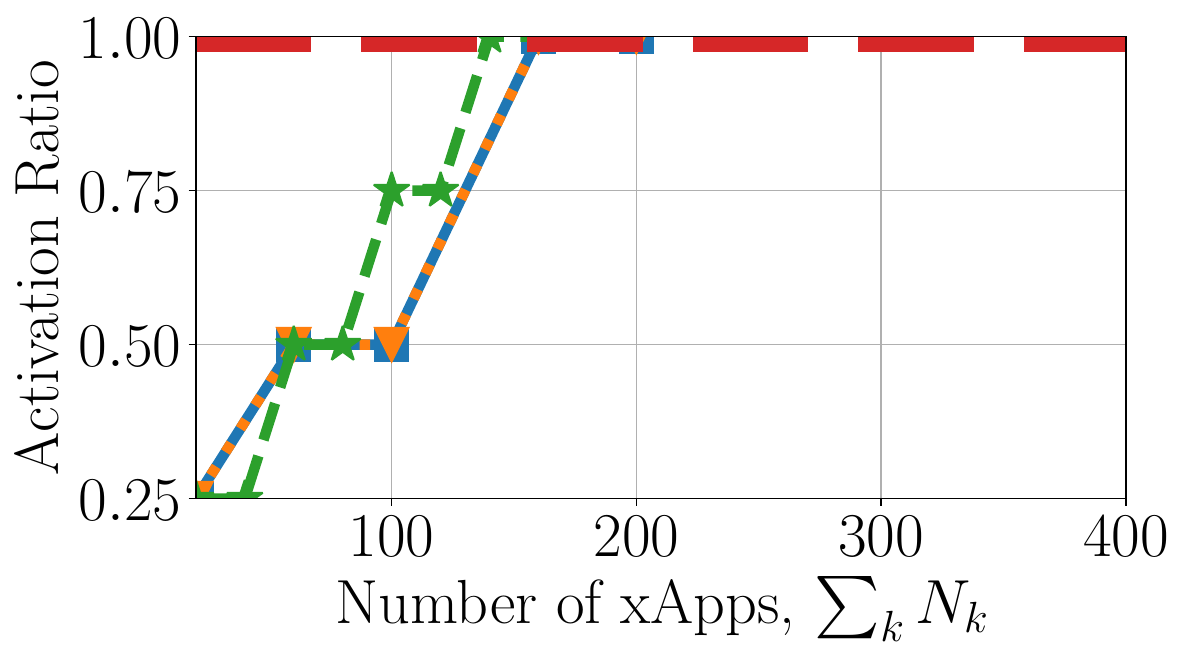}}}
    \\
    \hspace{0.3mm}
    \subfloat[][$\mathrm{C},\rho{=}1,\nu{=}1$]{\label{work06:fig:energy_kC_rho1_nu1}{\includegraphics[width=0.23\textwidth]{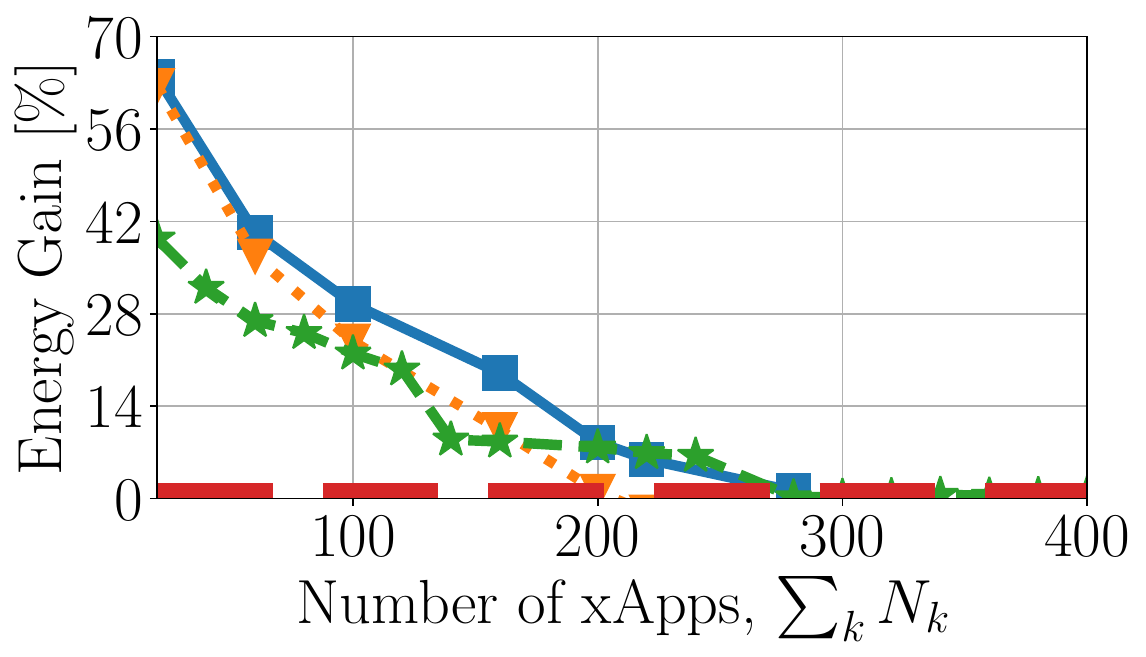}}}
    \hspace{2mm}
    \subfloat[][$\mathrm{C},\rho{=}1,\nu{=}1$]{\label{work06:fig:ratio_kC_rho1_nu1}{\includegraphics[width=0.23\textwidth]{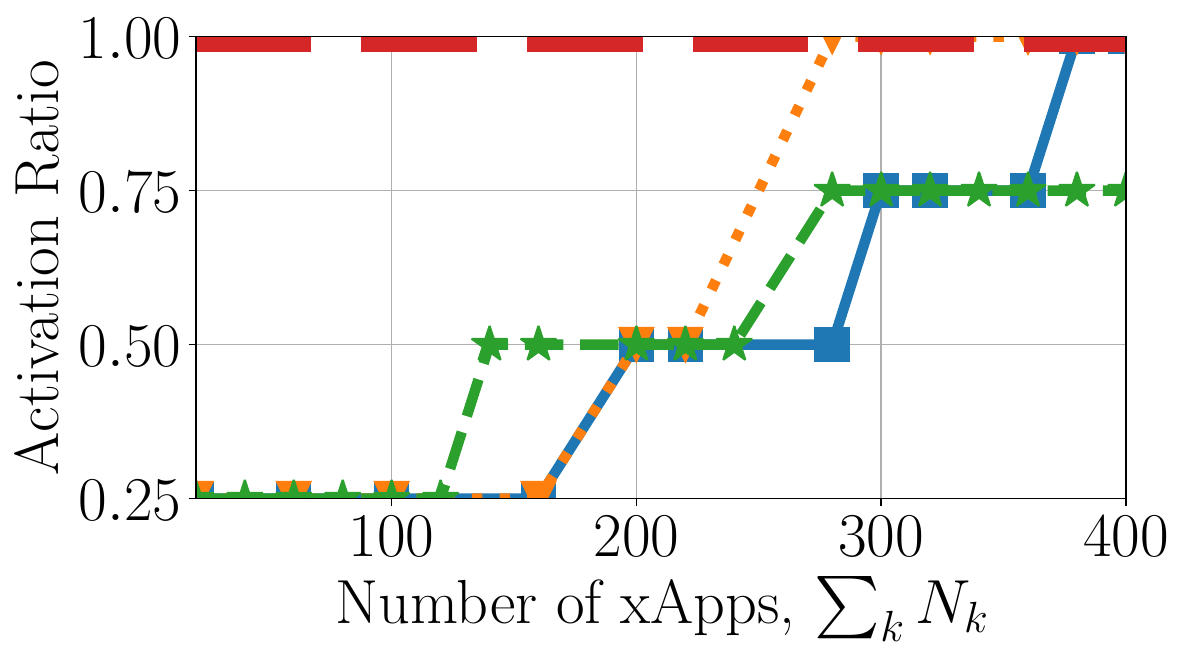}}}
    \hspace{2mm}
    \subfloat[][$\mathrm{D},\rho{=}1,\nu{=}1$]{\label{work06:fig:energy_kD_rho1_nu1}{\includegraphics[width=0.23\textwidth]{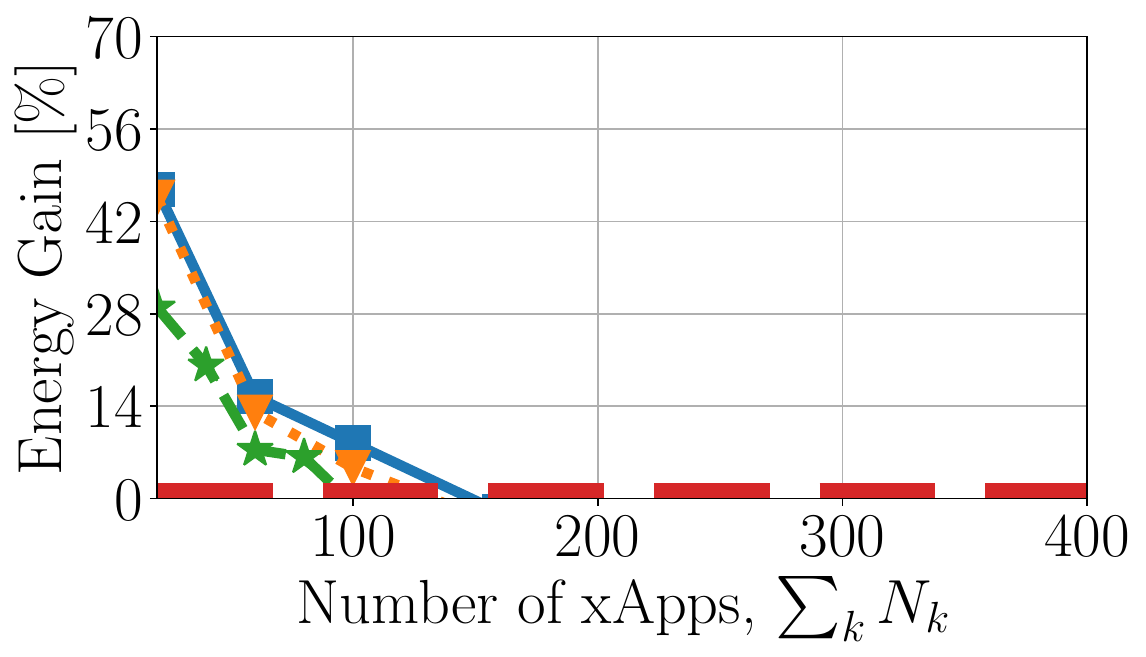}}}
    \hspace{2mm}
    \subfloat[][$\mathrm{D},\rho{=}1,\nu{=}1$]{\label{work06:fig:ratio_kD_rho1_nu1}{\includegraphics[width=0.23\textwidth]{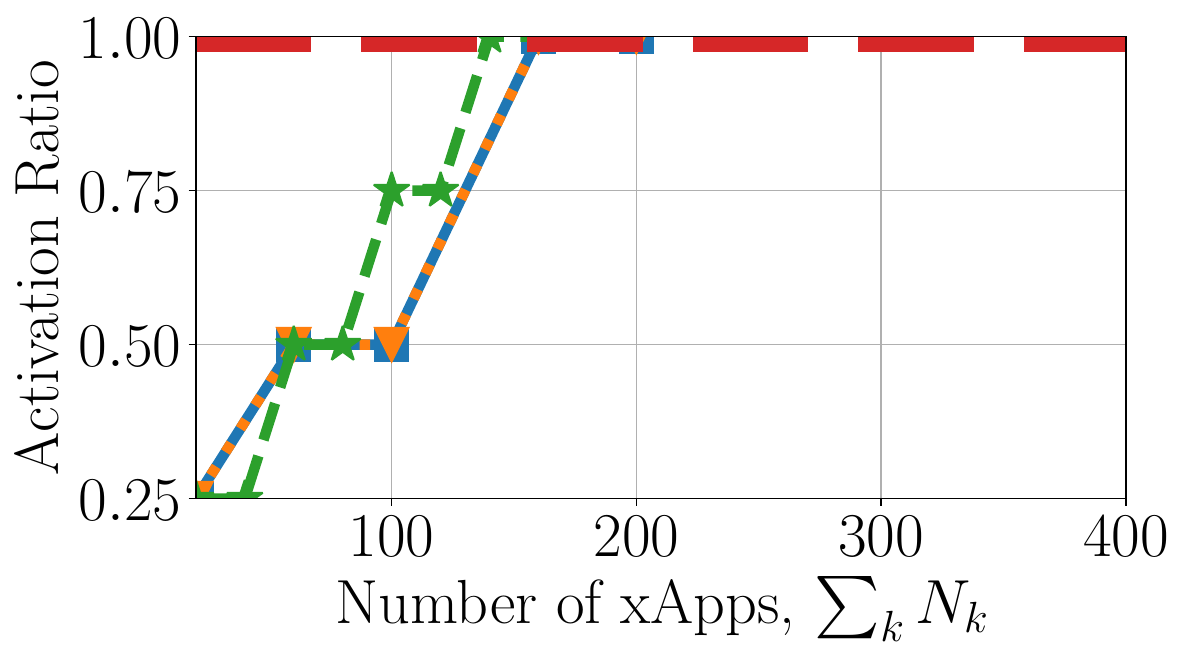}}}
    \vspace{2mm}
    \newline
    \includegraphics[width=0.75\columnwidth]{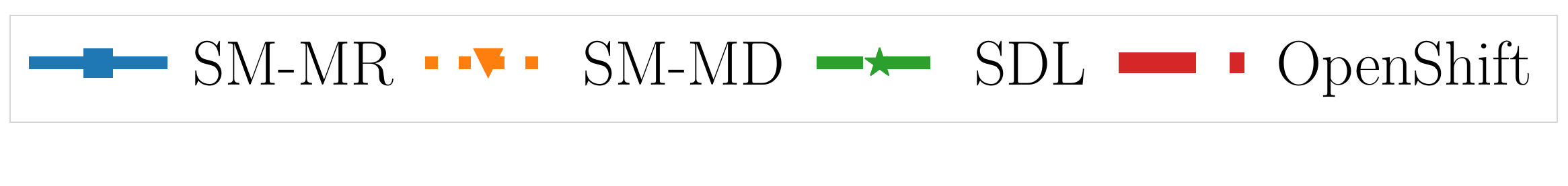}
    \caption{\rev{\novelty energy consumption reduction with respect to the OpenShift default scheduler and servers activation ratio for varying: (i) dominant xApp class (75\% distribution); (ii) xApp state size $\rho$; (iii) maintenance period $\nu$; (iv) migration strategy.}}
    \label{work06:fig:opt_energy_ratio}
\end{figure*}

\begin{figure*}[hbt!]
    \centering
    \subfloat[][$k{=}\mathrm{A}$, $\nu{=}1\,\mathrm{s}$]{\label{work06:fig:feasibility_sdl_kA_nu1}{\includegraphics[width=0.23\textwidth]{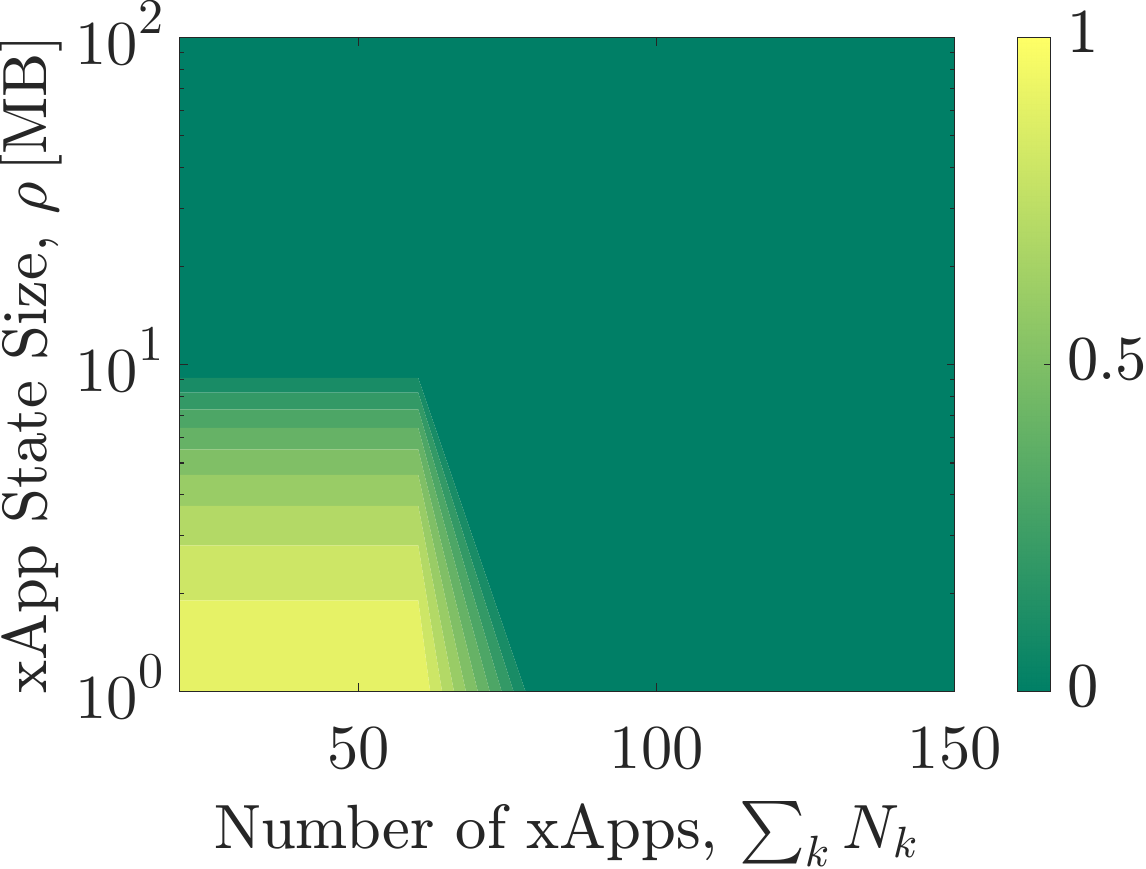}}}
    \hspace{2mm}
    \subfloat[][$k{=}\mathrm{B}$, $\nu{=}1\,\mathrm{s}$]{\label{work06:fig:feasibility_sdl_kB_nu1}{\includegraphics[width=0.23\textwidth]{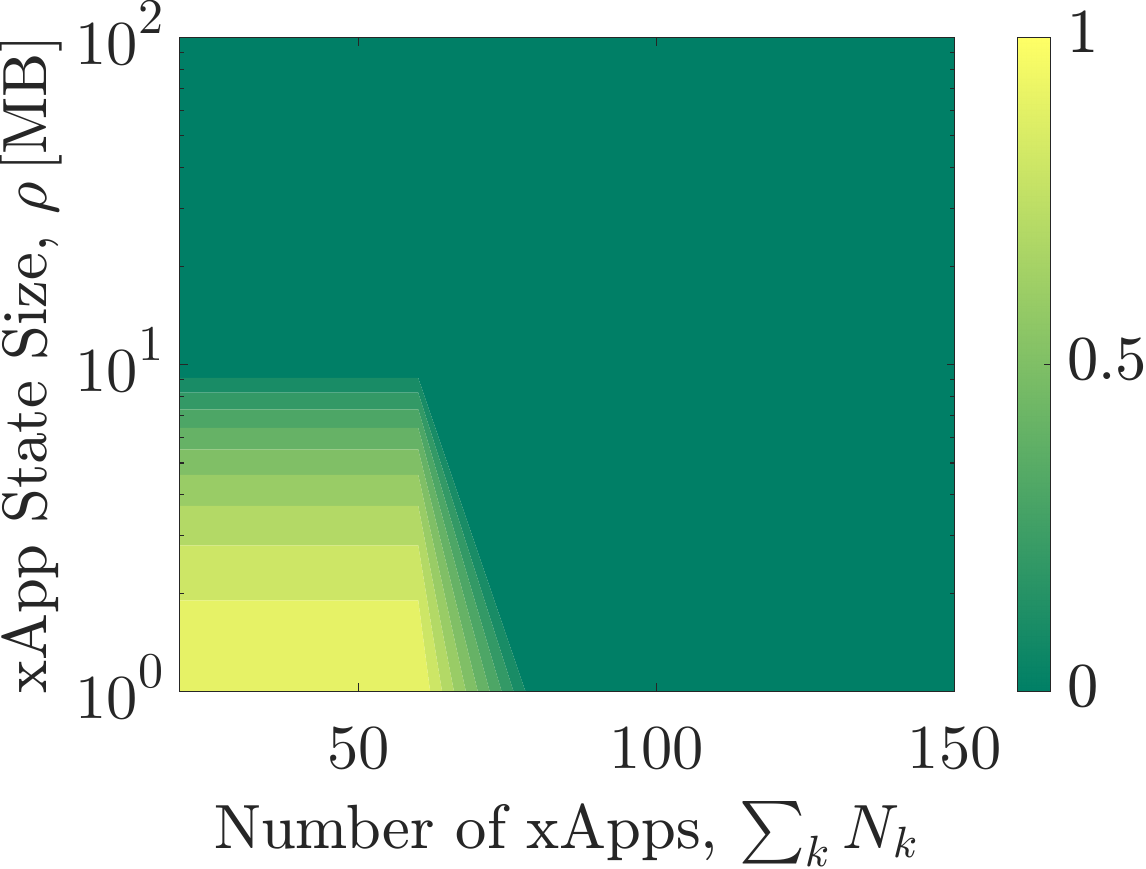}}}
    \hspace{2mm}
    \subfloat[][$k{=}\mathrm{C}$, $\nu{=}1\,\mathrm{s}$]{\label{work06:fig:feasibility_sdl_kC_nu1}{\includegraphics[width=0.23\textwidth]{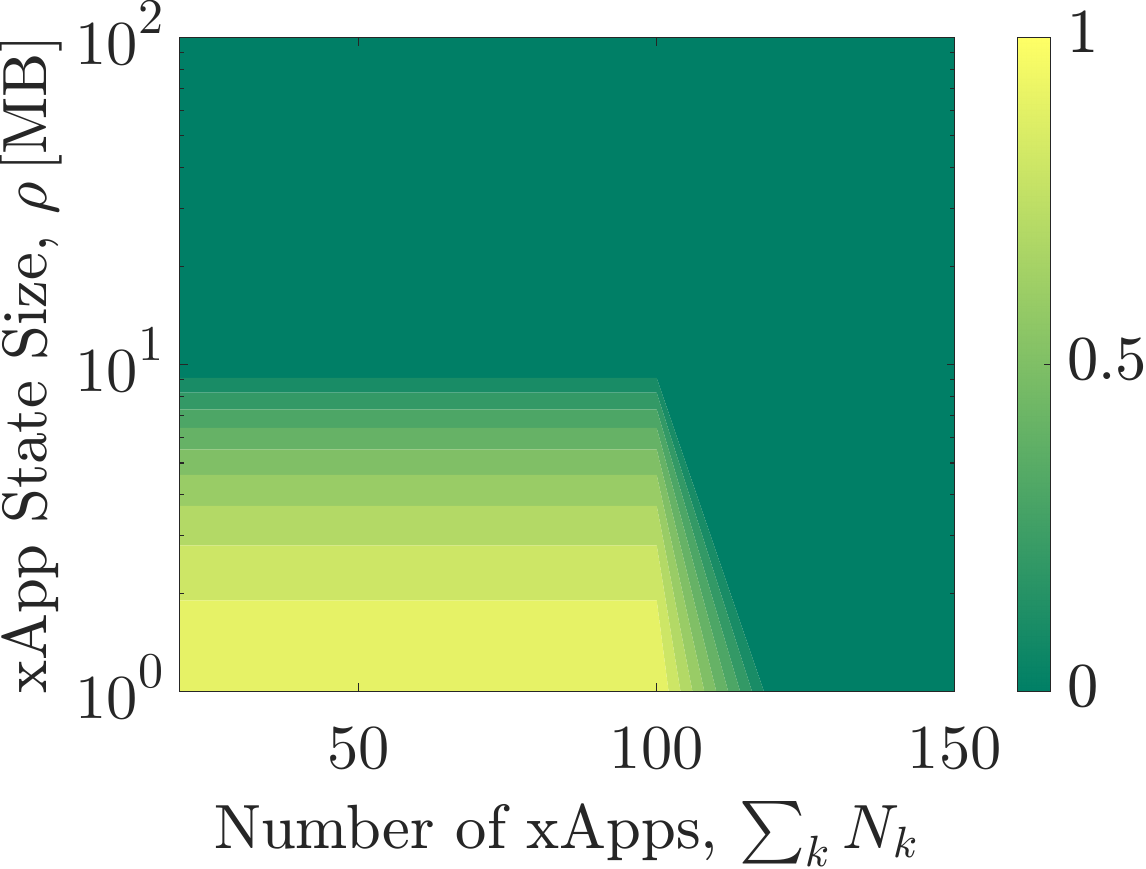}}}
    \hspace{2mm}
    \subfloat[][$k{=}\mathrm{D}$, $\nu{=}1\,\mathrm{s}$]{\label{work06:fig:feasibility_sdl_kD_nu1}{\includegraphics[width=0.23\textwidth]{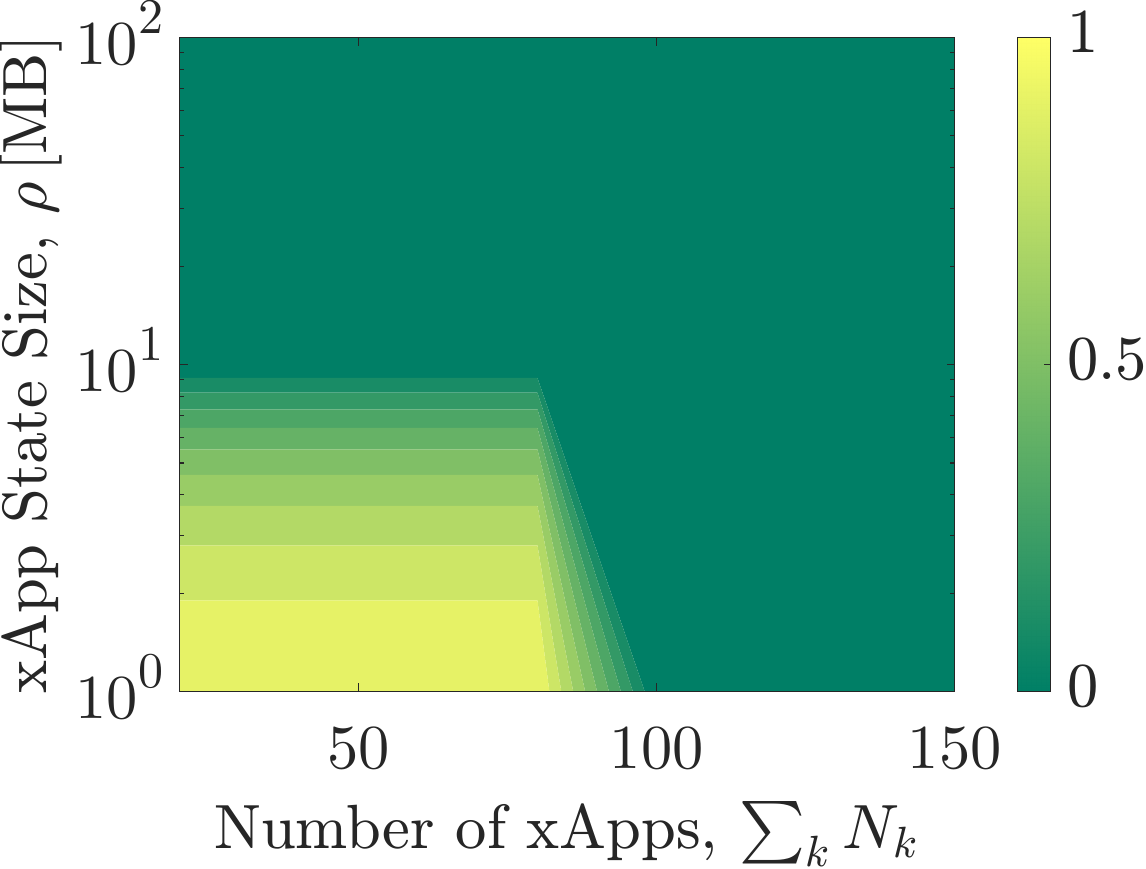}}}
    \\
    \subfloat[][$k{=}\mathrm{A}$, $\nu{=}120\,\mathrm{s}$]{\label{work06:fig:feasibility_sdl_kA_nu120}{\includegraphics[width=0.23\textwidth]{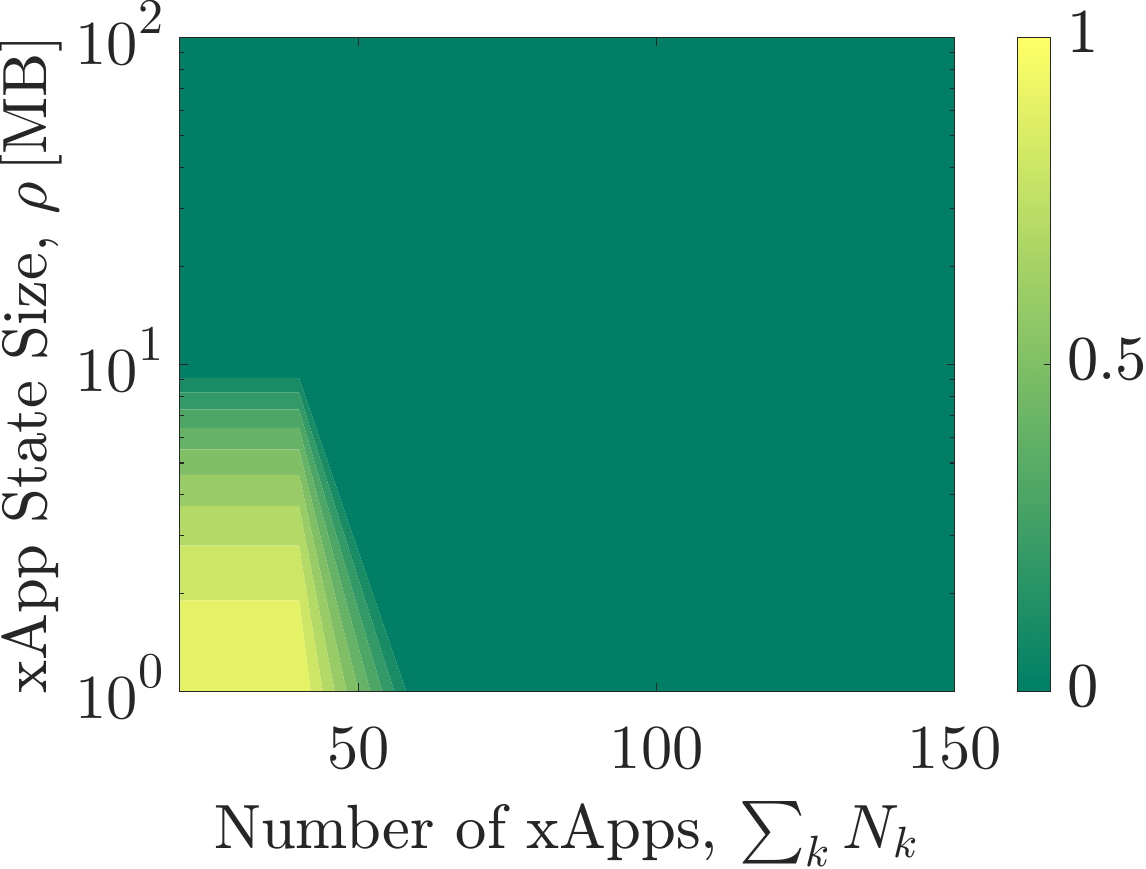}}}
    \hspace{2mm}
    \subfloat[][$k{=}\mathrm{B}$, $\nu{=}120\,\mathrm{s}$]{\label{work06:fig:feasibility_sdl_kB_nu120}{\includegraphics[width=0.23\textwidth]{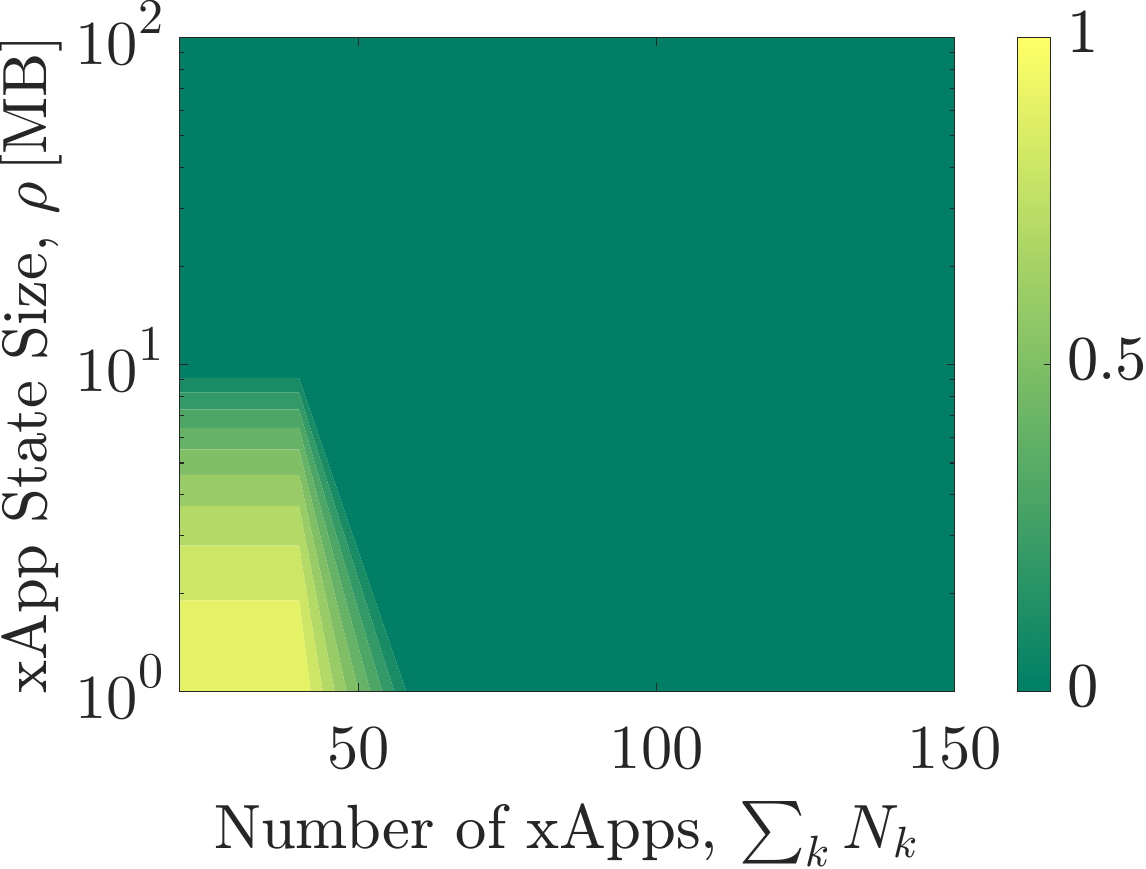}}}
    \hspace{2mm}
    \subfloat[][$k{=}\mathrm{C}$, $\nu{=}120\,\mathrm{s}$]{\label{work06:fig:feasibility_sdl_kC_nu120}{\includegraphics[width=0.23\textwidth]{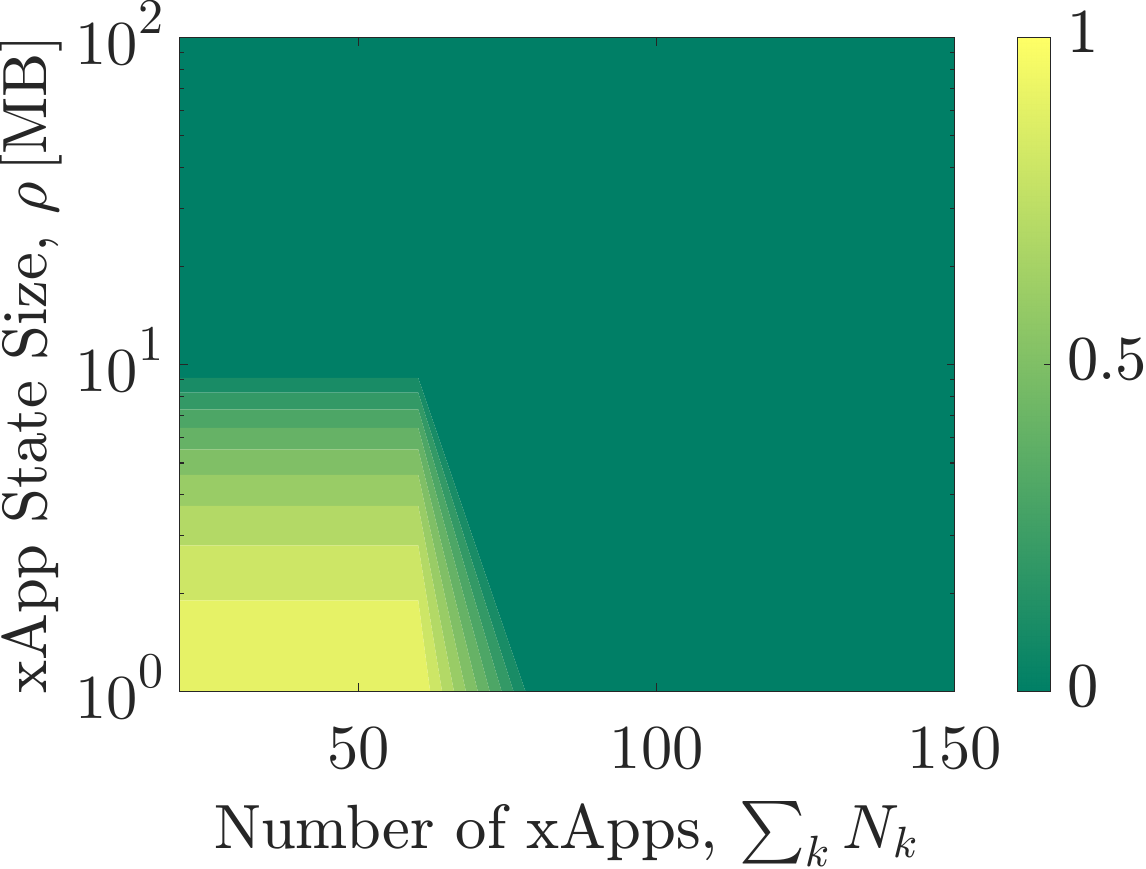}}}
    \hspace{2mm}
    \subfloat[][$k{=}\mathrm{D}$, $\nu{=}120\,\mathrm{s}$]{\label{work06:fig:feasibility_sdl_kD_nu120}{\includegraphics[width=0.23\textwidth]{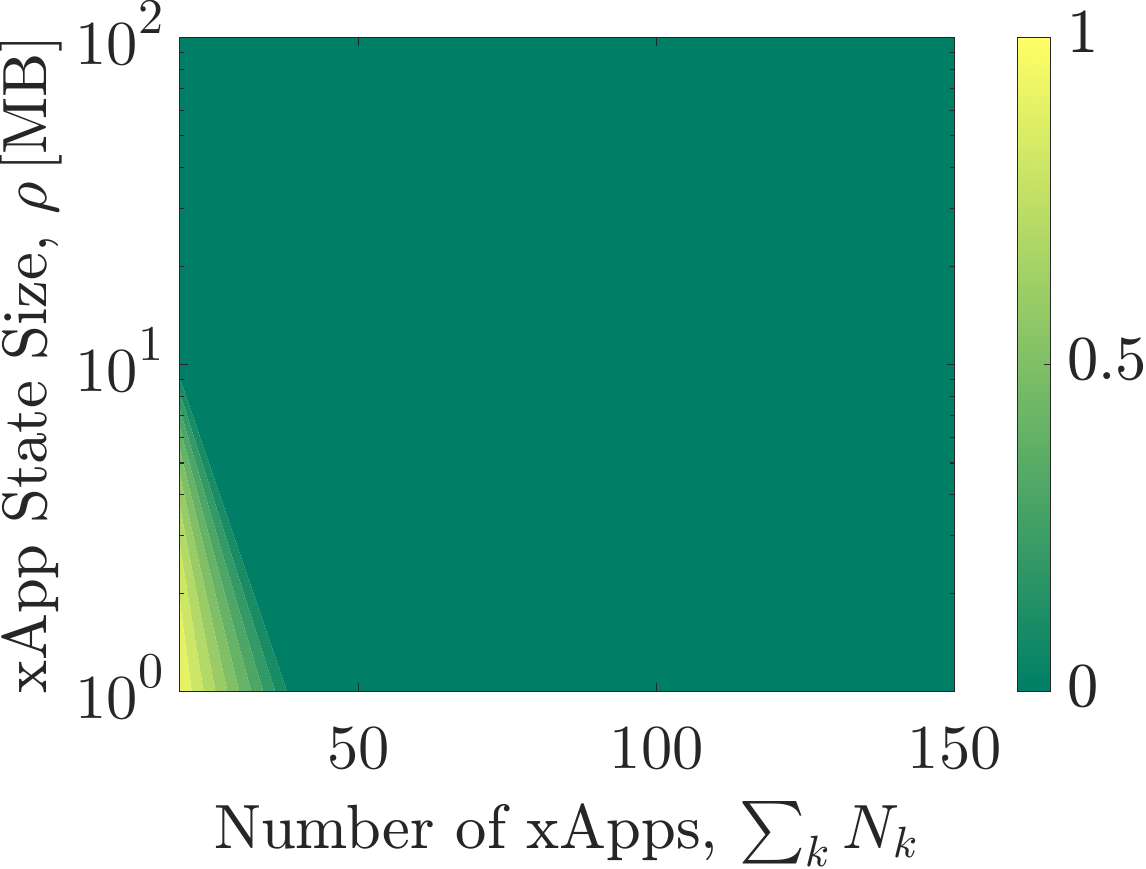}}}
    \caption{\rev{\novelty feasibility analysis for SDL-based migration and varying class $k$, state size $\rho$ and maintenance period $\nu$.}}
    \label{work06:fig:sdl_opt_feasibility}
    \vspace{-3mm}
\end{figure*}

\begin{figure}[ht!]
    \centering
    \subfloat[][SM-MR, $\forall k, \rho, \nu$]{\label{work06:fig:feasibility_sm_mr}{\includegraphics[width=0.23\textwidth]{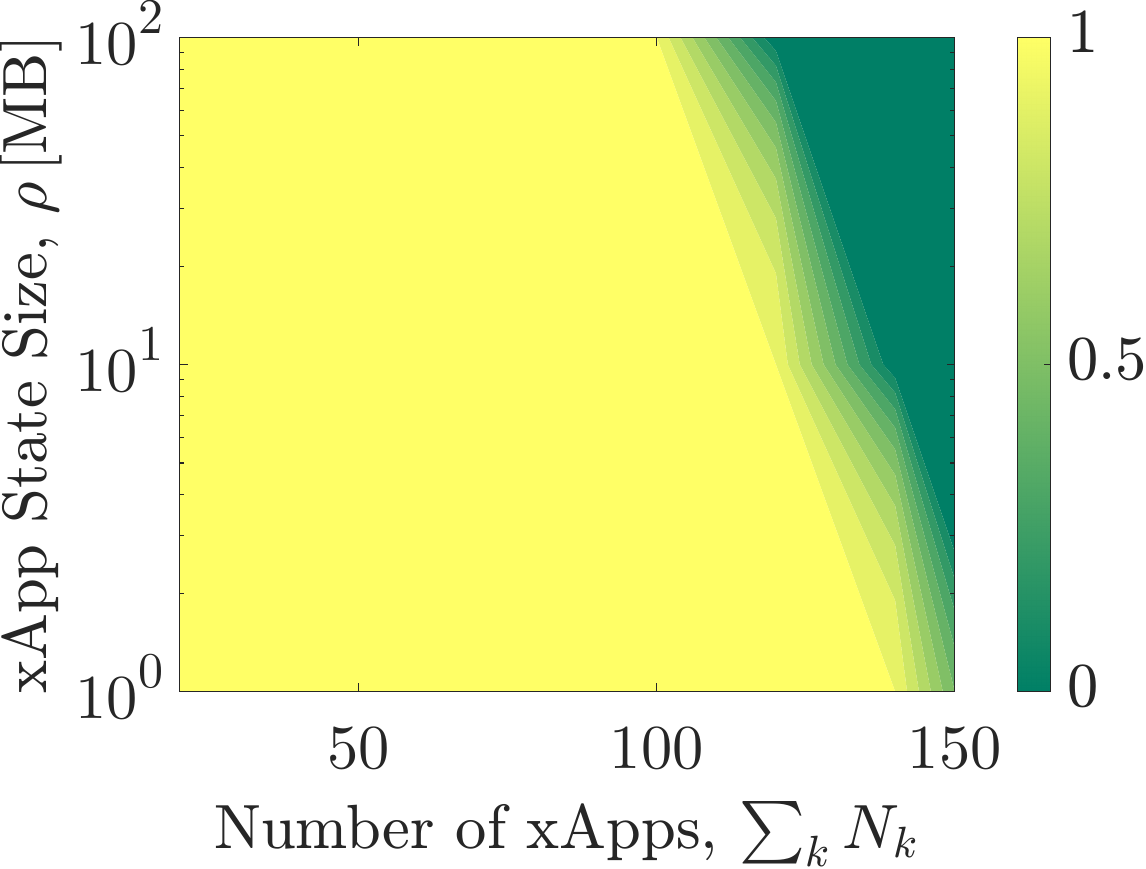}}}
    \hspace{2mm}
    \subfloat[][SM-MD, $\forall k, \rho, \nu$]{\label{work06:fig:feasibility_sm_md}{\includegraphics[width=0.23\textwidth]{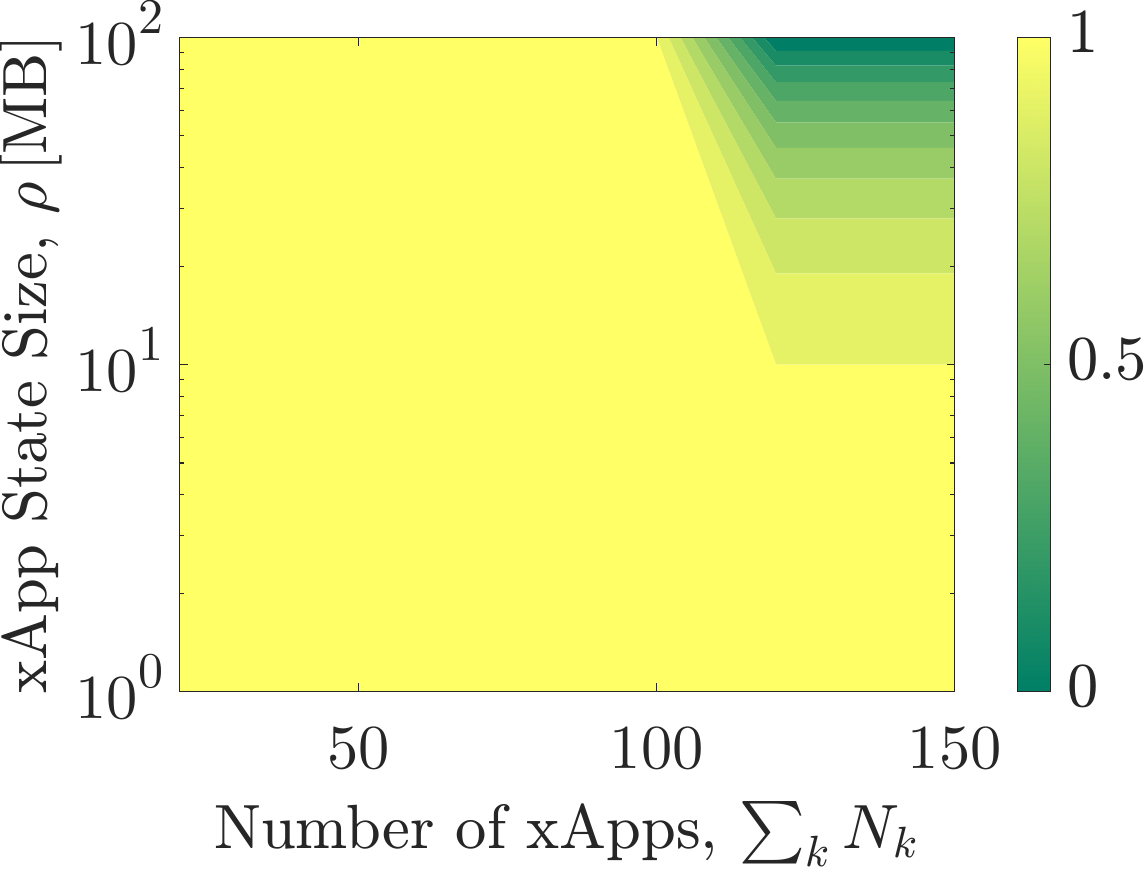}}}
    \caption{\rev{\novelty feasibility analysis under (a) SM-MR and (b) SM-MD for varying xApp state size $\rho$.}}
    \label{work06:fig:sm_opt_feasibility}
    \vspace{-3mm}
\end{figure}

Fig.~\ref{work06:fig:opt_energy_ratio} shows the energy gain and servers activation ratio as functions of the migration strategy and for varying configurations of dominant xApp class $k$, xApp state size $\rho$ and maintenance period $\nu$. \rev{We compute the energy gain with respect to a baseline in which all compute servers remain permanently active and xApps are placed according to the resource-based load balancing scheme that is natively implemented in OpenShift---a widely adopted approach in practice and frequently considered in the literature~\cite{vaquero2011dynamically,bauer2019chamulteon,gulati2011cloud}. We recall that, to the best of our knowledge, no prior work addresses xApp migration in the O-RAN context; therefore, the chosen baseline both reflects the state of the art and provides a well-grounded reference for quantifying the benefits introduced by our framework.} It can be observed that, by turning off compute servers that are not required during low traffic periods, \novelty attains a significant reduction in energy consumption. As the number of xApps grows, a higher number of active servers is needed, yielding an increased activation ratio and a reduced energy gain. Such gain approaches 0\% when the activation ratio is 1, i.e., same energy consumption as the baseline. Notably, both energy gain and activation ratio strongly depend on the configuration that is set: (i) as the dominant xApp class changes from low to high demanding, e.g., from A to B (Fig.~\ref{work06:fig:energy_kA_rho1_nu1} vs Fig.~\ref{work06:fig:energy_kB_rho1_nu1}) or from C to D (Fig.~\ref{work06:fig:energy_kC_rho1_nu1} vs Fig.~\ref{work06:fig:energy_kD_rho1_nu1}), the energy gain decreases with higher pace and a fewer number of xApps can be hosted due to constraint~\eqref{work06:eq:con8}, i.e., the one on the resource usage; (ii) looking at, e.g., Fig.~\ref{work06:fig:energy_kA_rho1_nu1}
and Fig.~\ref{work06:fig:energy_kA_rho100_nu1}, the larger $\rho$, the smaller the number of xApps that can be migrated compatibly with the maximum downtime (see constraint~\eqref{work06:eq:con9}); (iii) comparing, e.g., Fig.~\ref{work06:fig:energy_kA_rho1_nu1} and Fig.~\ref{work06:fig:energy_kA_rho1_nu120}, when the value of $\nu$ increases from 1\,s to 120\,s, the cost due to SDL maintenance is reduced, yielding a higher energy gain and lower activation ratio; and (iv) in general, comparing to \gls{sdl}, \gls{sm} strategies achieve higher values of energy gain (up to 64\%) as they do not require the additional cost to host and maintain the \gls{sdl} backend database.

\noindent
Figures~\ref{work06:fig:sdl_opt_feasibility} and~\ref{work06:fig:sm_opt_feasibility} show the feasibility region of, respectively, \gls{sdl} and \gls{sm} migration strategies for varying configurations of dominant xApp class $k$, xApp state size $\rho$ and maintenance period $\nu$ under $T^{\mathrm{max}}_{D_k}{=}300\,\mathrm{s}$ and $T^{\mathrm{max}}_{\mathrm{DF}}{=}1\,\mathrm{s}$. To compute such regions we enforce \eqref{work06:eq:con8} for \gls{sm}, and, \eqref{work06:eq:con10} and \eqref{work06:eq:con11} for \gls{sdl}.
Fig.~\ref{work06:fig:sdl_opt_feasibility} demonstrates that, due to scalability limits, the feasibility of \gls{sdl}-based migration strongly depends on the values of $\rho$ and $\nu$: (i) when $\rho$ increases, the maximum number of xApps that \gls{sdl} can host (compatibly with the near-RT \gls{ric} strict timing requirements) decreases, up to $\rho{=}100\,\mathrm{MB}$ for which no configuration is actually feasible, regardless of the number of xApps and the dominant class; and (ii) when $\nu$ increases, despite the higher energy gain observed in Fig.~\ref{work06:fig:opt_energy_ratio}, the maximum number of xApps significantly decreases, due to higher values of the defrag downtime that lead to near-RT \gls{ric} deadline violation.
On the other hand, Fig.~\ref{work06:fig:sm_opt_feasibility} shows that \gls{sm} is way more feasible, allowing also for the extreme scenario of $\rho{=}100\,\mathrm{MB}$, and SM-MD attains higher feasibility values thanks to migration downtime minimization. We recall that, despite its limited feasibility, \gls{sdl} is the only strategy that enables zero-downtime migration process. \gls{sm}, instead, implies a migration downtime that is way above the near-RT \gls{ric} deadline and needs to be accounted for (see Fig.~\ref{work06:fig:sm_vs_sdl_downtime_vs_num_xapps}).

Thus, we conclude that \novelty effectively addresses the trade-off among service availability, scalability, and energy consumption. In the case of large deployments all servers need to be active and \novelty has no significant impact on the energy consumption. On the other hand, when the traffic load is low and the number of xApps is small, e.g., at nighttime, \novelty allows to identify, for varying system configurations, which migration strategy is feasible and its effectiveness in reducing the overall energy consumption, yielding a cost reduction that is up to 64\%.

\section{Conclusions}\label{work06:sec:conclusions}
In this paper, we proposed \novelty, a data-driven orchestrator that jointly optimizes the activation of near-RT \gls{ric} compute nodes and the migration of stateful xApps to minimize the overall system energy consumption, while ensuring uninterrupted xApp control.
We first introduced the two key technologies for preserving the xApp internal state upon migration, i.e., \gls{sm} and \gls{sdl}, while accounting for the O-RAN context and time constraints.
Then, we leveraged our experimental testbed based on Red Hat OpenShift to perform a thorough temporal \glspl{kpi} and resource usage analysis under both migration strategies and varying use case scenarios, revealing pivotal trade-offs involving resource usage, scalability, and service availability.
Our results demonstrate that \novelty accurately identifies feasibility and effectiveness of each migration strategy and computes the optimal xApp allocations across the available compute nodes, yielding up to 64\% reduction of the system energy consumption.

\balance

\bibliographystyle{IEEEtran}
\bibliography{myBibliography}

\begin{IEEEbiographynophoto}
{Antonio Calagna} is a Postdoctoral Researcher with Politecnico di Torino, Italy. He received the B.Sc. degree in Electronics Engineering in 2019, the M.Sc. degree in Communication and Computer Networks Engineering in 2021, and the Ph.D. degree (cum laude) in Electrical, Electronics and Communications Engineering in 2025, all from Politecnico di Torino. His research interests include the orchestration and management of AI-driven edge services for next-generation mobile networks, with an emphasis on time-sensitive resource allocation and service optimization. He received the Marie Skłodowska-Curie Actions (MSCA) Seal of Excellence in 2025.
\end{IEEEbiographynophoto}
\vskip -2.5\baselineskip plus -1fil
\begin{IEEEbiographynophoto}
{Stefano Maxenti} is a Ph.D. Candidate at the Institute for the Wireless Internet of Things at Northeastern University, under Prof. Tommaso Melodia. He received a Bachelor's degree in Engineering of Computing Systems in 2020 and a Master of Science degree in Telecommunication Engineering from Politecnico di Milano, Italy. He is interested in AI applications for wireless communications and orchestration, integration, and automation of O-RAN networks.
\end{IEEEbiographynophoto}
\vskip -2.5\baselineskip plus -1fil
\begin{IEEEbiographynophoto}
{Leonardo Bonati} is an Associate Research Scientist at the Institute for the Wireless Internet of Things, Northeastern University, Boston, MA, USA. He received a Ph.D. degree in Computer Engineering from Northeastern University in 2022. His main research focuses on softwarized approaches for the Open Radio Access Network (RAN) of the next generation of cellular networks, on O-RAN-managed networks, and on network automation, orchestration, and virtualization. He was awarded the 2024 Mario Gerla Award for Research in Computer Science. Leonardo was TPC co-chair for the IEEE DTwin 2025 workshop, a co-chair for the track on Testbeds, Experimentation and Datasets for Communications and Networking of IEEE CCNC 2025, and a guest editor of the special issue of Elsevier Computer Networks on Advances in Experimental Wireless Platforms and Systems.
\end{IEEEbiographynophoto}
\vskip -2.5\baselineskip plus -1fil
\begin{IEEEbiographynophoto}
{Salvatore D'Oro} is a Research Associate Professor at Northeastern University, CTO and co-founder of zTouch Networks. He received his Ph.D. degree from the University of Catania and is an area editor of IEEE Vehicular Technology Magazine and Elsevier Computer Communications. He serves on the TPC of IEEE INFOCOM, IEEE CCNC \& ICC and IFIP Networking. He is one of the contributors to OpenRAN Gym, the first open-source research platform for AI/ML applications in the Open RAN. His research interests include optimization, AI \& network slicing for NextG Open RANs.
\end{IEEEbiographynophoto}
\vskip -2.5\baselineskip plus -1fil
\begin{IEEEbiographynophoto}
{Tommaso Melodia} is the William Lincoln Smith Chair Professor with the Department of Electrical and Computer Engineering at Northeastern University in Boston. He is also the Founding Director of the Institute for the Wireless Internet of Things and the Director of Research for the PAWR Project Office. He received his Ph.D. in Electrical and Computer Engineering from the Georgia Institute of Technology in 2007. He is a recipient of the National Science Foundation CAREER award. Prof. Melodia has served as Associate Editor of IEEE Transactions on Wireless Communications, IEEE Transactions on Mobile Computing, Elsevier Computer Networks, among others. He has served as Technical Program Committee Chair for IEEE INFOCOM 2018, General Chair for IEEE SECON 2019, ACM Nanocom 2019, and ACM WUWnet 2014. Prof. Melodia is the Director of Research for the Platforms for Advanced Wireless Research (PAWR) Project Office, a \$100M public-private partnership to establish four city-scale platforms for wireless research to advance the US wireless ecosystem in years to come. Prof. Melodia's research on modeling, optimization, and experimental evaluation of Internet-of-Things and wireless networked systems has been funded by the National Science Foundation, the Air Force Research Laboratory the Office of Naval Research, DARPA, and the Army Research Laboratory. Prof. Melodia is a Fellow of the IEEE and a Distinguished Member of the ACM.
\end{IEEEbiographynophoto}
\begin{IEEEbiographynophoto}
{Carla Fabiana Chiasserini} is currently a Full Professor with the Department of Electronics and Telecommunications Engineering at Politecnico di Torino, Italy, a WASP Guest Professor at Chalmers University of Technology, Sweden, and a Research Associate with the Italian National Research Council (CNR) and CNIT. She was a visiting researcher with UCSD, a visiting professor with Monash University, Technische Berlin University, and HPI at Potsdam University. Her research interests include 5G-and-beyond networks, NFV, mobile edge computing, connected vehicles, and distributed machine learning at the network edge.
\end{IEEEbiographynophoto}
\end{document}